\RequirePackage{silence}
\documentclass[fleqn,usenatbib,useAMS]{mnras}

\usepackage{newtxtext,newtxmath}

\usepackage{graphicx}
\usepackage{amsmath}
\DeclareMathOperator{\arcsinh}{arcsinh}
\usepackage{amsfonts}
\usepackage{float}
\usepackage{bm}
\usepackage{ae,aecompl}
\usepackage{array}
\usepackage{soul}
\usepackage{mathtools}
\usepackage{multirow}
\usepackage[T1]{fontenc}
\usepackage[utf8]{inputenc}
\usepackage{booktabs}
\usepackage{graphicx}
\usepackage{subcaption}
\usepackage[normalem]{ulem}
\usepackage{anyfontsize}
\usepackage{xcolor}

\DeclareRobustCommand{\appropto}{\mathrel{\vcenter{
		\offinterlineskip\halign{\hfil$##$\cr 
			\propto\cr\noalign{\kern2pt}\sim\cr\noalign{\kern-2pt}}}}}

\DeclareRobustCommand{\AU}{A_{\mathrm{U}}}
\DeclareRobustCommand{\AGC}{A_{\mathrm{GC}}}
\DeclareRobustCommand{\AUCMB}{A_{\mathrm{U}}^{\mathrm{CMB}}}
\DeclareRobustCommand{\AstarCMB}{A_{\star}^{\mathrm{CMB}}}
\DeclareRobustCommand{\AUETS}{A_{\mathrm{U}}^{\mathrm{ETS}}}
\DeclareRobustCommand{\AstarETS}{A_{\star}^{\mathrm{ETS}}}

\DeclareRobustCommand{\H0CMB}{H_0^{\mathrm{CMB}}}
\DeclareRobustCommand{\HCMBz}{H^{\mathrm{CMB}}(z)}
\DeclareRobustCommand{\OmegaM}{\Omega_{\mathrm{m}}}
\DeclareRobustCommand{\tf}{t_{\mathrm{f}}}
\DeclareRobustCommand{\LX22}{\mathcal{L}}
\DeclareRobustCommand{\alphaML}{\alpha_{_\mathrm{ML}}}
\DeclareRobustCommand{\alphaY}{\alpha_{_\mathrm{Y}}}
\DeclareRobustCommand{\Act}{\mathcal{A}_{\mathrm{ct}}}

\DeclareRobustCommand{\DFeH}{\Delta \left[ \mathrm{Fe}/\mathrm{H} \right]}
\DeclareRobustCommand{\DalphaFe}{\Delta \left[ \alpha/\mathrm{Fe} \right]}
\DeclareRobustCommand{\Jtz}{\tilde{J}_z}
\DeclareRobustCommand{\Jtzo}{\tilde{J}_{z,0}}
\DeclareRobustCommand{\DJtz}{\Delta \tilde{J}_z}

\defcitealias{Xiang_2022}{X22}

\title[The cosmic age from LAMOST DR7 and \emph{Gaia} stars]{The age of the Universe from a large sample of the oldest Galactic stars} 

\author[I. Banik et al.]{Indranil Banik$^{1}$\thanks{E-mail: \href{mailto:indranil.banik@port.ac.uk}{indranil.banik@port.ac.uk} (Indranil Banik); \newline \hspace*{1.9em}
\href{mailto:stephen.cookson@sca-uk.com}{stephen.cookson@sca-uk.com} (Stephen Cookson); \newline \hspace*{4em}
\href{mailto:harry.desmond@port.ac.uk}{harry.desmond@port.ac.uk} (Harry Desmond)}, Thenujaya Kudakolawa Kaluarachchige$^{2}$, Stephen Cookson$^{3}$ and Harry Desmond$^{1}$\\
$^{1}$Institute of Cosmology and Gravitation, University of Portsmouth, Dennis Sciama Building, Burnaby Road, Portsmouth PO1 3FX, UK\\
$^{2}$Scottish Universities Physics Alliance, University of Saint Andrews, North Haugh, Saint Andrews, Fife, KY16 9SS, UK\\
$^{3}$Independent researcher}

\date{Accepted XXX. Received YYY; in original form ZZZ}

\pubyear{\the\year{}}

\begin{document}
\label{firstpage}
\pagerange{\pageref{firstpage}--\pageref{lastpage}}
\maketitle

\begin{abstract} 
We estimate the age of the Universe using the Xiang \& Rix sample of 247,103 Milky Way stars with low-resolution spectroscopy from LAMOST~DR7 and \emph{Gaia}~eDR3 parallaxes. Stellar ages were estimated using YY isochrones up to 20~Gyr. We require consistency between YY ages and photometric FLAME ages based only on \emph{Gaia} data. To remove a population of stars with unusually low vertical action for their age, metallicity, and distance, we require old stars to be metal-poor and $\alpha$-enriched. Our final sample of 146,810 stars within 5~kpc provides consistent cosmic age estimates using several techniques of increasing rigour. Our main results use an MCMC reconstruction of the latent age distribution, though our iterative reconstruction is very similar. Applying an innovative approach to our MCMC reconstruction and its uncertainties, we find that the oldest star has an age of $A_\star = 13.80^{+0.20}_{-0.16}$~Gyr. We conservatively estimate a systematic uncertainty of $\pm 0.4$~Gyr due to the choice of quality cuts, isochrone model, and initial helium abundance. Allowing 0.2~Gyr for the first long-lived stars to form, our inferred $A_\star$ is consistent with the 13.6~Gyr expected in CMB-calibrated $\Lambda$CDM. This agreement casts doubt on solutions to the Hubble tension solely through new physics prior to recombination, which generally imply a cosmic age of $12.9 \pm 0.2$~Gyr to match low redshift probes. It is difficult for stellar modelling uncertainties to reconcile such a low age with our result given the low metallicities of the oldest stars in our sample and independent asteroseismic constraints.


\end{abstract}


\section{Introduction}
\label{Introduction}

Cosmology generally seeks to understand the history of the Universe by directly looking into the past, exploiting the finite speed of light $c$. This way of learning about our past is unique to the astronomical sciences. Most fields of study rely on fossil clues that preserve conditions from ancient times. In an astrophysical context, this would entail learning about the past by exploring ancient objects in the Solar neighbourhood. The most obvious such objects are stars. Indeed, studies of old stars have given rise to the field of Galactic archaeology \citep[e.g.][]{Belokurov_2013, Deason_2024, Almannei_2026}. While this usually concerns itself with the history of our Galaxy, it can also provide important information about the history of the Universe as a whole. In particular, the ages of the oldest stars provide a lower limit to $\AU$, the age of the Universe.

The first long-lived stars in the Milky Way are generally assumed to have formed approximately $\tf = 0.2$~Gyr after the Big Bang \citep{Cimatti_2023, Valcin_2025, Tomasetti_2026, Valcin_2026}. This timescale is consistent with both theoretical models of early structure formation and observations of galaxies at redshifts $z \ga 14$ with the \emph{JWST}, which demonstrate that star formation was already underway within the first few hundred Myr of cosmic history \citep{Carniani_2024}. Although these observations do not imply that the progenitor of the Milky Way was itself among the earliest galaxies, they show that the interval between the Big Bang and the formation of the first long-lived stars was short compared with $\AU$. Adopting $\tf = 0.2$~Gyr therefore provides a reasonable and conservative allowance for the time required to form the first long-lived stellar populations. This allows us to estimate that
\begin{eqnarray}
    \AU ~=~ A_\star + \tf \, ,
    \label{tf_allowance}
\end{eqnarray}
where $A_\star$ is the latent or true age of the oldest star in a large sample. Given that $A_\star > 10$~Gyr \citep{Knox_1999, Bedin_2025}, even reducing $\tf$ to zero would only decrease $\AU$ by $<2\%$. Conversely, larger values of $\tf$ would increase the inferred $\AU$, which we will see later would strengthen our main conclusion.

Many of the outstanding questions in cosmology can be addressed if we know two key quantities, namely $\AU$ and the Hubble-Lema\^{\i}tre constant $H_0$, the present value of the Hubble-Lema\^{\i}tre parameter $H \equiv \dot{a}/a$, where $a$ is the cosmic scale factor normalised to unity today, an overdot denotes a derivative with respect to time $t$ since the Big Bang, and 0 subscripts denote present-day quantities. The dimensionless combination $H_0 \AU$ tells us about the shape of the cosmic expansion history, or $a(t/\AU)$. In a matter-dominated universe, we expect that $H_0 \AU = 2/3$. Larger values require a corresponding reduction in the past expansion rate, with sufficiently high $H_0 \AU$ requiring an exotic component driving the Universe towards $\ddot{a} > 0$ at late times. This unexpectedly became apparent in the mid-1990s from combining $H_0$ measurements with the ages of old Galactic globular clusters \citep[GCs;][]{Bolte_1995, Jimenez_1996}. These observations led to the reintroduction of a cosmological constant into the field equations of General Relativity \citep[GR;][]{Einstein_1915, Einstein_1917}, forming a central component of the currently standard Lambda cold dark matter ($\Lambda$CDM) cosmological paradigm \citep*{Efstathiou_1990, Ostriker_Steinhardt_1995}.

Understanding stars is the oldest branch of astronomy, a fact readily apparent from the name of a field that today concerns itself with a wide variety of other issues. Our theory of stellar evolution is calibrated using radioactive dating of the oldest meteorite samples, which provide a `ground truth' that our Solar System has an age of $A = 4.567$~Gyr with negligible uncertainty \citep{Connelly_2012}. This is in line with asteroseismic age dating of the Sun \citep{Guenther_1998, Bonanno_2020, Betrisey_2024}. It is much easier to find the age of a star once it has left the main sequence and entered the subgiant phase, when stellar parameters change much more rapidly with age (see figure~1 of \citealt{Xiang_2022}, hereafter \citetalias{Xiang_2022}). This means the stars most relevant to determining $A_\star$ have a lifetime of around 14~Gyr, so their mass $M_\star \approx 0.8 \, M_\odot$, only very slightly below that of the Sun. Clearly, there is every reason to expect that such stars are well understood. Many of them are also well observed thanks to ongoing surveys like \emph{Gaia} \citep{Perryman_2001}, which provides accurate parallaxes and therefore absolute magnitudes in combination with photometric measurements.

In this contribution, we determine $A_\star$ from a large sample of old Galactic stars with heliocentric distance $D_\odot < 5$~kpc observed with the Large Sky Area Multi-Object Fiber Spectroscopic Telescope \citep[LAMOST;][]{Cui_2012, Zhao_2012}.\footnote{Also known as the  Guo Shoujing Telescope.} Spectra from LAMOST Data Release~7 (DR7) were combined with astrometric information from \emph{Gaia} early Data Release 3 \citep[eDR3;][]{Gaia_2021} by \citetalias{Xiang_2022} to derive ages for 247,103 stars that also pass additional quality cuts specifically targeting old stars. Besides the large sample size, another major advantage of their catalogue is its weak cosmological prior of $A_\star < 20$~Gyr, which comes from the 20~Gyr limit to the Yonsei--Yale (YY) isochrone grid \citep{Yi_2001, Demarque_2004}. This is crucial to our goal of inferring $A_\star$ with minimal biases of our own, especially cosmological biases. Since our results rely on standard stellar evolution codes, they are not applicable to models where fundamental constants of nature change over time or depend on the environment. They are also best understood as placing a lower limit on $\AU$, though the gap between its predicted value in some model and our inferred $A_\star$ needs to be reasonable in the context of that model.

A measurement of $\AU$ is important because it can be precisely predicted in a variety of cosmological models once their parameters are calibrated in other ways. By far the most precise constraints on the $\Lambda$CDM parameters come from the angular power spectrum of the 0.001\%-level anisotropies in the cosmic microwave background \citep[CMB;][]{Aiola_2020, Planck_2020, Tristram_2024, Calabrese_2025, SPT_2025, SPT_2026}. Using both terrestrial and space-based observations of the CMB, $\Lambda$CDM predicts a cosmic age of $\AUCMB = 13.8$~Gyr with negligible uncertainty \citep[see table~1 of][]{SPT_2026}. Allowing 0.2~Gyr for the first long-lived stars to form, we therefore expect to observe stars with ages up to $\AstarCMB = 13.6$~Gyr (Equation~\ref{tf_allowance}).

However, there is a very real possibility of finding a different $A_\star$. This is because of the Hubble tension, a statistically significant mismatch between the CMB-derived value of $H_0^\mathrm{CMB}$ and $H_0^\mathrm{local}$, the local measurement obtained from the rate at which redshift rises with distance in the local Universe \citep{Valentino_2021_problem, Valentino_2025, Cai_2026}. The latter exploits the fact that observations along our past lightcone observe cosmic epochs when $a$ was smaller than today. A wide variety of distance indicators in the local Universe give consistent results for $H_0^\mathrm{local}$. Sometimes this is based on a distance ladder approach \citep{Scolnic_2023, Breuval_2024, Riess_2024_consistency, Freedman_2025, Jensen_2025, Li_2026}. It is also possible to use only a single type of astrophysical object, for instance megamasers \citep{Pesce_2020, Barua_2025, Stiskalek_2026_maser}, Type~II supernovae \citep[SNe~II;][]{Vogl_2025}, Cepheids \citep{Stiskalek_2025_Cepheids}, or the tip of the red giant branch \citep[TRGB;][]{Stiskalek_2026_TRGB}. Estimates of $H_0^\mathrm{local}$ have remained stable for a generation, with the modern value very similar to what it was a quarter century ago \citep{Freedman_2001}. These factors have led to an emergent ``consensus'' that $H_0^\mathrm{CMB}$ and $H_0^\mathrm{local}$ are discrepant at $>7\sigma$ confidence, with $H_0^\mathrm{local}$ exceeding $H_0^\mathrm{CMB}$ by 9\% \citep{H0DN_2026}.

This discrepancy may have existed over the vast majority of cosmic history, with the entire $H(z)$ curve perhaps 9\% higher than $\HCMBz$, the $\Lambda$CDM expectation calibrated to the CMB. Such a broad excess over $\HCMBz$ is assumed in proposals that try to solve the Hubble tension through new physics prior to recombination. The idea is that once the proposed new physics is included, cosmological parameter inference from the CMB leads to a revised `prediction' for $H_0$ in line with $H_0^\mathrm{local}$.\footnote{These are not \emph{a priori} predictions, but rather estimates of $H_0$ obtained without using $H_0^\mathrm{local}$ as a constraint.} For instance, it is possible that primordial magnetic fields (PMF) increase the redshift of recombination from the standard 1090 to about 1110, reducing the sound horizon at recombination and thereby requiring a reduced comoving distance to it, which entails higher $H_0$ \citep*{Mirpoorian_2025}. Such early-time solutions to the Hubble tension struggle to maintain a good CMB fit when the new physics is large enough to address the Hubble tension \citep{Calabrese_2025, SPT_2026}. Moreover, fitting the CMB typically requires a higher baryon density than in $\Lambda$CDM, creating difficulty explaining the primordial light element abundances \citep{Aver_2026, Giovanetti_2026, Launders_2026, Pettini_2026, Poulin_2026}. Given these are also sensitive to the expansion rate, modifications to it must be quite small in the first few minutes of the Universe, even though significant modifications later on are essential to reduce the sound horizon at recombination.

A generic consequence of these proposals is that a 9\% faster expansion rate all the way back to the infancy of the universe would reduce the standard value of $\AUCMB$ to $13.8/1.09 = 12.7$~Gyr, assuming the expansion history retains the same shape \citep{Boylan_Kolchin_2021}. However, this assumption need not be correct in an early-time solution to the Hubble tension because that necessarily requires new physics prior to recombination. This prevents us from relying on the precise CMB-derived value of $\OmegaM$, the fraction of the cosmic critical density presently in the matter component. We must instead determine $H_0$ and $\OmegaM$ solely from probes in the late Universe assuming $\Lambda$CDM at late times. $\AU$ can be calculated by inverting equation~45 of \citet*{Haslbauer_2020}.
\begin{eqnarray}
    H_0 \AU ~=~ \frac{2 \arcsinh \sqrt{\frac{1 - \OmegaM}{\OmegaM}}}{3\sqrt{1 - \OmegaM}} \, .
    \label{AU_analytic}
\end{eqnarray}
Taking $H_0 = 73.17 \pm 0.86$~km/s/Mpc \citep{Breuval_2024} and $\OmegaM = 0.2975 \pm 0.0086$ \citep{DESI_2025} and assuming independent uncertainties due to the completely different nature of these constraints, early-time solutions to the Hubble tension predict a cosmic age of $\AUETS = 12.91 \pm 0.18$~Gyr. The predicted age of the oldest star then becomes $\AstarETS = 12.71 \pm 0.18$~Gyr. We adopt this estimate for the rest of this contribution. However, it may actually be an overestimate because $H_0 \AU \appropto \OmegaM^{-0.28}$ \citep*[equation~35 of][]{Poulin_2023} and most late Universe studies assuming flat $\Lambda$CDM generally give a higher $\OmegaM$. For instance, assuming instead that $\OmegaM = 0.330 \pm 0.015$ based on the Dark Energy Survey (DES) Dovekie supernova catalogue \citep{Popovic_2026}, we would get that $\AUETS = 12.54 \pm 0.22$~Gyr and thus $\AstarETS = 12.34 \pm 0.22$~Gyr. This decreases to $\AstarETS = 12.29 \pm 0.21$~Gyr if we also adopt the ``community consensus'' value of $H_0 = 73.50 \pm 0.81$~km/s/Mpc \citep{H0DN_2026}. This is $>1$~Gyr lower than $\AstarCMB$.


Alternatively, it is possible that the Hubble tension is a more recent or local phenomenon \citep{Perivolaropoulos_2024, Jia_2026, Pantos_2026}. The required adjustment to $\Lambda$CDM would then have a much smaller impact on $\AU$ because it would not integrate all the way back to early times. This is suggested by the fact that cosmological probes at intermediate epochs between recombination and today give parameters consistent with those we obtain from the CMB \citep*{Lin_2021_UCS, Banik_2025_cosmology}. The only anomaly in these studies seems to be $H_0^\mathrm{local}$, which is obtained from the lowest redshift data they consider. Solutions that modify the expansion history only at late times predict only minor departures from the standard $\AUCMB$ \citep{Najera_2026}. It is even possible that the background expansion history is not modified at all, with the Hubble tension instead arising due to outflow from a local void inflating redshifts in the nearby universe \citep{Haslbauer_2020, Banik_2025_BAO, Stiskalek_2025_void}. This scenario would have negligible impact on the predicted $\AU$, but it can be tested in other ways \citep[for a review, see][]{Banik_2026_void}. For comprehensive lists of proposals including the above, we refer the reader to \citet{Valentino_2021_problem} and the more recent \citet{Valentino_2025}. The latter also discusses various other cosmic tensions besides the Hubble tension. Better constraints on $\AU$ may well have broader implications for these problems as well, given the inter-related nature of cosmological parameters \citep*{Kable_2019}.

Our goal is to shed light on ongoing cosmic controversies by inferring $A_\star$ using old stars in the Milky Way whose ages were previously determined by \citetalias{Xiang_2022} without a cosmologically motivated ceiling at $\AUCMB$. In Section~\ref{Sample_selection}, we present the stellar sample we use for this purpose and the additional quality cuts that we apply. We justify some of our most important quality cuts in Section~\ref{AC_cuts_justification} because they are not commonly used in the literature. We then estimate $A_\star$ in three different ways from the individual stellar ages and their uncertainties (Section~\ref{Simplified_Astar_estimates}). To do a more rigorous analysis, we need to reconstruct the latent age distribution of the stars in our sample (Section~\ref{Reconstructed_age_distribution}). We do this using an iterative approach (Section~\ref{Population_prior_protocol}) and then with a Markov Chain Monte Carlo (MCMC) method (Section~\ref{MCMC_reconstruction}). The latter underpins our main results, which we present in Section~\ref{Results}. We obtain additional support for our inferred $A_\star$ by obtaining a similar value from a $102\times$ smaller sample of stars whose chemistry is indicative of formation early in Galactic history, leading to a narrower age spread that allows us to use a simpler model (Section~\ref{Early_chemistry_analysis}). We discuss the reliability of our results and their cosmological implications in Section~\ref{Discussion}. We then conclude in Section~\ref{Conclusions}.

\section{Sample selection and quality cuts}
\label{Sample_selection}

Our initial sample of 247,103 stars comes from \citetalias{Xiang_2022}, excluding a duplicate entry as discussed in Appendix~\ref{Duplicate_star}. The underlying observations are LAMOST~DR7 spectra \citep{Cui_2012, Zhao_2012} and \emph{Gaia}~eDR3 astrometric parallaxes \citep{Gaia_2021}. Stellar ages were obtained from a Bayesian fit to the effective temperature $T_\mathrm{eff}$ and absolute $K$-band magnitude $M_K$ using the YY isochrones \citep{Yi_2001, Demarque_2004}, which are based on the Yale Rotating Evolution Code \citep[\textsc{yrec};][]{Pinsonneault_2026}. The isochrone fits in \citetalias{Xiang_2022} consider the metallicity [Fe/H] and $\alpha$-element abundance [$\alpha$/Fe], where square brackets denote a decimal logarithmic abundance relative to the Solar value. While we can in principle find the isochrone that best fits a star at any evolutionary stage, this is much more accurate after the main sequence phase as stellar properties change much more rapidly. This led \citetalias{Xiang_2022} to focus on the subgiant region of the ($T_\mathrm{eff}$, $M_K$) diagram (see their figure~1).

\begin{figure}
    \centering
    \includegraphics[width=\linewidth]{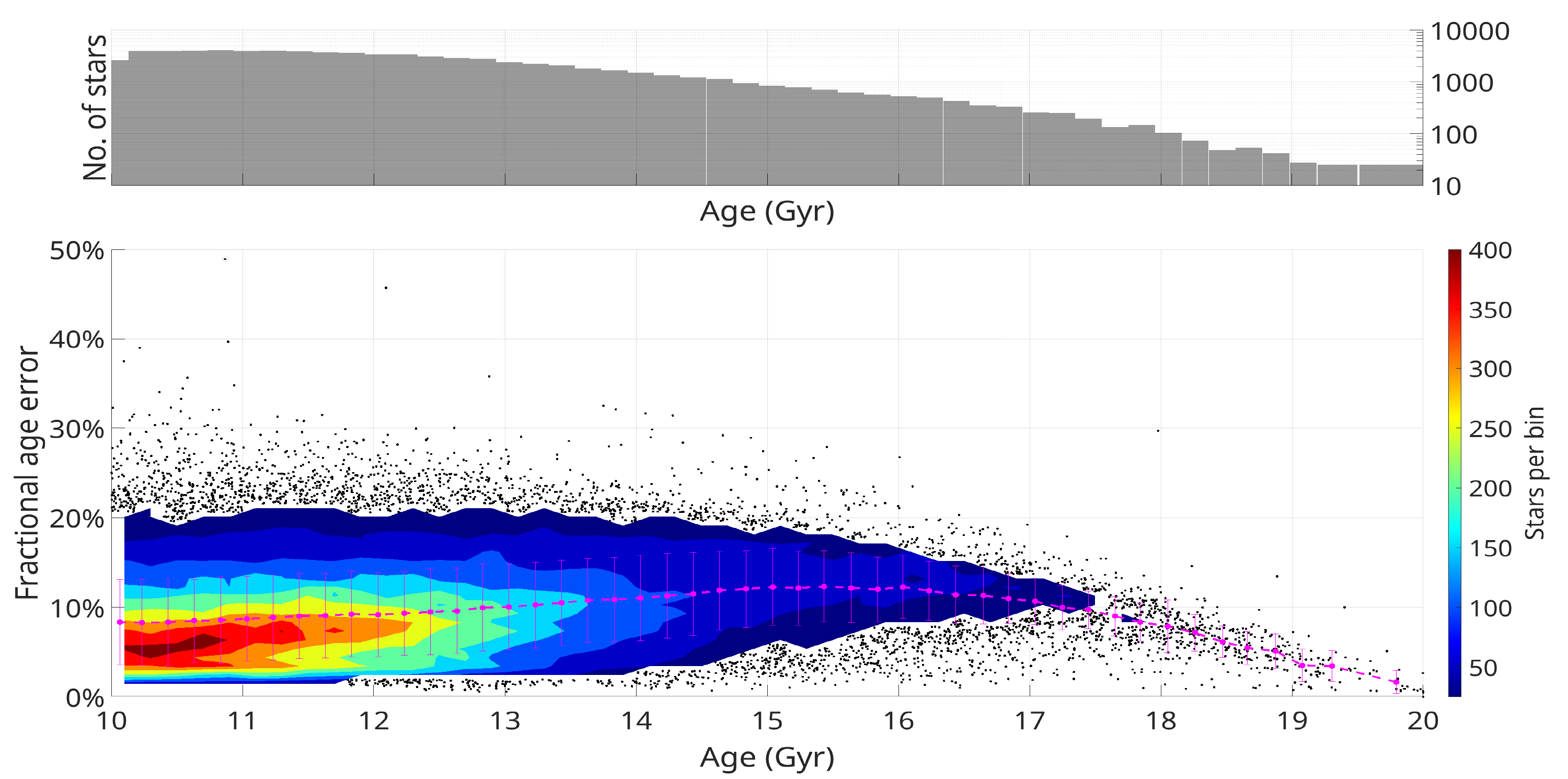} 
    \caption{\emph{Top}: The age distribution of all stars in the \citetalias{Xiang_2022} catalogue older than 10~Gyr. \emph{Bottom}: Fractional age uncertainties shown against observed ages for these stars. The high-density regions are shown as a heat map, with stars in lower density regions shown individually. The magenta points with uncertainties show the mean and standard deviation in $\widetilde{\sigma}_A$ in each age bin, for which we impose a floor on the number of stars and the bin width. The same age bins are used in both panels. Notice that $\widetilde{\sigma}_A$ starts decreasing at $A \ga 16$~Gyr.}
    \label{Age_error_turnover_none}
\end{figure}

\citetalias{Xiang_2022} applied various quality cuts to an initially larger sample, for instance that the spectrum should have a signal-to-noise ratio (SNR) of at least 20 and that $M_K < 0.5$, which causes an almost total loss of stars with $A < 1.5$~Gyr. Figure~\ref{Age_error_turnover_none} shows $A$ against its fractional uncertainty $\widetilde{\sigma}_A$ for stars with $A > 10$~Gyr ($\sigma$ denotes an uncertainty and $\widetilde{\sigma}$ a fractional uncertainty). We expect that $\widetilde{\sigma}_A$ should generally rise with $A$ because older stars within their selected subgiant sample are usually fainter, thus requiring a longer exposure to obtain a high SNR spectrum. Fainter stars are also likely to have a larger parallax uncertainty, further increasing $\widetilde{\sigma}_A$. Moreover, we will at some point reach the $A > A_\star$ regime. Stars can certainly be observed with $A > A_\star$ due to measurement errors, which must therefore be larger at higher $A - A_\star$. Figure~\ref{Age_error_turnover_none} confirms this general expectation for $A \la 15$~Gyr -- but not for older stars. The clear downturn in $\widetilde{\sigma}_A$ beyond 16~Gyr is almost certainly unrealistic. We estimate that this is caused by $\approx 0.1\%$ of the sample having a problematic age determination. We therefore apply a few additional quality cuts to the \citetalias{Xiang_2022} sample, as discussed in the rest of this section. We find that the stars contributing to the downturn in $\widetilde{\sigma}_A$ at high $A$ cannot be removed based on \citetalias{Xiang_2022} reporting an asymmetric age likelihood, even though this can be a useful cut in other samples of old stars \citep[section 3.1 of][]{Tomasetti_2026}.

Our quality cuts must be carefully designed to avoid introducing implicit cosmological biases about what might be a reasonable value for $A_\star$. Moreover, our new cuts must extend those already considered by \citetalias{Xiang_2022}. Those authors apply a wide range of cuts, but these focus on properties of the star itself. If an additional star is observed, it would be accepted or rejected based on its properties, but this would have no bearing on whether any of the existing stars pass the quality cuts.

The large sample size opens the door to population-level quality cuts. The basic idea is to check the distribution of some quantity $X$ and remove stars where its value is an outlier from the overall distribution, potentially indicating an unreliable measurement. However, extreme values of $X$ might also be physical. We therefore extend this concept to 2 parameters $X$ and $Y$, with stars rejected if they lie in a very low density part of the $(X, Y)$ plane. If $X$ and $Y$ generally lie close to a trendline, we can quantify this trendline and the scatter around it, which we can use to reject stars far from the trendline. The properties of the trendline must then be iteratively recalculated until convergence is achieved. We find that this `trendline outlier rejection' approach significantly improves the \citetalias{Xiang_2022} sample. If there is no trendline but there is a clear `edge' to the $(X, Y)$ distribution, we argue that stars well beyond this edge are more likely to have problematic age estimates. These novel techniques cannot be applied to very small samples, which probably explains why they are not commonly used when inferring $A_\star$ from a catalogue of old stars \citep[e.g.][]{Tomasetti_2026}. Crucially, these cuts are not circular because they use chemistry and cross-catalogue age agreement, axes which are physically expected to correlate with age, yet are distinct from the age extremum we wish to measure. They are also conservative along the sample axis: relaxing them only raises the inferred $A_\star$, so they cannot manufacture tension with $\AstarETS$ (Section~\ref{Results}).

\subsection{Parallax precision and consistency}
\label{Parallax_error_cut}

A critical ingredient in dating a star is its absolute magnitude. This can be inferred without an accurate parallax $\varpi$, but the results will be more reliable if one is available. We therefore require $\varpi$ to have an uncertainty $\sigma_\varpi$ of at most 10\%, i.e., $\sigma_\varpi / \varpi < 0.1$.

\begin{figure}
    \centering
    \includegraphics[width=\linewidth]{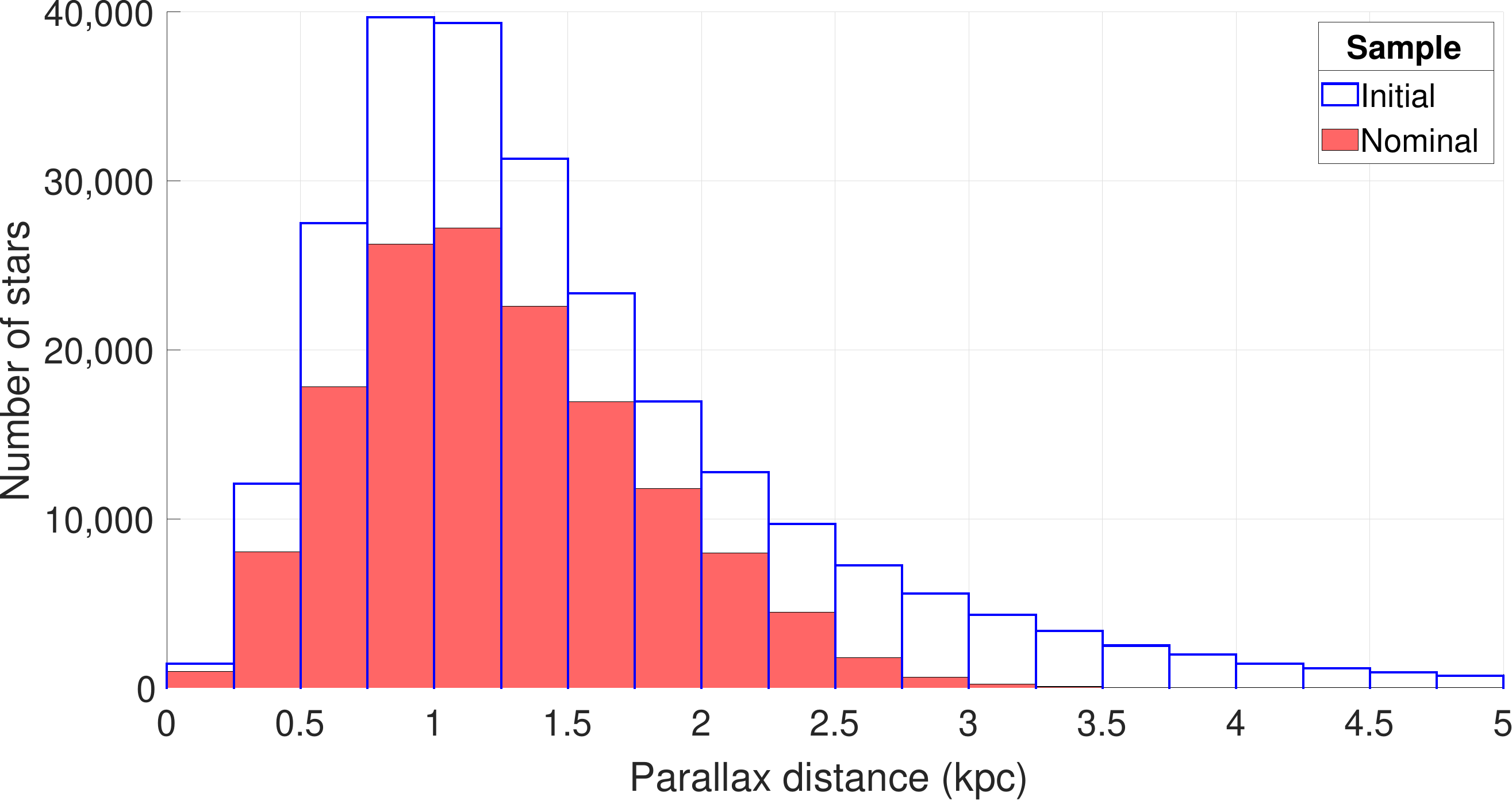}
    \caption{The distance distribution of the stars in our initial sample from \citetalias{Xiang_2022} (unfilled blue bars) and in our nominal sample after applying the quality cuts of Section~\ref{Sample_selection} (filled red bars). Distances are obtained from inverting the \emph{Gaia} parallax after accounting for the parallax zero point. Due to our requirement of a precise parallax, inverting it to obtain a distance should be accurate for our nominal sample \citep{Bailer_Jones_2018}. Since this is restricted to stars within 5~kpc, results are shown here only within this range.}
    \label{d_distribution}
\end{figure}

Beyond this cut, we also exclude stars with very small $\varpi$ to mitigate the impact of systematics that become significant at low parallax. We therefore require $\varpi > 0.2$~mas, limiting ourselves to stars within 5~kpc of the Sun. We expect that there are enough stars within 5~kpc to reliably measure $A_\star$, making it unnecessary to go out further and risk inaccurate astrometry influencing our results. The vast majority of the stars in our nominal sample are far closer, as shown in Figure~\ref{d_distribution}. The subgiants we consider are therefore unlikely to be associated with the Galactic bulge, given also the high Galactic latitudes targeted by LAMOST \citep{Liu_2020, Yan_2022}. We note that it cannot observe the Galactic centre because this remains too low above the horizon at the northern Xinglong station that hosts LAMOST.

\subsubsection{Consistency of spectroscopic and trigonometric parallaxes}

The distance to a star can in principle be determined from spectroscopic measurements and broadband colours. This leads to a spectroscopic `parallax', which we can combine with the \emph{Gaia} trigonometric parallax given the latter is required to have a precision of better than 10\%. For this approach to be valid, the spectroscopic and photometric parallaxes should be consistent. In the LAMOST catalogue, it is possible to check this using the parameters $\texttt{MK\_SPEC}$ and $\texttt{MK\_GEO}$, the absolute magnitude found from the apparent magnitude using the spectroscopic and geometric distance, respectively. We define the internal distance tension $\chi_{_d}$ as follows:
\begin{eqnarray}
    \chi_{_d} ~=~ \frac{\texttt{MK\_SPEC} - \texttt{MK\_GEO}}{\sqrt{\sigma_{\texttt{MK\_SPEC}}^2 + \sigma_{\texttt{MK\_GEO}}^2}} \, ,
\end{eqnarray}
with the LAMOST catalogue listing the uncertainties by adding $\texttt{\_ERR}$ after the variable name.

We find that the distribution of $\chi_{_d}$ is approximately Gaussian, but the mean is somewhat offset from zero and the variance is not exactly unity. We therefore define a normalised distance tension $\widetilde{\chi}_{_d}$ as follows:
\begin{eqnarray}
    \widetilde{\chi}_{_d} ~=~ \frac{\chi_{_d} - 0.145}{1.147} \, .
\end{eqnarray}
We require that $\left| \widetilde{\chi}_{_d} \right| < 3$ to remove stars where the \emph{Gaia} parallax and spectroscopy disagree on the distance by more than is usually the case.

\subsection{\emph{Gaia} binary indicators}
\label{Gaia_binary_cuts}

Since the stellar parameters were derived assuming an isolated star, a close binary companion could lead to inaccurate results. \citetalias{Xiang_2022} already considered some quality cuts targeted at removing binaries, for instance that the renormalised unit weight error (RUWE) is at most 1.2. We complement their cuts with several additional cuts based on indications in the \emph{Gaia} catalogue that a star is part of a binary. The most obvious of these is the \emph{Gaia} catalogue parameter $\texttt{non\_single\_star}$. We require this to be zero as any non-zero value indicates a star for which binary motion is needed to fit the data. We also make use of several other \emph{Gaia} parameters that are suggestive of a binary companion.

\subsubsection{Stars resolved into multiple peaks}

We require the \emph{Gaia} catalogue parameter $\texttt{ipd\_frac\_multi\_peak} \leq 1$ \citep[see section~2.4.3 of][]{Banik_2024_WBT}. This is the percentage of \emph{Gaia} focal plane transits in which a star appears resolved into multiple peaks. It is therefore well suited to detecting a marginally resolved binary, which is ideal for projects aiming to exclude a possible binary rather than include a high confidence binary.

A genuinely isolated star should have $\texttt{ipd\_frac\_multi\_peak} = 0$, but it might occasionally appear resolved into multiple images just by chance. We allow this to happen once, motivated by the likely significant increase in sample size compared to requiring $\texttt{ipd\_frac\_multi\_peak} = 0$ \citep[see figure~3 of][]{Banik_2024_WBT}. We suggest that since a genuine binary is likely to appear resolved into multiple peaks more often, this slight relaxation is unlikely to allow a genuine binary to `slip in'. Indeed, those authors and \citet*{Pace_2022} both use a higher threshold of 2.

\subsubsection{Stars with elongated images}

Even if a binary is never resolved into multiple point sources, we might still expect its image to appear elongated. This would cause a single point source to provide a poor description of the image. Importantly, \emph{Gaia} observations have a much different sensitivity along the scan direction than perpendicular to it. If a source is not point-like, we might expect the scan angle to affect the goodness of fit of a single point source solution. We can only usefully measure any dependence on scan angle if there are a large number of \emph{Gaia} observations with a range of scan directions. We therefore require the \emph{Gaia} parameter $\texttt{visibility\_periods\_used} \geq 10$, ensuring at least ten scans.

The impact of the scan angle is quantified by the \emph{Gaia} parameter $\texttt{ipd\_gof\_harmonic\_amplitude}$, which we allow to be at most 0.15. This is because there is a sharp peak in the distribution at zero followed by a rapidly declining tail, which presumably corresponds to measurement noise. The decline becomes much more gradual around 0.15, suggesting a transition to a second component of the $\texttt{ipd\_gof\_harmonic\_amplitude}$ distribution. We suggest that this more extended tail is associated with a second component in the source. This need not be physically associated with the star, since a chance background alignment could have a similar impact. It is helpful to remove the star even in this case due to the observational difficulties it would cause. We note that the LAMOST stars are generally at high Galactic latitudes, so a chance alignment seems less plausible than a faint companion if there is indeed a faint secondary source.

\subsection{Reddening}
\label{Reddening}

As part of their parameter estimates, \citetalias{Xiang_2022} found the selective extinction $E \left( B - V \right)$. While this can be thought of empirically as a colour excess, astrophysically it should come from dust. We therefore require $E \left( B - V \right) \geq 0$. The negative inferred reddening for a small number of stars could indicate problems with the parameter estimates.

\subsection{Consistency of the \texorpdfstring{\emph{Gaia}}{Gaia} photometry}
\label{C_star_cut}

The \emph{Gaia} photometry is obtained in three passbands, namely a broad $G$ band and the blue and red photometers ($BP$ and $RP$), whose fluxes we denote by $F_G$, $F_\mathrm{BP}$, and $F_\mathrm{RP}$, respectively. While $F_G$ comes from a small aperture tailored to the point spread function, $F_\mathrm{BP}$ and $F_\mathrm{RP}$ are obtained by integrating a slitless prism spectrum over a much larger aperture of $3.5\arcsec \times 2.1\arcsec$ \citep{Evans_2018}. Any additional flux falling within that larger window is therefore counted in $F_\mathrm{BP}$ and $F_\mathrm{RP}$, but not in $F_G$. The standard diagnostic for this is the catalogue parameter \texttt{phot\_bp\_rp\_excess\_factor}, which is defined as follows:
\begin{eqnarray}
    C ~\equiv~ \frac{F_\mathrm{BP} + F_\mathrm{RP}}{F_G} \, .
    \label{C_definition}
\end{eqnarray}
We expect this to be a smooth function of colour alone, provided we have an isolated, well-behaved source. Thus, checking $C$ against the smooth trend with colour reveals whether the relative values of $F_\mathrm{G}$, $F_\mathrm{BP}$, and $F_\mathrm{RP}$ are mutually consistent with what we expect for a source of that colour. A star deviating significantly from this trend should be excluded because it could indicate a nearby contaminating source. This may affect the stellar parameters given that the LAMOST fibre diameter of $3.3\arcsec$ is comparable to the \emph{Gaia} aperture.

\citet{Riello_2021} showed that the colour dependence of $C$ follows a reference locus $C_\mathrm{ref}$ given by a piecewise polynomial in the observed colour $x \equiv m_\mathrm{BP} - m_\mathrm{RP}$, where $m$ indicates an apparent magnitude and the subscript denotes the \emph{Gaia} passband. Every star in our sample has $0.58 < x < 1.99$, so we only use the middle branch of their equation~6. This provides the following empirical relation:
\begin{eqnarray}
    C_\mathrm{ref} ~=~ 1.162004 + 0.011464 \, x + 0.049255 \, x^2 - 0.005879 \, x^3.
    \label{C_ref}
\end{eqnarray}
The branches at $x < 0.5$ and $x \geq 4$ are never reached for our sample.

The residual $C_\star \equiv C - C_\mathrm{ref}$ is in principle stored as the \emph{Gaia} catalogue parameter \texttt{c\_star}, but we find that it is unavailable for 25,131 stars (10\% of our sample) despite these stars all having valid entries for $C$, $m_\mathrm{BP}$, $m_\mathrm{RP}$, and $m_\mathrm{G}$. We therefore calculate $C_\star$ using Equations~\ref{C_definition} and~\ref{C_ref}. Where a check is possible, our result lines up with the catalogue \texttt{c\_star}. \citet{Riello_2021} also give the width of the $C_\star$ distribution as a function of $m_\mathrm{G}$, finding that the width grows steeply towards fainter sources (see their equation~18).

We find that the \citet{Riello_2021} prescription requires two modifications before it can be applied to our sample. The first concerns the width. Their calibration is necessarily a global one covering the full range of source types and crowding conditions in the \emph{Gaia} catalogue, whereas our stars are a restricted and comparatively well-observed subset. The robust scatter of $C - C_{\mathrm{ref}}$ in our sample rises from 0.0013 at $m_{\mathrm{G}} = 10-12$ to 0.0023 at $m_{\mathrm{G}} = 15-16$, much narrower than their published width of 0.0070 and 0.0153, respectively. The recommended cut at triple the published width is therefore ineffective on our sample, presumably due to the published width describing a much more heterogeneous population than ours. We therefore retain the functional form of the scatter model in equation~18 of \citet{Riello_2021}, but halve the magnitude-dependent term and then scale the overall result by 0.3. We call this revised scatter $\sigma_{\widetilde{C}}$, which depends only on $m_{\mathrm{G}}$.


The second modification is to the zero point, motivated by our finding that $C_\star$ is not centred on zero. Instead, $C_\star$ retains a systematic dependence on reddening, colour, and metallicity -- the last being especially important. The median $C_\star$ at fixed colour $x$ runs from $-0.0023$ at $E(B - V) < 0.03$ to $+0.0007$ at $E(B - V) > 0.12$, a trend comparable to the width itself. If uncorrected, this would cause the cut to be centred on a value $1\sigma$ off for highly reddened stars, leading to a rejection of stars with a typical $C$ given their other properties. We argue that this is a genuine extinction effect rather than crowding. At fixed Galactic latitude $\left| b \right|$, the residual moves by 0.0030 across the range of $E(B - V)$ that we cover, but the change at fixed $E(B - V)$ is only 0.0011 across the whole range of $\left| b \right|$. Moreover, the excess is smaller towards the Galactic plane, opposite to what we would expect for contaminating flux from a crowded field. We also find that the correction must be made using the observed rather than dereddened colour. Using the latter in Equation~\ref{C_ref} makes matters considerably worse, increasing the swing in $C_\star$ from 0.0025 to 0.018. This confirms that the effect follows the light actually entering the $BP$ and $RP$ filters.

At fixed observed colour and reddening, the residual retains a strong dependence on metallicity, running from $+0.0026$ for stars with [Fe/H] $< -1$ to $-0.0041$ for stars with [Fe/H] $> 0$. This swing of 0.0067 is nearly triple the width of the distribution, making it much the largest of the three effects. We argue this is due to line blanketing: at fixed observed colour, a metal-poor star has a different spectral energy distribution than a metal-rich one within the blue and red passbands. This causes a different fraction of the flux to enter the photometric windows. \citet{Riello_2021} could not have included this term because no metallicity is available for a general \emph{Gaia} source, but we have one for every star in our sample thanks to the LAMOST spectra. We therefore add a correction $\Delta C$ which is linear in all three variables:
\begin{eqnarray}
    \Delta C &=& -0.001679 - 0.003162 \, x + 0.020944 \, E \left( B - V \right) \nonumber \\
    && - \, 0.005736 \left[ \mathrm{Fe}/\mathrm{H} \right].
    \label{C_offset}
\end{eqnarray}
The coefficients are obtained by least squares on the median of $C - C_\mathrm{ref}$ within each of 181 cells of a grid in colour, reddening, and metallicity, with each cell holding around 1000 stars. This fit reduces the scatter of those cell medians from $2.4 \times 10^{-3}$ to $3.8 \times 10^{-4}$, so the three linear terms account for nearly all the structure which the \citet{Riello_2021} fitting formula necessarily leaves in a sample such as ours. Our corrected excess factor and its width are then as follows:
\begin{eqnarray}
    \widetilde{C} &\equiv& C - C_\mathrm{ref} \left( x \right) - \Delta C, \\
    \sigma_{\widetilde{C}} &=& 0.30 \left( 0.0059898 + 0.5 \times 8.817481 \times 10^{-12} \, m_{\mathrm{G}}^{7.618399} \right) \nonumber \\
    && \times \left[ 1 - 0.0751 \, E \left( B - V \right) \right] \, .
    \label{C_width}
\end{eqnarray}
The term in Equation~\ref{C_width} within round brackets applies our two rescalings to the published width \citep[equation~18 of][]{Riello_2021}. The final term in square brackets accounts for a mild narrowing of the $C_\star$ distribution at high reddening, reducing $\sigma_{\widetilde{C}}$ by only 1.5\% at $E(B - V) = 0.2$. We neglect the observational uncertainty in $\widetilde{C}$ as this is almost always $\ll \sigma_{\widetilde{C}}$ at the same $m_{\mathrm{G}}$.

Following \citet{Riello_2021}, we apply a $3\sigma$ quality cut as follows:
\begin{eqnarray}
    \left| \widetilde{C} \right| ~<~ 3 \, \sigma_{\widetilde{C}} \, .
    \label{C_star_cut_equation}
\end{eqnarray}
We find that 6378 of the 247,103 stars fail this cut.



\subsection{The Kiel diagram}
\label{Kiel}

Since we are interested in old subgiants, we restrict the region of the Kiel diagram. Following \citet{Nepal_2024} and equation~A1 of \citet{Tomasetti_2026}, we require the following:
\begin{eqnarray}
    -0.0003 \, T_\mathrm{eff} + 4.8 < \log_{10} g < 4.1, ~~T_\mathrm{eff} > 500 \, \log_{10} g + 3000,
    \label{Kiel_cut}
\end{eqnarray}
where $T_\mathrm{eff}$ is in K and $g$ is the surface gravity in cm/s\textsuperscript{2}. Subgiants have lower $g$ than main sequence stars, so the ceiling on $g$ aims to remove the latter.

\subsection{The age--metallicity relation}
\label{Age_metallicity_relation}

\begin{figure}
    \centering
    \includegraphics[width=\linewidth]{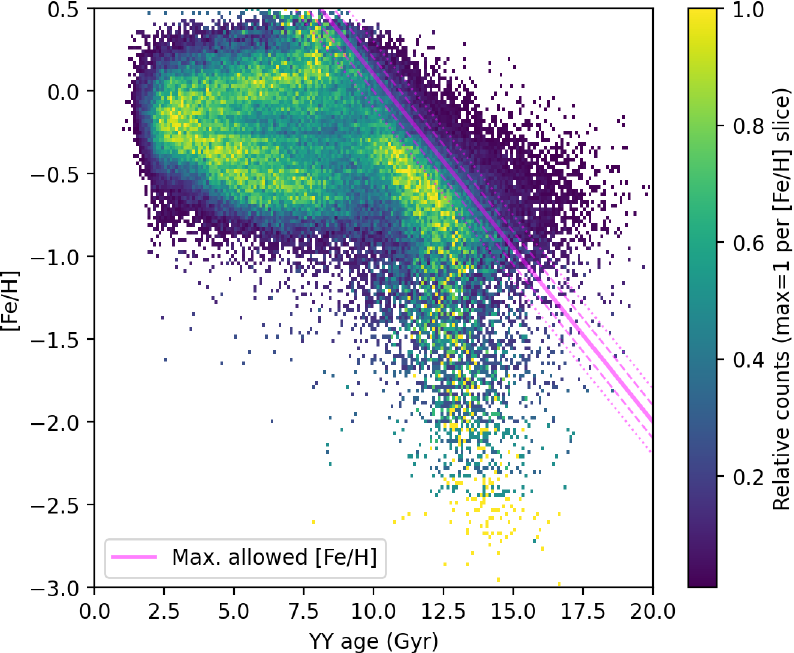} 
    \caption{Heat map showing the age--metallicity relation of our sample prior to cuts related to the chemistry and comparison with FLAME ages. Counts in each cell are normalised separately for each [Fe/H] slice relative to the cell with the highest value, better highlighting the features \citepalias{Xiang_2022}. The solid magenta line shows our nominal quality cut (Equation~\ref{Age_metallicity_cutoff_line}). Stars above it are excluded because a very old star is unlikely to be metal-rich. The dashed (dotted) magenta lines offset vertically by $\pm 0.1$ ($\pm 0.2$) dex show alternative quality cuts that we consider in Section~\ref{Results}.}
    \label{age_feh_figure}
\end{figure}

The problematic stars evident at the bottom right of Figure~\ref{Age_error_turnover_none} most likely have a lower age. If they are genuinely very old, they should have a very low metallicity, especially given that our sample is too nearby to include the Galactic bulge (Figure~\ref{d_distribution}) and the LAMOST survey targets rather high Galactic latitudes, with no ability to observe the Galactic centre. The typical age--metallicity relation of our sample is shown in Figure~\ref{age_feh_figure}, which includes the quality cuts discussed above. We normalise the counts separately for each [Fe/H] slice relative to the maximum for that slice, meaning each non-empty slice always has a cell with a normalised count of unity, which we indicate as a yellow colour. This way of showing the age--metallicity relation follows figure~2 of \citetalias{Xiang_2022}, which we find makes the details much clearer. In particular, it reveals a striking bimodality, which those authors attribute to a combination of radial migration and the distinct episodes of star formation which led to the Galactic thin and thick discs \citep{Gilmore_1983, Fuhrmann_1998, Juric_2008, Jayaraman_2013, Hayden_2015, Wu_2021}.

For our purposes, the important aspect of Figure~\ref{age_feh_figure} is the downward-sloping `ridgeline' above which there are very few stars. These stars may well have incorrectly estimated parameters given they are not in line with the overall trend for our sample. We therefore manually choose the solid magenta line as our nominal quality cut, excluding stars above it due to an unrealistically high metallicity for the reported high age. Our quality cut is as follows:
\begin{eqnarray}
    \left[ \mathrm{Fe}/\mathrm{H} \right] ~<~ 2.2 - 0.21 \left( \frac{A}{\mathrm{Gyr}} \right).
    \label{Age_metallicity_cutoff_line}
\end{eqnarray}
We stress that this still allows stars into our sample with $A \gg \AUCMB$, but only if [Fe/H] is very low. The dashed and dotted magenta lines show alternative quality cuts with vertical offsets of $\pm 0.1$ and $\pm 0.2$~dex, respectively. We use these in some analysis variants to check the robustness of our results (Section~\ref{Results}). We provide further justification for our age--metallicity cut and the similar age--[$\alpha$/Fe] cut in Section~\ref{AC_cuts_justification} as such cuts are not common in the literature.

An interesting aspect of Figure~\ref{age_feh_figure} is the almost vertical tail towards lower metallicity at $A \approx 14$~Gyr, which is most likely related to the fact that the Galaxy lacked metals initially. Since [Fe/H] uses a log-scale, we expect it to reach down to very negative values at $A \approx A_\star$. This precludes us from imposing a lower limit on [Fe/H] at high $A$. In fact, considering only the most metal-poor stars could actually help as these are likely the oldest stars. We do just such an analysis in Section~\ref{Early_chemistry_analysis}.

\subsection{The age--[\texorpdfstring{$\alpha$}{alpha}/Fe] relation}
\label{Age_alpha_Fe_relation}

Just as we expect old stars to have a low metallicity, so also we expect them to have a high [$\alpha$/Fe]. This is because stars which formed early in Galactic history would contain heavy elements synthesized by core-collapse supernovae, but not yet by Type~Ia supernovae (SNe~Ia). The latter produce a much larger fraction of iron-peak elements, but they only arise after a roughly 1~Gyr delay due to the need for a period of accretion from a binary companion \citep{Tinsley_1979}.

\begin{figure} 
    \centering
    \includegraphics[width=\linewidth]{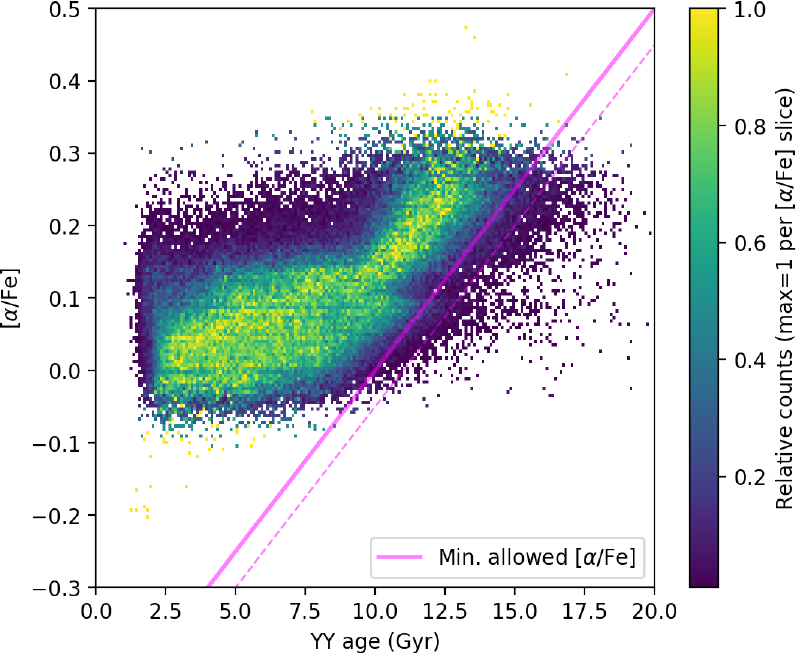} 
    \caption{Similar to Figure~\ref{age_feh_figure}, but now showing [$\alpha$/Fe] against age. The solid magenta line shows our nominal quality cut (Equation~\ref{Age_alpha_Fe_cutoff_line}), below which stars are excluded. The dashed magenta line offset down by 0.05~dex shows an alternative quality cut that we consider in Section~\ref{Results}.}
    \label{age_alpha_figure}
\end{figure}

We can use this delay as the basis for another quality cut using the age--[$\alpha$/Fe] relation, which we show in Figure~\ref{age_alpha_figure}. In this case, we require the following:
\begin{eqnarray}
    \left[ \alpha/\mathrm{Fe} \right] ~>~ -0.5 + 0.05 \left( \frac{A}{\mathrm{Gyr}} \right).
    \label{Age_alpha_Fe_cutoff_line}
\end{eqnarray}
This is shown as a solid magenta line. We also consider a variant where the cutoff line is lowered by 0.05~dex, which goes into a significantly less dense part of the age--[$\alpha$/Fe] diagram.

We note that in regions of intense star formation and thus a high rate of SNe~Ia, the [$\alpha$/Fe] ratio can start dropping within just $40-50$~Myr \citep{Matteucci_2001}. Although such short timescales are more typical for elliptical galaxies rather than disc galaxies like our own with an extended star formation history, there are indeed some old $\alpha$-poor Galactic stars \citep{Nepal_2024, Borbolato_2025}. The general trend in Figure~\ref{age_alpha_figure} shows that this is atypical, so apart from a few isolated regions like the Galactic bulge (which lies far outside the considered region; see Figure~\ref{d_distribution}), we expect that the oldest Galactic stars should have a high [$\alpha$/Fe]. Requiring this may lose some old $\alpha$-poor stars whose ages are correctly estimated, but since our sample is still quite large (see Appendix~\ref{Sample_size_revised_AC_cuts}), we argue that it is better to accept the loss of some stars whose ages might be reliable only if they formed in unusual environments. Moreover, we will see in Section~\ref{Results} that our inferred $A_\star$ does not depend much on precisely where we place our age-dependent floor on [$\alpha$/Fe].

\subsection{Comparison with FLAME ages}
\label{FLAME_comparison}

As a final test of whether a star is likely to have reliable data in the \citetalias{Xiang_2022} catalogue, we compare their reported YY ages with ages obtained from only \emph{Gaia}~DR3 \citep{Gaia_2023} using the FLAME package \citep{Pichon_2007}.\footnote{\href{https://gea.esac.esa.int/archive/documentation/GDR3/index.html}{https://gea.esac.esa.int/archive/documentation/GDR3/index.html}} We note that Gaia~DR3 was not yet available at the time of publication of \citetalias{Xiang_2022}, allowing us to cross-check their YY ages against photometric FLAME ages, which are listed in the \emph{Gaia} catalogue as $\texttt{age\_flame}$ \citep{Creevey_2023, Creevey_2026}. About 80\% of the stars in \citetalias{Xiang_2022} have a valid FLAME age in \emph{Gaia}~DR3. The remaining stars can simply be excluded, since losing 20\% of the sample is not a severe setback. However, this could bias our results because FLAME ages were artificially restricted to at most 13.5~Gyr. Since this removes the oldest stars, requiring a FLAME age for every star in \citetalias{Xiang_2022} could indirectly bias our $A_\star$ inference low.

\begin{figure}
    \centering
    \includegraphics[width=\linewidth]{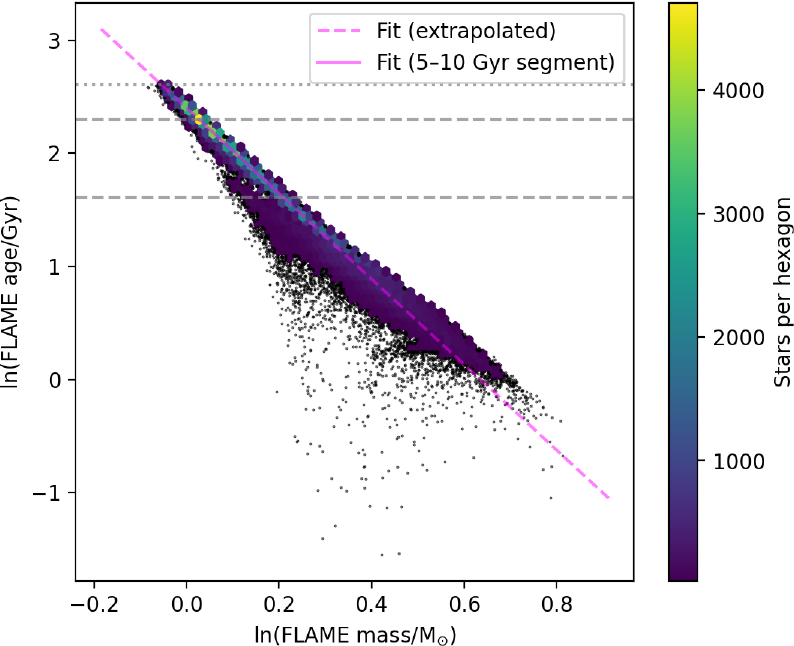} 
    \caption{Heat map showing the FLAME age--mass relation for stars in our sample prior to the quality cut comparing FLAME ages with the \citetalias{Xiang_2022} spectroscopic ages using YY isochrones. Stars are shown individually in sparsely populated cells. Notice that the vast majority of the stars lie close to a line (brighter coloured cells). The solid magenta line shows our linear fit using $3\sigma$ trendline outlier rejection over the $5-10$~Gyr range. These limits are shown with horizontal dashed grey lines. The dashed magenta line shows the extrapolation of our linear fit to lower and higher ages. The horizontal dotted grey line at 13.5~Gyr shows the artificial limit on published FLAME ages (see the text). For stars where the FLAME age is not known but the FLAME mass is known, we extrapolate our linear fit to impute the FLAME age.}
    \label{flame_lnage_lnmass} 
\end{figure}

We identified that this issue of deleted FLAME ages affects 3302 stars, which have otherwise valid FLAME data -- except for the blank age. It is possible to infer this using the other available FLAME data on such stars. The approach we follow is to exploit the very tight age--mass relation, which we show on Figure~\ref{flame_lnage_lnmass} using logarithmic axes. Most of the stars lie in a very narrow band, which is expected for subgiant stars near the end of their life given stars follow a tight age--lifetime relation. We use linear regression to quantify the FLAME age--mass relation, focusing on stars with $A = 5-10$~Gyr \citep[equations~18 and 19 of][]{Banik_2018_escape}. The upper limit is chosen to avoid edge effects associated with the removal of stars older than 13.5~Gyr, while the lower limit is motivated by the possibility that the age--mass relation departs from a power-law form over a sufficiently wide range. Since a small number of outliers are apparent, we compute the orthogonal distance $d_\perp$ of each star from the regression line. We find the standard deviation of $d_\perp$ and remove stars whose $d_\perp$ differs by $>3\sigma$ from the mean value. The linear regression is then repeated iteratively until convergence is achieved. We find that the age--mass relation is as follows:
\begin{eqnarray}
    \ln \left( \frac{A}{\mathrm{Gyr}} \right) ~=~ 2.399 - 3.779 \ln \left( \frac{M_\star}{M_\odot} \right) \quad \mathrm{(FLAME)},
    \label{FLAME_age_mass_relation}
\end{eqnarray}
where $A$ is the FLAME age and $M_\star$ is the FLAME mass. We use this to impute the FLAME age of any star where this is unavailable but the FLAME mass is available, neglecting uncertainty in the fit coefficients given the very tight linear regression (Figure~\ref{flame_lnage_lnmass}) and the very small fraction of stars in our nominal sample with ages imputed in this way.\footnote{Our nominal sample has only 392 of the 3302 stars where we imputed the FLAME age. These 392 stars have FLAME masses at the low end of the range for the full sample. Since these imputed FLAME ages carry no assigned uncertainty and lie preferentially near the high-age end relevant to $A_\star$, they should be treated with some caution, though their small number limits their impact.} If neither age nor mass is available from FLAME, we assume the star was not analysed by FLAME at all. We remove such stars from our sample as we cannot cross-check YY and FLAME ages. This likely leads to a loss of some old stars due to their faintness, but our initial sample undoubtedly has the same issue. We do not claim that the latent age distribution of the stars in our sample is representative of the Solar neighbourhood.

FLAME ages are derived with much more limited information than the YY ages derived by \citetalias{Xiang_2022} using ground-based spectra from LAMOST~DR7. These spectra provide much more precise chemical abundances, which are then used to inform the isochrone fitting. FLAME ages assume a Solar [$\alpha$/Fe] given the lack of a measurement. This assumption is likely to become increasingly inaccurate at high ages (Figure~\ref{age_alpha_figure}). Moreover, FLAME itself provides barely any constraints on the metallicity, making the prior completely dominant \citep{Fouesneau_2022}. This means that although YY and FLAME ages should certainly be correlated, the trendline might differ from the line of equality.

\begin{figure}
    \centering
    \includegraphics[width=\linewidth]{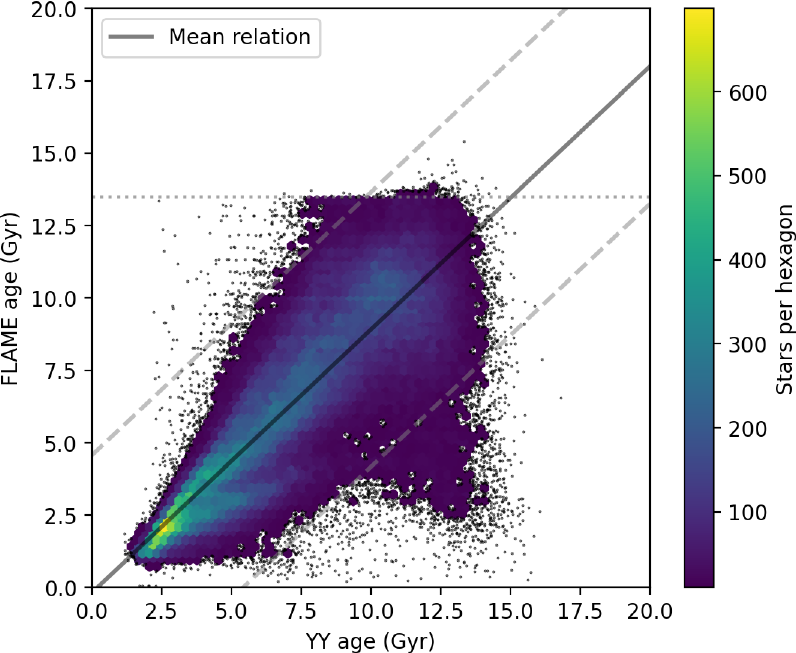} 
    \caption{Heat map showing the relation between FLAME ages (if necessary imputed as shown in Figure~\ref{flame_lnage_lnmass}) and ages in the \citetalias{Xiang_2022} catalogue prior to our quality cut involving comparison between the two. This uses $3\sigma$ trendline outlier rejection, with the resulting trendline shown in solid grey. Stars are included in our nominal sample only if they lie between the dashed grey lines. The dotted grey horizontal line at 13.5~Gyr highlights the limit beyond which FLAME ages were deleted. Any stars beyond this limit have an imputed FLAME age via Equation~\ref{Age_YY_FLAME_relation}.}
    \label{yy_vs_flame_age_figure} 
\end{figure}

To account for this possibility and avoid a problematic direct comparison between YY and FLAME ages, we fit the relation between them using a similar trendline outlier rejection approach to that used in deriving Equation~\ref{FLAME_age_mass_relation}. We recover the following best-fitting relation:
\begin{eqnarray}
    A_{\mathrm{FLAME}} ~=~ 0.909 \, A_{\mathrm{YY}} - 0.183 \, \mathrm{Gyr}.
    \label{Age_YY_FLAME_relation}
\end{eqnarray}
This is shown as the solid grey line on Figure~\ref{yy_vs_flame_age_figure}. The dashed grey lines illustrate the maximum allowed $3\sigma$ orthogonal scatter of 3.510~Gyr. Only stars between these dashed lines are included in our nominal analysis. We also run a variant where we do not consider FLAME ages, but in this case there are signs that the sample now contains some stars with inaccurate ages in the regime most relevant to finding $A_\star$ (see the bottom right panel of Figure~\ref{Age_error_turnover_variants}).


\begin{figure}
    \centering
    \includegraphics[width=\linewidth]{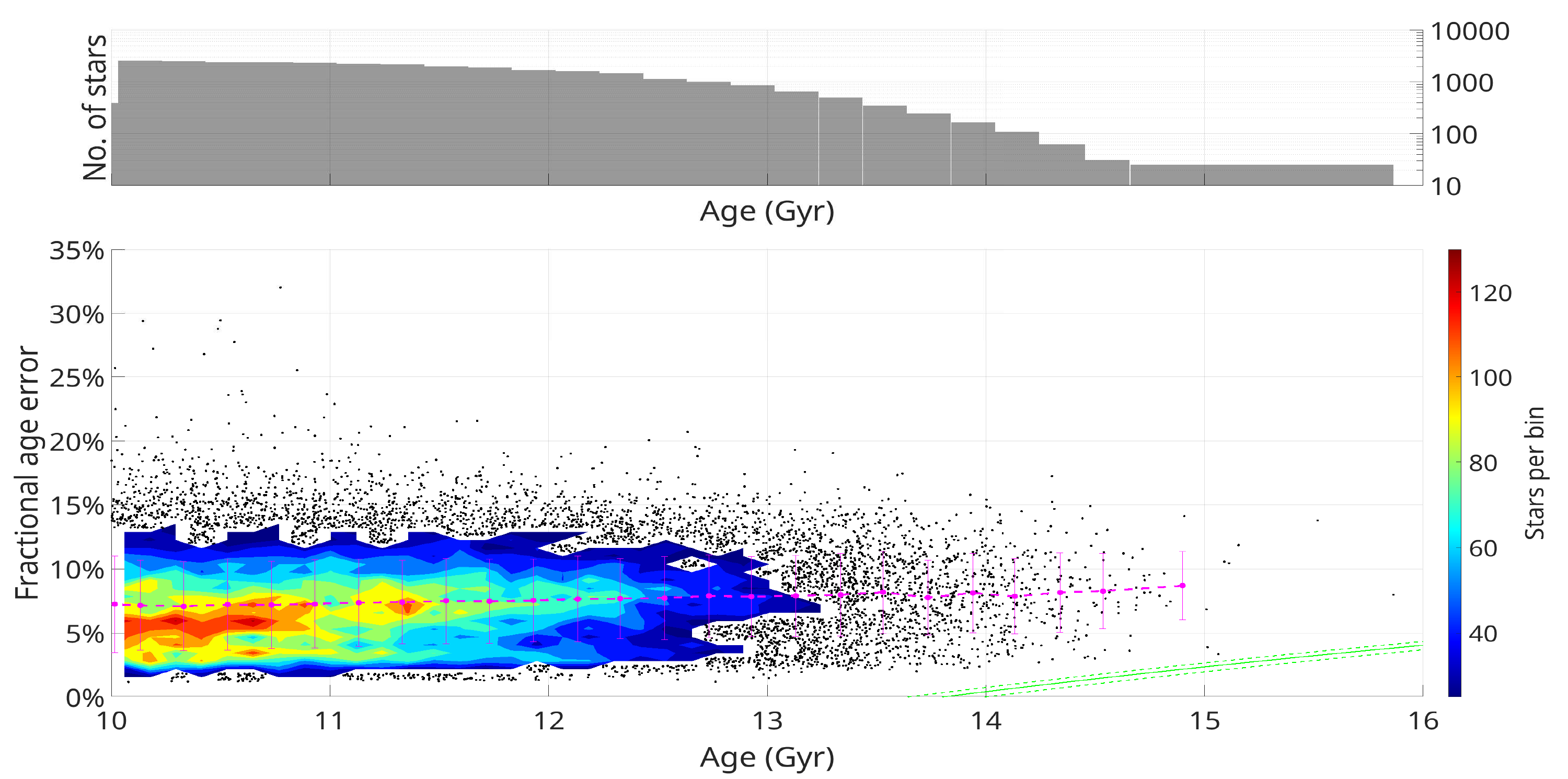} 
    \caption{Similar to Figure~\ref{Age_error_turnover_none}, but for our nominal sample after all our quality cuts. The solid green line shows the minimum $\widetilde{\sigma}_A$ required for consistency at $3.4\sigma$ with the most likely $A_\star$ determined for this sample (Section~\ref{Results}). This threshold is approximately the maximum deviation we expect for Gaussian age errors in a sample whose effective size is 1\% that of our nominal sample (Section~\ref{Individual_age_likelihoods}). The dashed green lines also show $A - 3.4 \, \sigma_A = A_\star$, but put $A_\star$ at the $1\sigma$ lower or upper limit of its posterior. Stars should therefore not go below the green lines, but should come close to them. Given this is the case and that there is no longer any downturn in $\widetilde{\sigma}_A$ towards higher $A$, we suggest that our quality cuts provide a clean sample with 146,810 of the original 247,103 stars.}
    \label{Age_error_turnover_nominal}
\end{figure}

We are now in a position to revisit the anomalous downturn in $\widetilde{\sigma}_A$ at high ages evident in Figure~\ref{Age_error_turnover_none}. This is almost certainly due to measurement problems with a small fraction of stars. We estimate this fraction at 0.1\% based on there being 228 stars in our initial sample of 247,103 which have $A - 5\sigma_A > 13.705$~Gyr, the maximum plausible age in $\Lambda$CDM corresponding to the minimum plausible formation time of $\tf = 0.1$~Gyr. Despite not explicitly excluding stars in the \citetalias{Xiang_2022} catalogue on the basis of being inconsistent with $\AUCMB$, our nominal quality cuts are able to remove all such stars. We stress that this is a consistency check rather than a design goal, as our cuts were defined \emph{a priori} on chemistry and cross-catalogue agreement, axes which are distinct from the age extremum. Crucially, our nominal sample has a clean rising trend in $\widetilde{\sigma}_A$ with $A$ up to the highest ages (Figure~\ref{Age_error_turnover_nominal}). This healthy behaviour suggests that our nominal sample with $N_\mathrm{tot} = 146,810$~stars within 5~kpc has been adequately cleaned and is suitable for cosmological analysis. There is no single star with very high $A$ and low $\widetilde{\sigma}_A$ that might unduly influence our results. To minimise the risk of confirmation biases, we pre-specified the most important quality cuts before we developed several key stages in the $A_\star$ inference, and thus well before any cosmological conclusions could be drawn from the inference.

We can also check for internal consistency with our inferred $A_\star$, which we find using the techniques discussed in Section~\ref{MCMC_reconstruction}. The green lines on Figure~\ref{Age_error_turnover_nominal} show the minimum age uncertainty required for consistency with $A_\star$. These lines are shown only for $A > A_\star$ because a star with lower age is consistent with $A_\star$ regardless of $\sigma_A$. Given the large sample size, we expect that $\gg 1\sigma$ age overestimates are possible. We draw the green lines at $A - A_\star = 3.4 \, \sigma_A$ because we expect the maximum deviation for Gaussian errors to be about $3.4\sigma$ with an effective sample size of $0.01 \, N_\mathrm{tot}$, where the factor of 0.01 is our estimated fraction of stars relevant to a determination of $A_\star$ (this is discussed further in Section~\ref{Individual_age_likelihoods}). As expected, some stars get close to the solid green line, but none go below it. The stellar ages are thus all consistent with the recovered $A_\star$ at $3.4\sigma$, which we estimate is the appropriate threshold for our large sample. Our nominal analysis is internally consistent in this respect.

\section{The necessity of the age--chemistry (AC) cuts}
\label{AC_cuts_justification}

Our initial sample of 247,103 stars suffers from a pathology whereby fractional age errors stop their expected increase towards older ages once $A \ga 15$~Gyr and then noticeably decrease at $A \ga 16$~Gyr (Figure~\ref{Age_error_turnover_none}). Given that older stars are less massive and therefore intrinsically fainter, this behaviour is indicative of inaccurate age likelihoods at high reported ages. This problem is not present in our nominal sample (Figure~\ref{Age_error_turnover_nominal}). This must be due to our quality cuts, the most innovative of which are cuts in the age--metallicity and age--[$\alpha$/Fe] plane (Sections~\ref{Age_metallicity_relation} and \ref{Age_alpha_Fe_relation}, respectively). We refer to these collectively as our AC cuts. We explore their relevance in more detail in this section.

\begin{figure}
    \centering
    \includegraphics[width=\linewidth]{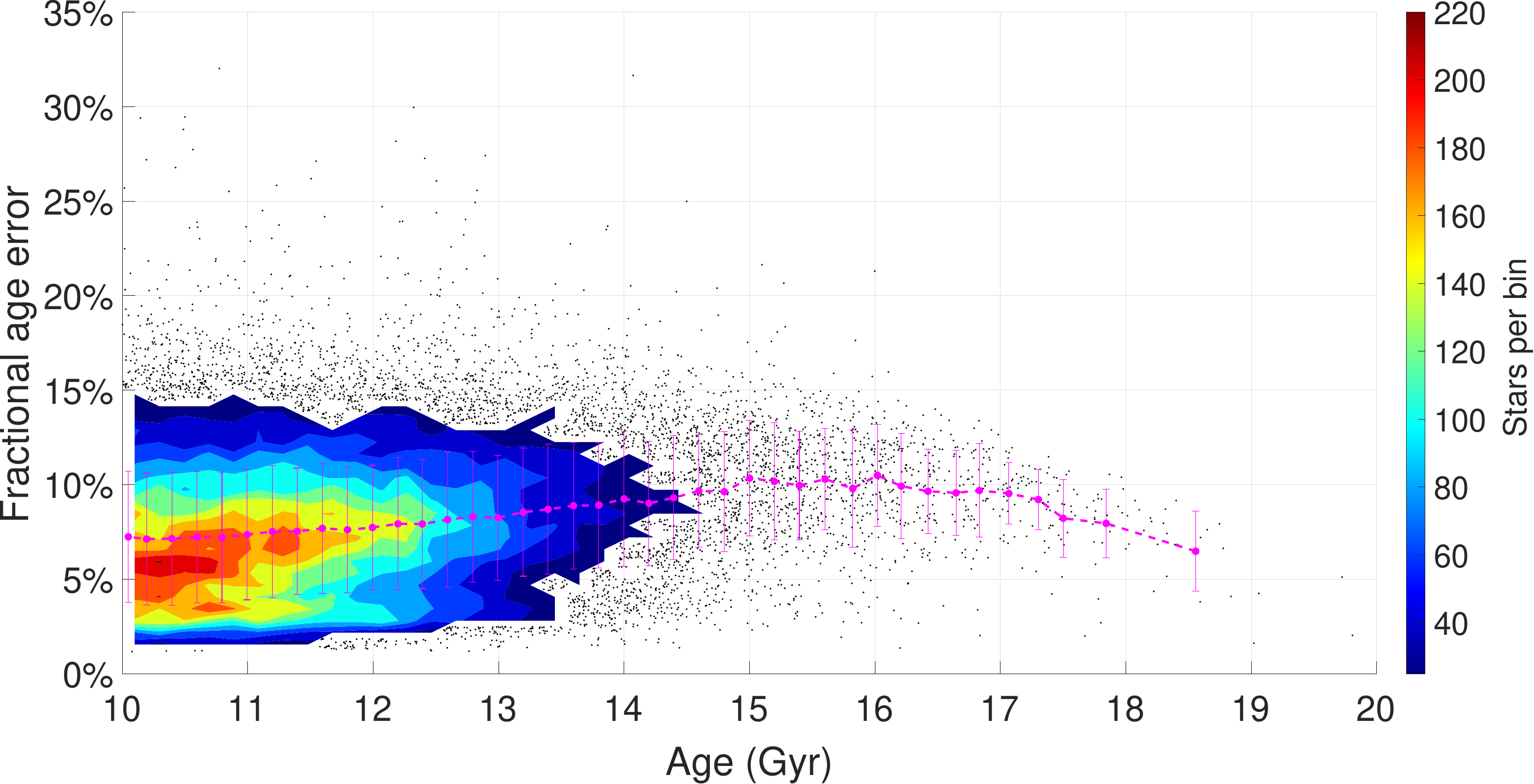} 
    \caption{Similar to the bottom panel of Figure~\ref{Age_error_turnover_nominal}, but without our AC cuts from Sections~\ref{Age_metallicity_relation} and \ref{Age_alpha_Fe_relation}. The anomalous downturn in $\widetilde{\sigma}_A$ towards higher $A$ reappears. Our quality cuts are thus unable to adequately clean the sample without the AC cuts, preventing us from reliably inferring $A_\star$.}
    \label{Age_error_turnover_no_age_chemistry}
\end{figure}

We begin by simply removing the AC cuts and implementing all our remaining quality cuts. The resulting $\left( A, \widetilde{\sigma}_A \right)$ distribution once again shows the anomalous downturn in $\widetilde{\sigma}_A$ at high ages (Figure~\ref{Age_error_turnover_no_age_chemistry}). We find that the stars removed by the AC cuts share a similar distance distribution to our nominal sample and are located on the expected relation between [Fe/H] and [$\alpha$/Fe]. There is also little sign of an undetected companion in the \emph{Gaia} colour excess parameter (Section~\ref{C_star_cut}) or other binary indicators, the most common of which we already use as quality cuts (Section~\ref{Gaia_binary_cuts}). This strongly suggests that we cannot reliably clean our initial sample without the AC cuts. We therefore do not infer $A_\star$ without these cuts. In the rest of this section, we provide further justification for them.

\subsection{The vertical action as an age proxy}
\label{Jz_age_proxy}

To gain further evidence that the stars removed by the AC cuts indeed have inaccurate ages, we need an independent estimate of their age. We already compare the spectroscopic YY ages with FLAME ages (Section~\ref{FLAME_comparison}). While this does appear to help somewhat (see the bottom right panel of Figure~\ref{Age_error_turnover_variants}), it is not sufficient to achieve a clean sample -- even in combination with all the remaining non-AC quality cuts.

We therefore focus on the vertical action $J_z$, which characterises the amplitude of a star's oscillations perpendicular to the Galactic disc \citep{Binney_2012}. These oscillations are expected to have a larger amplitude for older stars, which should be `dynamically hotter' than younger stars due to the progressive impact of random stellar encounters and disc heating processes like transient spiral waves \citep{Sellwood_2014, Aumer_2016, Aumer_2016_AVR, Ting_2026}. $J_z$ is defined as follows:
\begin{eqnarray}
    J_z \equiv \frac{1}{2\pi} \oint v_z \, \mathrm{d}z \, ,
\end{eqnarray}
where $v_z$ is the vertical velocity and the integration is done over a full vertical oscillation. $J_z$ is an integral of motion in a stationary axisymmetric potential, making it adiabatically invariant to changes in the potential over timescales longer than a single orbit \citep{Binney_2008}. Since we only observe a star at one point along its orbit, we have to make assumptions regarding the Galactic potential to calculate $J_z$ \citep{Binney_2012, Ting_2019}. While this can introduce inaccuracies, our aim is to use $J_z$ merely as an age proxy to check if the stars removed by the AC cuts likely have overestimated ages. This makes our results sensitive only to relative values of $J_z$ -- a fixed rescaling of all the $J_z$ values would not affect the results in this section beyond the zero point.

$J_z$ is based purely on the astrometry and radial velocity (RV), both of which should be quite accurate thanks to \emph{Gaia} astrometry and the LAMOST spectra, respectively. It is quite unlikely that whatever observational difficulties caused an inaccurate age estimate would also cause inaccuracies in the parallax or proper motion that shift the calculated $J_z$ along the expected trend with age. We therefore use the $J_z$ values listed in the \citetalias{Xiang_2022} catalogue. Their methods section indicates that $J_z$ was found using \textsc{galpy} \citep{Bovy_2015_GALPY}, which was also used to calculate the similarly defined radial action $J_r$ and azimuthal action $J_\phi$.

To use $J_z$ as an age proxy, we need to fit it to the most important variables that it depends on, using some sample where we consider the YY ages to be reliable. The sample we use to `train' our $J_z$ model is our nominal sample restricted to ages of $2-14.5$~Gyr, given the small number of stars outside this range. Besides depending on age, we also expect $J_z$ to depend on metallicity, both because the metallicity gradient means metal-poor stars were typically born at larger radii where the disc is thicker, and because the metal-poor tail at any age contains accreted material. This means we generally expect lower $J_z$ for metal-rich stars, which are the ones preferentially removed by our AC cuts. If these stars do indeed have low $J_z$, part of the reason is simply a general trend with metallicity, which we must account for. We also expect that $J_z$ depends on $D_\odot$ because a dynamically hotter star is typically further from the Galactic disc -- and thus from us. We therefore fit the following functional form to the stars in our training sample:
\begin{eqnarray}
    \Jtz ~=~ \Jtzo + F_A \left( A \right) + F_{\mathcal{M}} \left( \mathcal{M} \right) + F_D \left( D_\odot \right),
    \label{Jtz_relation}
\end{eqnarray}
where $\Jtz \equiv \log_{10} J_z$ with $J_z$ in kpc~km/s, $\mathcal{M} \equiv \left[ \mathrm{Fe}/\mathrm{H} \right]$, $A$ is in Gyr, and $D_\odot$ is in kpc. The three $F$ functions capture the dependence of $\Jtz$ on the subscripted quantity. Since only the sum of these functions is relevant, we absorb additive constants into the zero point $\Jtzo$ and require the functions to vanish at a reference point. This is the centre of the $\tanh$ function used for $F_A$, Solar metallicity for $F_{\mathcal{M}}$, and the Solar position for $F_D$.

We find that we can express each $F_i$ ($i = A, \mathcal{M}, D$) as a simple algebraic function that provides an adequate fit (as discussed further in Appendix~\ref{Vertical_action_model}). Vertical heating is expected to occur primarily within the Galactic disc, where dynamically hotter stars spend less of their time \citep{Ting_2026}. We therefore expect vertical heating to saturate, motivating a $\tanh$ functional form for $F_A$. We find that a good model for $F_D$ is a cubic. No single polynomial is adequate for $F_{\mathcal{M}}$. Averaged over the range, the metallicity dependence is about $3\times$ steeper at $\mathcal{M} \la -0.6$ compared to higher $\mathcal{M}$. We attribute this to the metal-poor stars being predominantly in the thick disc, probably including a component of accreted material with a different heating history. We therefore use two quadratics joined at $\mathcal{M} = -0.6$. We require $F_{\mathcal{M}}$ itself to be continuous at this transition, but allow its gradient to change abruptly. We fit the break metallicity along with the remaining parameters. Although we could not constrain its location very precisely, a break is clearly needed if we stick to a simple parabolic dependence on either side. We thereby obtain the following parameters:
\begin{eqnarray}
    \Jtzo &=& 0.245, \nonumber \\
    F_A &=& 0.540 \tanh \left( \frac{A - 6.07}{5.00} \right), \nonumber \\
    F_{\mathcal{M}} &=& \begin{cases}
                -0.118 \mathcal{M} + 0.339 \mathcal{M}^2 \, \quad &\left( \mathcal{M} > -0.6 \right), \\
                +0.164 + 0.106 \mathcal{M} + 0.257 \mathcal{M}^2 \quad &\left( \mathcal{M} < -0.6 \right),
                        \end{cases} \nonumber \\
    F_D &=& 0.754 \, D_\odot - 0.255 \, D_\odot^2 + 0.042 \, D_\odot^3 \, ,
    \label{Jz_terms}
\end{eqnarray}
valid over $4 < A/\mathrm{Gyr} < 14.5$, $-1.5 < \mathcal{M} < 0.45$, and $D_\odot < 3.6$~kpc. We avoid fitting $F_A$ to stars with $A = 2-4$~Gyr to avoid the results being dominated by young stars, given our aim is to have accurate results for old stars. Figure~\ref{jz_partials} shows that our functional forms for the $F_i$ provide adequate fits to the residuals obtained if we only fit $\Jtz$ as a function of the remaining parameters. In particular, the break at sub-Solar metallicity and the saturation towards higher ages are both clearly apparent. This saturation makes it difficult to directly estimate the age from $\Jtz$, so we instead focus only on the general tendency for older stars to be dynamically hotter at a given metallicity and $D_\odot$.

Our main aim is to check if the stars removed by the AC cuts likely have overestimated ages, as we suspect from their high metallicity and/or low [$\alpha$/Fe]. We therefore define the AC-rejected stars as those stars which are not present in our nominal sample but are present if we relax the AC cuts, while stars in our nominal sample are referred to as `retained' and pass the AC cuts. We then use Equation~\ref{Jz_terms} to define the residual $\DJtz$, the difference between a star's catalogued $\Jtz$ and the value predicted by our relation (subtracting the latter). We cannot compare $\Jtz$ directly between the retained and AC-rejected stars because they are by construction different in metallicity. Since more metal-rich stars typically have a lower vertical action, any tendency for the AC-rejected stars to have lower $\Jtz$ would be partly due to this trend. Only after accounting for it can we get reliable results.

In the crucial $A = 13-14.5$~Gyr range most relevant to inferring $A_\star$, the retained stars have $D_\odot = 1.48^{+0.55}_{-0.54}$~kpc while the AC-rejected stars have $D_\odot = 1.58^{+0.46}_{-0.55}$~kpc.\footnote{Median values and percentiles $15.87-84.13$, which is equivalent to the $1\sigma$ range for a Gaussian distribution.} $F_D$ therefore nearly cancels in a comparison between the two populations, so our $F_D$ model is not that important. In particular, the results below barely change if we replace our cubic model for $F_D$ with a linear model -- even though this is a very bad fit. In a comparison where we already restrict to narrow age intervals, $F_A$ is likewise not that important. This is especially true given the saturation at high ages: the slope of $0.096$~dex/Gyr over $A = 4 - 11$~Gyr shallows to just $0.010$~dex/Gyr by 13.75~Gyr. The most important dependence of $\Jtz$ for our analysis is therefore on metallicity, in which the retained and AC-rejected samples differ by construction.

\begin{figure}
    \centering
    \includegraphics[width=\linewidth]{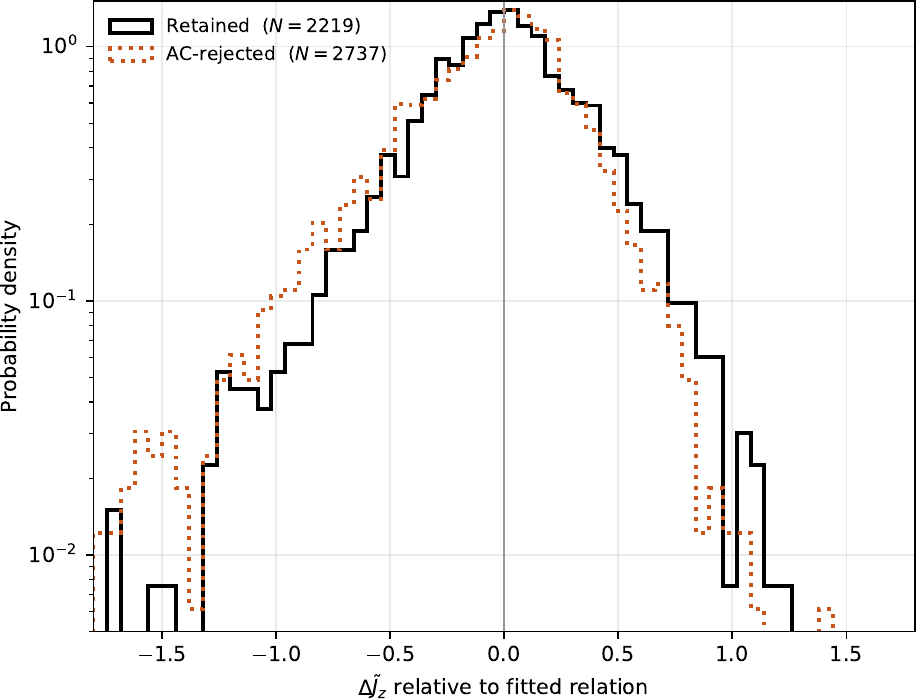} 
    \vspace{4pt}
    \includegraphics[width=\linewidth]{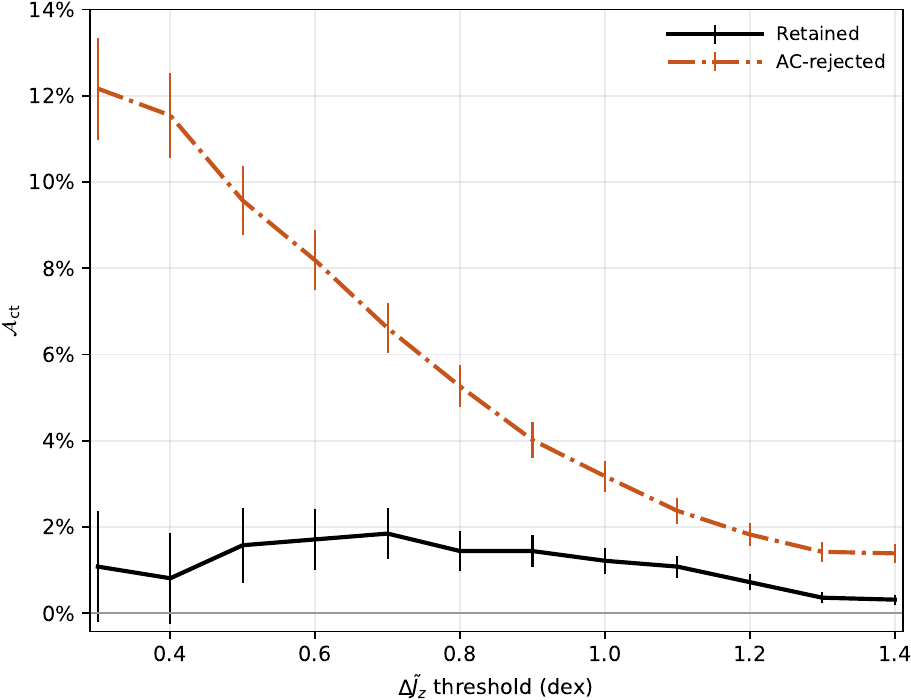} 
    \caption{\emph{Top}: Distribution of the residual $\DJtz$ about Equation~\ref{Jz_terms} for the retained stars that pass all our quality cuts (solid black) and the AC-rejected stars (dotted red), shown for stars with $A = 13-14.5$~Gyr that are most relevant to $A_\star$. The vertical scale is logarithmic to better highlight the tails, since the medians agree to 0.013~dex. \emph{Bottom}: The cold-tail asymmetry $\Act$ of the retained stars (solid black) and the AC-rejected stars (dot-dashed red) as a function of the threshold adopted in Equation~\ref{cold_tail_definition}. Poisson uncertainties are shown as solid vertical lines in the same colour.}
    \label{jz_distributions}
\end{figure}

The top panel of Figure~\ref{jz_distributions} shows the distribution of $\DJtz$ for the retained and AC-rejected stars with ages of $13-14.5$~Gyr. The two distributions have a similar mode, but the AC-rejected stars have a skewed distribution of $\Jtz$ residuals from the expected relation (Equation~\ref{Jz_terms}). We associate this with a contaminating sub-population of dynamically colder stars whose ages have likely been overestimated by a few Gyr. If this is the case for a few percent of the stars, they would hardly move the mean or median of the whole distribution, but they might show up in its tail. We therefore define the cold-tail asymmetry $\Act$ as the fraction of stars in some population of size $N$ for which $\Jtz$ lies further than some specified threshold below the expected value given the age, metallicity, and distance of each star less the fraction lying further than that same threshold above the expected value. Adopting 0.7~dex for the threshold, this definition becomes the following:
\begin{eqnarray}
    \Act ~\equiv~ \frac{N \left( \DJtz < -0.7 \right) - N \left( \DJtz > +0.7 \right)}{N} \, .
    \label{cold_tail_definition}
\end{eqnarray}
Working with a measure of asymmetry has the major advantage that symmetric measurement scatter contributes equally to both counts and cancels exactly, so neither a larger observational error nor a broader intrinsic distribution can manufacture a signal. $\Act$ is sensitive only to a genuinely one-sided displacement.

The bottom panel of Figure~\ref{jz_distributions} shows $\Act$ as a percentage for different choices of the threshold used to define it. We find that for the retained stars with $A = 13-14.5$~Gyr, $\Act$ peaks when the threshold $\approx 0.7$~dex. We therefore expect our cold-tail statistic to be most sensitive to a dynamically cold sub-population if we adopt this threshold, so we use it in what follows unless stated otherwise. We find that $\Act = 1.85 \pm 0.58\%$ for the retained stars, but the AC-rejected stars have a much higher $\Act = 6.61 \pm 0.58\%$, a difference of $5.8\sigma$. The errors stated here are conservative because they assume Poisson statistics for simplicity -- using binomial statistics instead would further enhance the significance. A very large contrast between the two populations is apparent for essentially any choice of threshold used to define $\Act$. At a lower threshold of 0.5~dex, the retained stars have $\Act = 1.58 \pm 0.87\%$, while the AC-rejected stars have $\Act = 9.57 \pm 0.81\%$, which is almost $6\times$ larger. If we use a larger threshold of 1.1~dex to focus on the behaviour further into the tail regions, $\Act = 1.08 \pm 0.25\%$ for the retained stars, while the AC-rejected stars have $\Act = 2.37 \pm 0.30\%$. If we stick to the 0.7~dex threshold and instead expand the age range downwards to cover $12-14.5$~Gyr, we find that $\Act = 7.13 \pm 0.47\%$ for the 4477 AC-rejected stars against $2.55 \pm 0.31\%$ for the 8446 retained stars, a difference of $8.2\sigma$ in place of the $5.8\sigma$ above. We cannot meaningfully extend the age range upwards because there are only 36 retained stars at $A = 14.5 - 15$~Gyr and only 8 stars at even higher ages, preventing a useful comparison with the more numerous AC-rejected stars. Since these stars have higher $\Act$ than the retained stars regardless of the specific threshold used to define $\Act$ in Equation~\ref{cold_tail_definition}, it is not possible to use a ceiling on $\left| \DJtz \right|$ as a quality cut. This is because even if we adopt a very restrictive threshold that removes the vast majority of the sample, it is still not possible to clean it as well as we can achieve with our AC cuts.

\begin{figure}
    \centering
    \includegraphics[width=\linewidth]{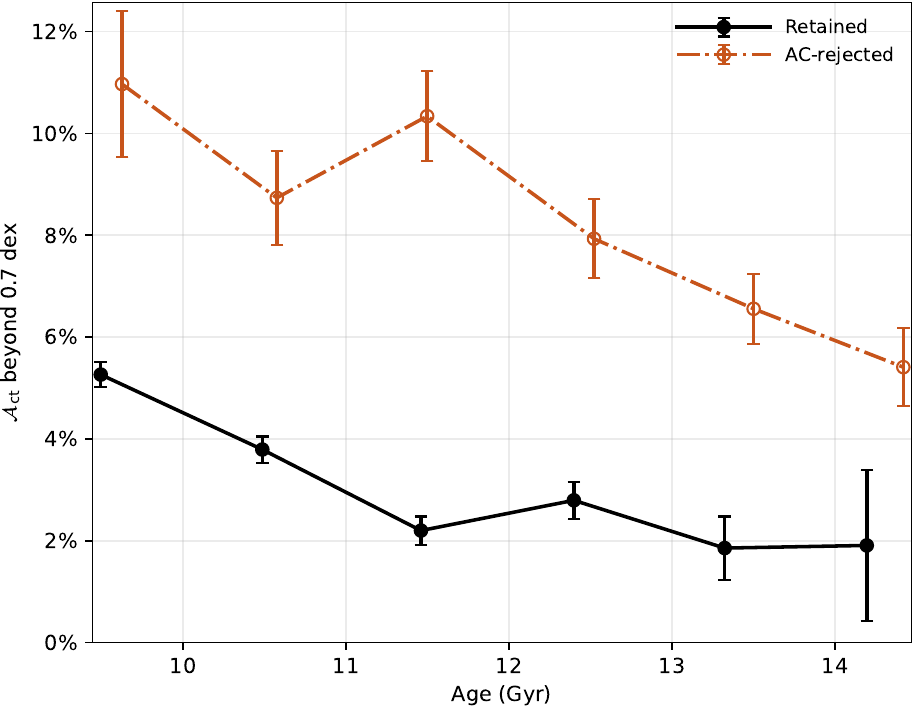} 
    \includegraphics[width=\linewidth]{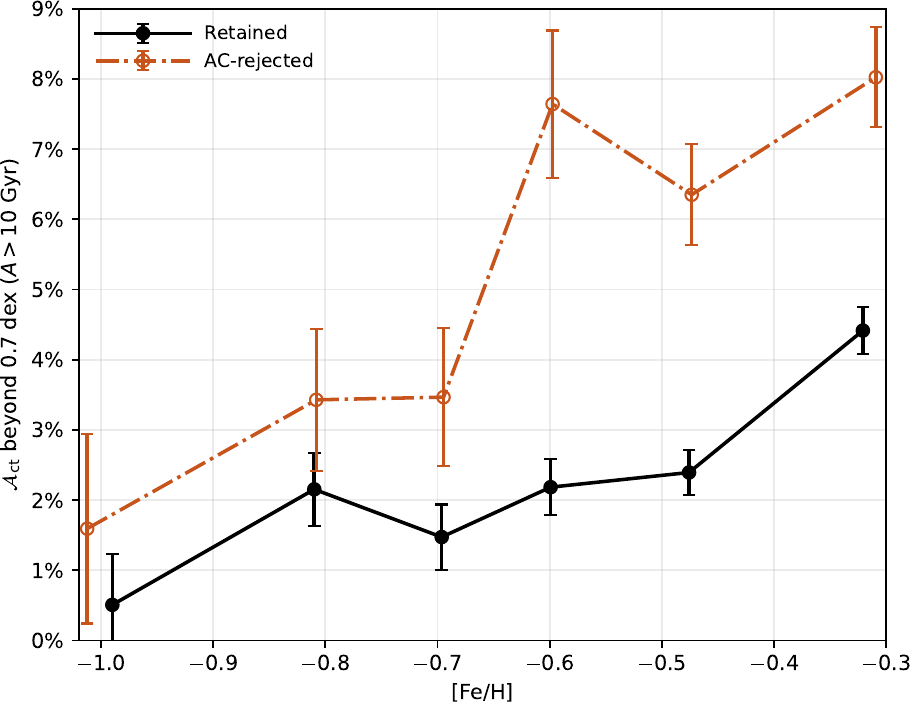}
    \caption{The cold-tail asymmetry $\Act$ (Equation~\ref{cold_tail_definition}) for the stars in our nominal sample (solid black points) and those removed by the AC cuts (open red points), shown against the reported age (\emph{top}) and metallicity (\emph{bottom}). The top panel covers $A = 9-15$~Gyr, while the bottom panel covers stars older than 10~Gyr.}
    \label{jz_cold_tail_vs_age}
\end{figure}

To gain further insight into the dynamically cold sub-population with likely overestimated ages, we can find how $\Act$ depends on YY age and metallicity. The results are shown in Figure~\ref{jz_cold_tail_vs_age}. Its top panel shows that for the retained stars, $\Act$ actually \emph{decreases} from $5.26 \pm 0.25\%$ at $9-10$~Gyr to $1.86 \pm 0.62\%$ at $13-14$~Gyr, though the results are basically flat around 2\% for $A \ga 11$~Gyr. The somewhat lower contamination in the age range most relevant to a determination of $A_\star$ shows that our AC cuts are able to clean our sample most effectively where it matters most. Without these cuts, our sample would include the AC-rejected stars. These have $\Act = 10.97 \pm 1.44\%$ at $9-10$~Gyr and $6.56 \pm 0.69\%$ at $13-14$~Gyr, which in both cases is far higher than for our nominal sample. The ratio of $\Act$ in the AC-rejected stars to that in the retained stars rises from about 2.3 at $10-11$~Gyr to 3.6 at $13-14.5$~Gyr, suggesting that our AC cuts are most effective near $A_\star$.

One may wonder if the AC cuts can be replaced by a simple ceiling on the metallicity, since the oldest stars should be metal-poor. We explore this in the bottom panel of Figure~\ref{jz_cold_tail_vs_age}, which shows $\Act$ against metallicity. Since the AC cuts remove stars above a diagonal line in the age--metallicity plane (Equation~\ref{Age_metallicity_cutoff_line}), it is difficult to disentangle age and metallicity. We can achieve this to some extent by limiting the comparison to $A > 10$~Gyr, as done here. Without this restriction, the median age within a metallicity bin can differ by as much as 6~Gyr, but we close the gap to just $1.7-2.0$~Gyr with the 10~Gyr age floor. We do not impose an age ceiling because the retained sample automatically runs out of stars towards high ages. An age ceiling on the AC-rejected sample would remove precisely those stars where we have reason to believe the ages were overestimated. Bearing in mind the difficulties of the comparison, our results show that the AC-rejected stars have a higher $\Act$ than the retained stars in every metallicity bin. A pure metallicity ceiling is thus not a very effective replacement for the AC cuts. But since the retained stars have no detectable cold-tail asymmetry at $[\mathrm{Fe/H}] \la -0.9$, it may be useful to have a metallicity ceiling here in addition to our nominal quality cuts. Our results suggest that this would lead to a particularly clean sample, though it would inevitably be far smaller. We follow this approach in Section~\ref{Early_chemistry_analysis}.

\begin{figure}
    \centering
    \includegraphics[width=\linewidth]{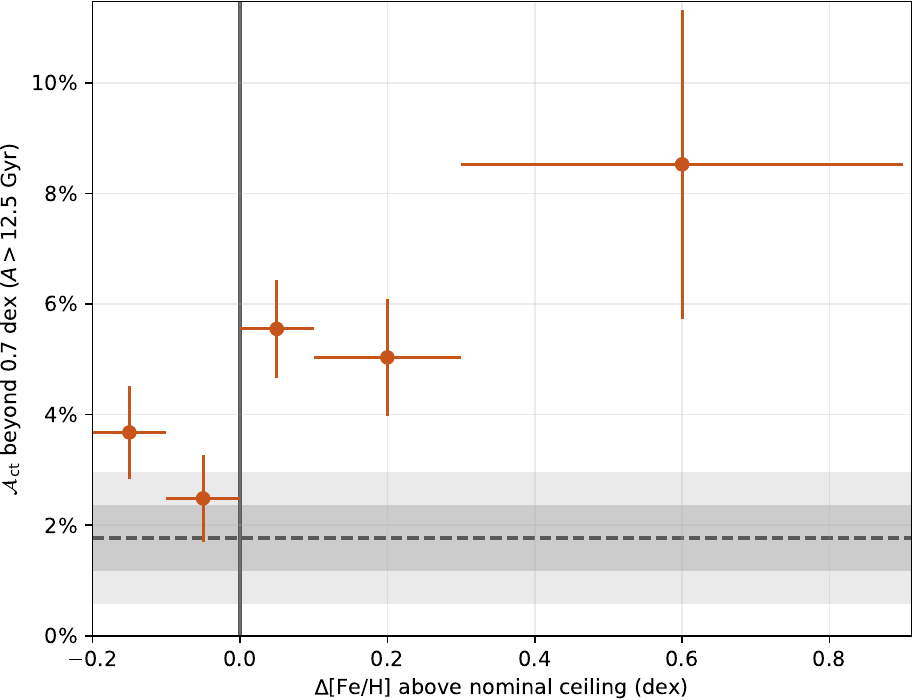} 
    \caption{The cold-tail asymmetry $\Act$ for stars older than 12.5~Gyr which pass every cut except the metallicity ceiling (Equation~\ref{Age_metallicity_cutoff_line}), shown against the offset $\DFeH$ above it (solid red points with Poisson uncertainties). The $\Jtz$ model (Equation~\ref{Jtz_relation}) is fitted to stars with $\DFeH < -0.2$, below the plotted range. This means none of the stars in the age--metallicity strips for which $\Act$ is plotted here were used to train the $\Jtz$ model used to calculate the $\DJtz$ values entering into $\Act$. The horizontal dashed black line shows $\Act$ for stars older than 12.5~Gyr that were used to train the model, with the shaded grey bands showing the $1\sigma$ and $2\sigma$ uncertainties. Notice the sharp increase in $\Act$ upon crossing the nominal cutoff line, which we show as a thick grey vertical line.}
    \label{jz_cold_tail_vs_FeHoffset}
\end{figure}

The age--metallicity distribution suggests that instead of a simple ceiling on [Fe/H], we should reject stars above a diagonal line parallel to the high-density downward-sloping ridgeline evident on Figure~\ref{age_feh_figure}. Although its slope is fairly clear from the figure, the intercept is less so. We can set it based on $\Act$ in strips parallel to the nominal cutoff line. The results are shown in Figure~\ref{jz_cold_tail_vs_FeHoffset} for stars older than 12.5~Gyr which pass every other cut. To ensure the $\Jtz$ model is not based on stars whose $\Act$ is plotted here, we fit the $\Jtz$ model (Equation~\ref{Jtz_relation}) only to stars which pass all our quality cuts but also have $\DFeH < -0.2$, where $\DFeH$ is the amount by which a star's [Fe/H] exceeds the nominal ceiling at its age (Equation~\ref{Age_metallicity_cutoff_line}). In the training sample where we expect reported ages to be more reliable, $\Act = 1.77 \pm 0.60\%$, which may simply represent inherent asymmetry of the $\DJtz$ distribution rather than a population with overestimated ages. In the $\DFeH = (-0.2, -0.1)$ strip, $\Act$ rises slightly to $3.68 \pm 0.84\%$, but it then decreases to $2.48 \pm 0.79\%$ for $\DFeH = (-0.1, 0)$, the age--metallicity strip just below the nominal cutoff line. This is within $1\sigma$ of the training sample, though the discrepancy in both cases is $<2\sigma$. When we consider the next age--metallicity strip of the same width immediately above the nominal age--metallicity cutoff line, $\Act$ rises sharply to $5.55 \pm 0.89\%$, which exceeds the training sample by $3.5\sigma$. At even higher $\DFeH$, though we are forced to use wider $\DFeH$ bins due to a smaller sample size, $\Act$ is still clearly higher than for the training sample and the strips outside it yet below our nominal cutoff line ($-0.2 < \DFeH < 0$). Our vertical action-based proxy for a contaminating population with overestimated ages therefore rises sharply precisely upon crossing our nominal cutoff line in the age--metallicity diagram.

Since the AC-rejected stars necessarily occupy a distinct part of the parameter space compared to the retained stars used to train our model for $J_z$, it is possible that the model is less accurate for the AC-rejected stars. It is also possible that they have larger uncertainties in $J_z$. To mitigate against the possibility that the AC-rejected $\Act$ is inflated by a broader and/or offset distribution compared to the retained sample, we can simply subtract the mean value of $\DJtz$ and normalise the distribution to its own standard deviation, considering the two samples separately. In this case, the tail region used to compute the cold-tail asymmetry must be defined as a fixed number of standard deviations from the mean. The results of this approach are shown in Appendix~\ref{Dimensionless_cold_tail}. The behaviour of this dimensionless version of $\Act$ remains very similar, including especially the sharp jump in the cold-tail asymmetry upon crossing $\DFeH = 0$. This provides further justification for our placement of the age--metallicity ceiling, which is the AC quality cut that our final $A_\star$ inference is most sensitive to (Section~\ref{Results}).

So far in this section, we have considered the AC cuts as a single quality cut, but in fact it consists of separate age-dependent cuts on [Fe/H] and [$\alpha$/Fe]. To disentangle the effects of these cuts, we need to classify the AC-rejected stars based on which of these two cuts they fail. Table~\ref{jz_cold_tail_table} shows $\Act$ for stars with ages of $13-14.5$~Gyr, split according to whether they pass or fail the age--metallicity cut (Equation~\ref{Age_metallicity_cutoff_line}) and/or the age--[$\alpha$/Fe] cut (Equation~\ref{Age_alpha_Fe_cutoff_line}).\footnote{Some 122 AC-rejected stars (799 at all ages) are not shown here because they are indirectly excluded by the AC cuts due to the 7\% tighter trendline it causes between YY and FLAME ages. These stars are not retained in our nominal sample despite passing both AC conditions, so they cannot be represented in Table~\ref{jz_cold_tail_table}. Their $\Act = -0.8 \pm 2.7\%$, which combined with their small number means they do not contribute to the cold-tail asymmetry for the AC-rejected stars.} The retained stars in this age range have $\Act = 1.85 \pm 0.58\%$, similar to the result evident in the top panel of Figure~\ref{jz_cold_tail_vs_age}. This rises to $5.71 \pm 1.68\%$ for the stars rejected solely by the [$\alpha$/Fe] cut. Importantly, this increase cannot be assigned to a difference in metallicity because the median [Fe/H] is very similar in both cases ($-0.79$ for the 2219 retained stars against $-0.82$ for the 315 stars that fail the [$\alpha$/Fe] cut but pass the [Fe/H] cut). Given also the narrow age range used for Table~\ref{jz_cold_tail_table}, it is clear that both cells in its top row occupy much the same position in the age--metallicity plane. We find that this $\Act$ excess is clearly present in stars with ages down to 10~Gyr. In the $12-13$~Gyr age range, $\Act = 2.79 \pm 0.37\%$ for the retained stars and $6.38 \pm 2.13\%$ for the stars rejected only by the [$\alpha$/Fe] cut. Stars at younger ages are less important to setting $A_\star$, but the larger sample size makes the excess even clearer.

\begin{table}
    \centering
    \begin{tabular}{lcc} 
        \hline
        [Fe/H] & \multicolumn{2}{c}{[$\alpha$/Fe] floor} \\
        ceiling & Pass & Fail \\ \hline
        \multirow{2}{*}{Pass} & $1.85 \pm 0.58\%$ & $5.71 \pm 1.68\%$ \\
                              & ($N = 2219$)    & ($N = 315$) \\ [3pt]
        \multirow{2}{*}{Fail} & $4.97 \pm 0.68\%$ & $12.41 \pm 1.43\%$ \\
                              & ($N = 1631$)    & ($N = 669$) \\ \hline
    \end{tabular}
    \caption{The cold-tail asymmetry $\Act$ for stars with $A = 13-14.5$~Gyr passing all our quality cuts besides the AC cuts, classified according to whether they pass or fail the two AC cuts individually (i.e., the [Fe/H] ceiling and [$\alpha$/Fe] floor). We indicate the sample size $N$ in brackets. The top left entry represents our nominal sample, while the other three entries show different ways of being rejected by the AC cuts. A further 122 AC-rejected stars in this age range are not listed here as they pass both AC conditions but fail the FLAME age consistency test once it is recalibrated on the cleaner sample yielded by the AC cuts (see the text).}
    \label{jz_cold_tail_table}
\end{table}

We can go one step further by considering all stars with $A = 11-14.5$~Gyr and weight the retained stars to match the ($A$, [Fe/H]) distribution of the stars rejected only due to their low [$\alpha$/Fe]. In other words, every retained star at these ages is given a weight equal to the number of [$\alpha$/Fe]-rejected stars in the corresponding ($A$, [Fe/H]) bin divided by the number of retained stars in that bin. The weighted retained stars have $\Act = 2.29 \pm 0.99\%$, while the [$\alpha$/Fe]-rejected stars have $\Act = 8.43 \pm 1.18\%$, which is $4\sigma$ higher. This reinforces our claim that the [$\alpha$/Fe]-rejected stars cannot have a higher cold-tail asymmetry due to differences in their age--metallicity distribution.

The final row of Table~\ref{jz_cold_tail_table} shows results for stars with $A = 13-14.5$~Gyr which fail the age--metallicity cut (Equation~\ref{Age_metallicity_cutoff_line}). These stars are not in our retained sample, but we can still distinguish whether they pass the age--[$\alpha$/Fe] condition (Equation~\ref{Age_alpha_Fe_cutoff_line}). The 1631 stars which pass it have $\Act = 4.97 \pm 0.68\%$, which is already considerably higher than the $1.85 \pm 0.58\%$ of the 2219 retained stars. The 669 stars which fail both AC conditions have $\Act = 12.41 \pm 1.43\%$, which is far higher than for stars that pass at least one of the AC cuts. This is to be expected if failing either AC cut independently selects for stars with a higher incidence of inaccurate age likelihoods.

\begin{figure}
    \centering
    \includegraphics[width=\linewidth]{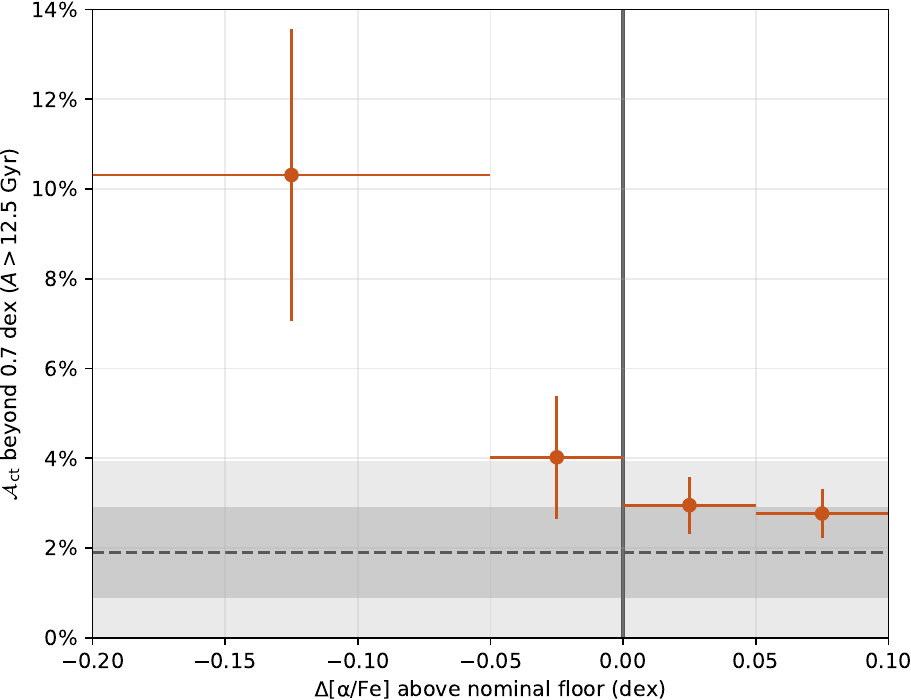} 
    \caption{Similar to Figure~\ref{jz_cold_tail_vs_FeHoffset}, but now showing $\Act$ as a function of offset from the [$\alpha$/Fe] floor (Equation~\ref{Age_alpha_Fe_cutoff_line}). The results here are for stars older than 12.5~Gyr which pass every cut except this floor, plotted against the offset $\DalphaFe$ above it. Note that the floor removes stars to the left of the solid grey vertical line at $\DalphaFe = 0$, which is opposite to Figure~\ref{jz_cold_tail_vs_FeHoffset}. The $\Jtz$ model (Equation~\ref{Jtz_relation}) used here is a fit to stars with $\DalphaFe > 0.05$, since a higher threshold leaves too few training stars at the oldest ages. The rightmost bin is therefore not independent of the training data, but the other bins are. The dashed horizontal line shows $\Act$ for stars older than 12.5~Gyr with $\DalphaFe = (0.1, 0.2)$. The shaded grey bands show the $1\sigma$ and $2\sigma$ uncertainties.}
    \label{jz_cold_tail_vs_alphaFeoffset}
\end{figure}

Following our approach in Figure~\ref{jz_cold_tail_vs_FeHoffset}, we can estimate the appropriate intercept for the age-dependent [$\alpha$/Fe] floor by computing $\Act$ in strips parallel to the nominal floor. We parametrise these strips using $\DalphaFe$, the amount by which the [$\alpha$/Fe] of a star exceeds the nominal floor at its age (Equation~\ref{Age_alpha_Fe_cutoff_line}). The results are shown in Figure~\ref{jz_cold_tail_vs_alphaFeoffset} for stars older than 12.5~Gyr which pass every other cut, including the age--metallicity cut. Since few old stars lie far above the nominal [$\alpha$/Fe] floor, we train our $\Jtz$ model using stars with $\DalphaFe > 0.05$. Stars with $\DalphaFe = (0.1, 0.2)$ at these ages have $\Act = 1.90 \pm 1.02\%$, which we can think of as a control. This is consistent with $\Act$ in both bins of width 0.05~dex immediately above the nominal [$\alpha$/Fe] floor, suggesting that these bins are not much contaminated. The equally wide bin below the nominal [$\alpha$/Fe] floor has $\Act = 4.02 \pm 1.38\%$. Though this is twice the control value, the large uncertainties mean the discrepancy is only $1.2\sigma$. Even so, this bin has a higher $\Act$ than any bin above the floor, suggesting that contamination sets in at or near the nominal floor. The data cannot be more precise about exactly where this occurs because 80\% of the old stars removed by the floor lie within 0.05~dex of it, which is less than twice the typical [$\alpha$/Fe] uncertainty of 0.03~dex, smearing out any genuine jump in $\Act$ with decreasing $\DalphaFe$. We can largely avoid this issue by going even further below the nominal floor. The bin covering $\DalphaFe = (-0.2, -0.05)$ has $\Act = 10.31 \pm 3.26\%$, which exceeds the control by $2.5\sigma$. This suggests that while it may be safe to lower our nominal [$\alpha$/Fe] floor by 0.05~dex, relaxing it any further would likely introduce contamination. We err on the side of caution with our [$\alpha$/Fe] quality cut, but we will see later that lowering the floor by 0.05~dex barely affects the inferred $A_\star$.

Although our results using $J_z$ help justify the AC cuts, we cannot directly use $J_z$ as a quality cut due to the 0.37~dex scatter about our $\Jtz$ model, which corresponds to several Gyr in age. The $\Act$ parameter is therefore best thought of as a population-level measure of sample quality rather than a star-level diagnostic of whether the age likelihood is inaccurate. We also stress that since our $\Jtz$ model is calibrated using YY ages, it is not sensitive to a coherent shift in ages due to systematic deficiencies with the YY isochrone model. Uncertainties in the YY treatment of stellar evolution are discussed further in Section~\ref{Stellar_modelling_uncertainties}.

Since $\Act$ is only a few percent, the great majority of the AC-rejected stars at ages relevant to $A_\star$ are not dynamically distinguishable from retained stars, so it seems likely that they are genuinely old. Since the downturn in fractional age errors towards high ages in our initial sample is caused by at best only a few thousand stars (see Figure~\ref{Age_error_turnover_none}) and our AC cuts remove 9561 stars, it is to be expected that they remove far more stars than is necessary to clean our sample. The AC cuts are therefore necessarily a blunt instrument, discarding a large population of genuinely old stars to remove a small number of younger stars incorrectly assigned a high age. That being said, the loss of 9561 stars still leaves a very large nominal sample of size 146,810, of which 8490 are older than 12~Gyr and 2263 are older than 13~Gyr (see Appendix~\ref{Sample_size_revised_AC_cuts}). The AC cuts are thus successful in leaving a viable sample without a pathological downturn in the fractional age uncertainty at high ages or a significant excess of stars with $\DJtz \ll 0$.

\subsection{Why the AC cuts help}
\label{Why_AC_cuts_help}

We have seen that the stars removed by the AC cuts probably contain a population with overestimated ages. Since the age estimates rely on a high-quality spectrum, it is natural to explore whether the problem might be related to the SNR of the spectrum. \citetalias{Xiang_2022} already impose that $\mathrm{SNR} \geq 20$, so we do not use the SNR in any of our quality cuts. Even so, the AC cuts can still influence the SNR indirectly via the apparent magnitude. We account for this using magnitude-weighting.

We compare the retained and AC-rejected samples defined previously using the ratio of their SNR distributions at matched
percentile.
\begin{eqnarray}
    \widetilde{S}_x ~\equiv~ \frac{\mathrm{SNR}_x^{\text{AC-rejected}}}{\mathrm{SNR}_x^{\mathrm{retained}}} \, ,
    \label{S_tilde_definition}
\end{eqnarray}
where the superscript indicates the sample and the subscript denotes which percentile we are considering. $\widetilde{S}_{50}$ is thus the ratio of medians, while $\widetilde{S}_x < 1$ means the AC-rejected stars are the more poorly observed. We compare percentiles 10, 50, and 90 so that we can check the behaviour of both the overall SNR distribution and its tails. Uncertainties on $\widetilde{S}_x$ are obtained by bootstrap resampling. The two samples are disjoint by construction, so we separately resample the retained and AC-rejected populations. We draw 4000 pairs of samples with replacement, each sample having the same size as the parent population. We then apply Equation~\ref{S_tilde_definition} to each pair and quote the median and $1\sigma$ Gaussian equivalent percentile range of the resulting distribution of $\widetilde{S}$ ratios. Any difference from unity must be partly due to the AC cuts selecting on average somewhat fainter stars (median $m_\mathrm{G} = 15.00$ against 14.77, respectively, at ages of $13-14$~Gyr). To account for this, the retained sample is weighted to match the $m_\mathrm{G}$ distribution of the AC-rejected stars. This is fairly straightforward here given the large sample sizes and the significant overlap in the $m_\mathrm{G}$ distributions.

\begin{figure}
    \centering
    \includegraphics[width=\linewidth]{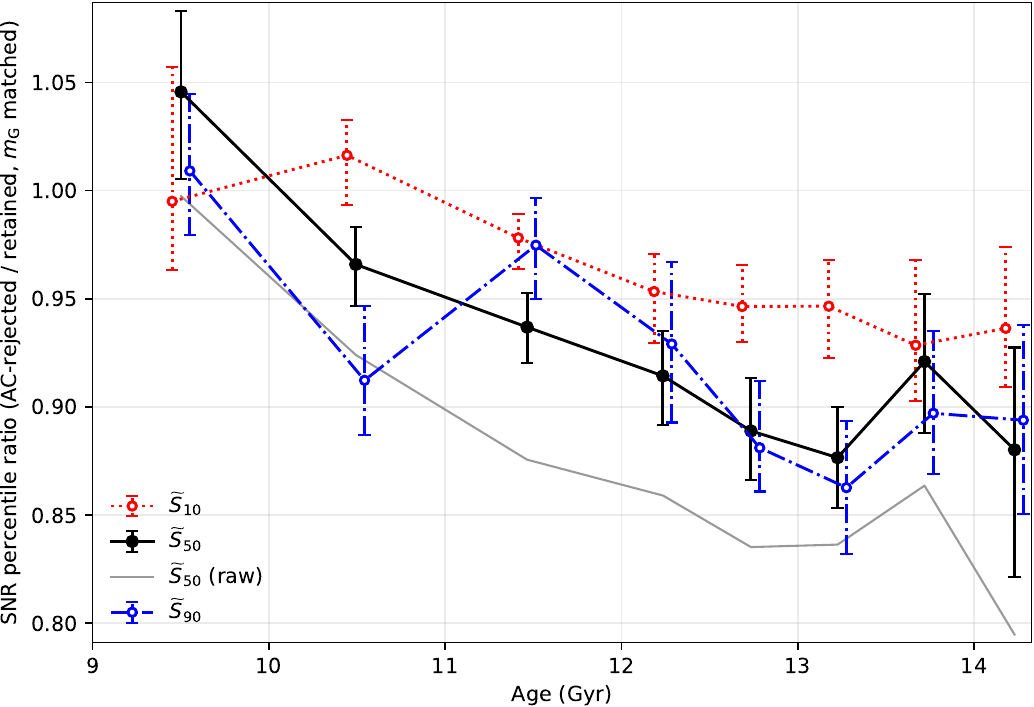} 
    \caption{The ratio $\widetilde{S}_x$ (Equation~\ref{S_tilde_definition}) between the SNR at percentile $x$ in the AC-rejected stars relative to that in the retained stars weighted to match the magnitude distribution of the AC-rejected stars. We show results for $x = 10$ (dotted red), $x = 50$ (the median; solid black), and $x = 90$ (dot-dashed blue). The thin grey line without uncertainties shows results without magnitude-weighting the retained stars. In the other cases, vertical lines show the uncertainty as $1\sigma$ Gaussian equivalent percentiles, found using 4000 bootstrap resamples of the two populations independently since they have no overlap (see the text). The results are plotted at the median age of the stars used for each bin, but since this is the same regardless of $x$, we dither $\widetilde{S}_{10}$ and $\widetilde{S}_{90}$ by $\pm 0.05$~Gyr for clarity (the median is not dithered). The SNR distributions underlying the results shown here are presented in Figure~\ref{snr_histograms}.}
    \label{snr_ratio_vs_age}
\end{figure}

Figure~\ref{snr_ratio_vs_age} shows the SNR ratio at percentiles $x = 10$, 50, and 90 against age, plotted at the median age for each sample and age bin. We show results at $A = 9-14.5$~Gyr because there are hardly any AC-rejected stars at lower ages, while the retained sample runs out towards higher ages. In the first age bin covering $9-10$~Gyr, all three SNR percentiles are consistent within or just beyond their $1\sigma$ uncertainty. Against this baseline, all three $\widetilde{S}$ values drop steadily towards higher ages, with the median ratio reaching a low of $\widetilde{S}_{50} = 0.88 \pm 0.02$ at ages of $13 - 13.5$~Gyr. The AC-rejected stars do indeed have a typically lower SNR, despite accounting for differences in $m_\mathrm{G}$. Since all three $\widetilde{S}$ values change with age in a similar manner, the whole SNR distribution shifts between the retained and AC-rejected populations at fixed age and $m_\mathrm{G}$. The one slight exception is $\widetilde{S}_{10}$, where the drop is less pronounced than for the other percentile ratios. We attribute this to the SNR floor at 20, which causes the low percentiles to pile up near the floor. The full SNR distributions in the age bins used here are shown in Figure~\ref{snr_histograms}, demonstrating that the typically lower SNR of the AC-rejected sample is not due to an easily removed tail of stars with low SNR.

The importance of magnitude-weighting the retained stars can be found by comparing the solid black line to the thin grey line on Figure~\ref{snr_ratio_vs_age}. Both lines show $\widetilde{S}_{50}$, but in the latter case without magnitude-weighting the retained stars. The magnitude-weighting is responsible for only about 1/3 of the reduction in median SNR between the retained and AC-rejected stars without any magnitude weighting, demonstrating that differences in $m_\mathrm{G}$ should be accounted for but cannot explain the typically lower SNR of the AC-rejected stars. We may also worry that since the AC-rejected stars are by construction more metal-rich, they are redder at fixed $T_\mathrm{eff}$, making them slightly fainter in the LAMOST $g$ band at fixed $m_\mathrm{G}$. We can control for this effect by noting that only one of the AC cuts involves metallicity. Accounting for differences in both magnitude and metallicity barely affects the results, as discussed in Appendix~\ref{snr_distributions_appendix}. The results there highlight that the apparent magnitude is by far the most important factor to control for when comparing the SNR between the retained and AC-rejected populations.

The results in Figure~\ref{snr_ratio_vs_age} clearly show that the AC-rejected stars at the ages most relevant to $A_\star$ typically have a lower SNR than the retained stars. However, the differences are small compared to the overall width of the SNR distribution. To quantify this, we note that there is a 57\% probability of a randomly chosen retained star at ages of $13-14$~Gyr having a higher SNR than a randomly chosen AC-rejected star. This is only slightly above the 50\% level we expect from randomly sampling the same distribution. We also find that the AC-rejected stars at $A \approx A_\star$ have a fractional age error that is in line with what we might expect from their SNR and fractional parallax error. This means that although we can remove some stars with $A \ga 17$~Gyr based on an unusually low $\widetilde{\sigma}_A$ for the uncertainties in the inputs used to determine $A$, we are still left with a downturn in $\widetilde{\sigma}_A$ at high ages. In any case, a low $\widetilde{\sigma}_A$ cannot directly be considered a sign of poor quality data because lower uncertainties are normally a good thing. It is therefore clear that a floor on SNR beyond that already imposed by \citetalias{Xiang_2022} cannot serve as a replacement for the AC cuts. Instead, we use the SNR percentile ratio at matched $m_\mathrm{G}$ merely to suggest that a lower SNR makes it more likely for the estimated age to be inaccurate.

The significant overlap between the SNR distributions of the retained and AC-rejected populations suggests that most `low' SNR spectra lead to quite valid age estimates, presumably due to the floor of 20 already in place \citepalias{Xiang_2022}. The age estimates are very inaccurate in a small fraction of cases, which is smaller still at higher SNR. At the low spectral resolution of $R = 1800$ in LAMOST~DR7, a lower SNR exacerbates difficulties with line blending and obscures crucial age-sensitive features like pressure broadening, which is important to an estimate of the surface gravity. Uncertainties in $T_\mathrm{eff}$ compound the issue: an error of 100~K corresponds to roughly 1~Gyr in age for $A > 13$~Gyr. At lower SNR, these observational limitations become significantly more pronounced.

Since estimating the age of a star requires precisely positioning it on the colour--magnitude diagram relative to the YY isochrones, it is perhaps inevitable that in a survey as large as LAMOST, observational difficulties would lead to inaccurate age estimates in a small fraction of cases. By contrast, the metallicity and $\alpha$-enhancement can be derived directly from many lines in the spectrum, making them far more robust \citep{Hill_2022}. This makes it extremely valuable to check if the estimated age and chemistry are in line with the overall trend for the population as a whole, which is the basis for our AC cuts. While a lower SNR does appear to be a risk factor that makes an inaccurate age more likely, a failure to pass our AC cuts is a symptom of an inaccurate age estimate. Focusing on symptoms rather than risk factors is vastly more effective, explaining why our AC cuts work far better than any of the alternatives that we considered.

\section{Simple \texorpdfstring{$A_\star$}{A*} estimates from extreme value statistics}
\label{Simplified_Astar_estimates}

The almost complete lack of stars in our nominal sample with $A > 15$~Gyr indicates that $A_\star$ is almost certainly lower (Figure~\ref{Age_error_turnover_nominal}). At the same time, $A_\star$ appears to exceed 13~Gyr due to the large number of stars that are apparently older, in many cases with quite small $\widetilde{\sigma}_A$. We therefore expect that $A_\star \approx 14$~Gyr, where Figure~\ref{Age_error_turnover_nominal} does indeed reveal a fairly sharp reduction in the age distribution. An age limit around 14~Gyr is also suggested by the low-metallicity tail in Figure~\ref{age_feh_figure}. However, the large sample size implies that several $3\sigma$ or even $4\sigma$ age overestimates are to be expected. While these would in many cases only affect stars much younger than $A_\star$, sometimes such overestimates would occur for stars with a latent age close to $A_\star$. This means we cannot simply read off $A_\star$ from Figure~\ref{Age_error_turnover_nominal}. We therefore estimate $A_\star$ using methods with increasing degrees of rigour, culminating in our nominal two-stage MCMC analysis. In all cases, it is crucial to take into account the measurement uncertainties, which we expect inevitably cause some stars to have an observed $A > A_\star$ by several times $\sigma_A$.

\subsection{Individual age likelihoods}
\label{Individual_age_likelihoods}

The Bayesian isochrone analysis in \citetalias{Xiang_2022} determines a posterior inference $\LX22(A)$ for the age of each star based on observations of it, but without considering other stars. We treat this as an observational likelihood for $A$, to which we can then apply other priors not based on cosmology. $\LX22(A)$ is stored on a grid with equally spaced `knots' covering $A = 0.1-8$~Gyr inclusive in steps of 0.1~Gyr, followed by a coarser grid covering $8-20$~Gyr inclusive in steps of 0.2~Gyr. So far, we have discussed the shape of $\LX22(A)$ in terms of a single uncertainty $\sigma_A$. However, a Gaussian assumption for the error distribution might be inaccurate when inferring $A_\star$, since this necessarily implies a focus on the tail of the individual $\LX22(A)$ distributions. In the rest of this contribution, we will therefore focus on the full 140-knot $\LX22(A)$.

Our first estimate of $A_\star$ focuses on trying different values and finding the star with the greatest $\LX22(A > A_\star)$, or the highest observational likelihood that $A > A_\star$. In a large sample, there should certainly be stars where the observational $\LX22(A)$ indicates that $A > A_\star$ at high confidence. After all, stars with a true age close to $A_\star$ will sometimes have an observed $A > A_\star$ by several standard deviations, with the maximum expected deviation being larger for a larger sample. But since only a small fraction of stars have a latent age close to $A_\star$, the relevant sample size in this context is not the full sample size of $N_\mathrm{tot} = 146,810$. This is because we expect the vast majority of the stars to be younger than $A_\star$ by an amount much larger than $\sigma_A$, so these stars will not be observed to have $A > A_\star$, even with a statistically rare age overestimate.

A more rigorous analysis requires a reconstruction of the latent age distribution, which we do in Section~\ref{Reconstructed_age_distribution}. For now, we borrow its result that the latent age distribution is approximately zero for $A < 2$~Gyr, flat over the period $2-10$~Gyr, and then linearly declines to 0 at 14~Gyr. The assumed end point is not crucial to this argument because we only use this toy model to estimate the fraction of stars within `striking distance' of $A_\star$, i.e., the fraction for which statistically plausible age overestimates might lead to an observed $A > A_\star$. We estimate this at 1\% based on the above toy model and the estimate of \citetalias{Xiang_2022} that $\widetilde{\sigma}_A = 7.5\%$, representing about 1~Gyr at an age of 14~Gyr (c.f. Figure~\ref{Age_error_turnover_nominal}). If these $A_\star$-relevant stars had Gaussian age errors, the maximum plausible age overestimate would be about $3.4\sigma$. Relaxing the Gaussian assumption, the more generalised argument is that there should be one star which satisfies
\begin{eqnarray}
    \LX22 \left( \leq A_\star \right) ~\leq \frac{1}{0.01 \, N_\mathrm{tot}} \, ,
    \label{Ntot_inv_condition}
\end{eqnarray}
where $\LX22(\leq x)$ denotes the cumulative distribution function of the observationally determined $\LX22(A)$ at $A = x$ and 0.01 is the fraction of the full sample within striking distance of $A_\star$. To estimate $A_\star$ with this approach, we find the number of stars which satisfy Equation~\ref{Ntot_inv_condition} for a range of possible $A_\star$. We expect the number of such stars to decrease with increasing $A_\star$, so the idea here is to check when there is only one star passing this condition.

\begin{figure}
    \centering
    \includegraphics[width=\linewidth]{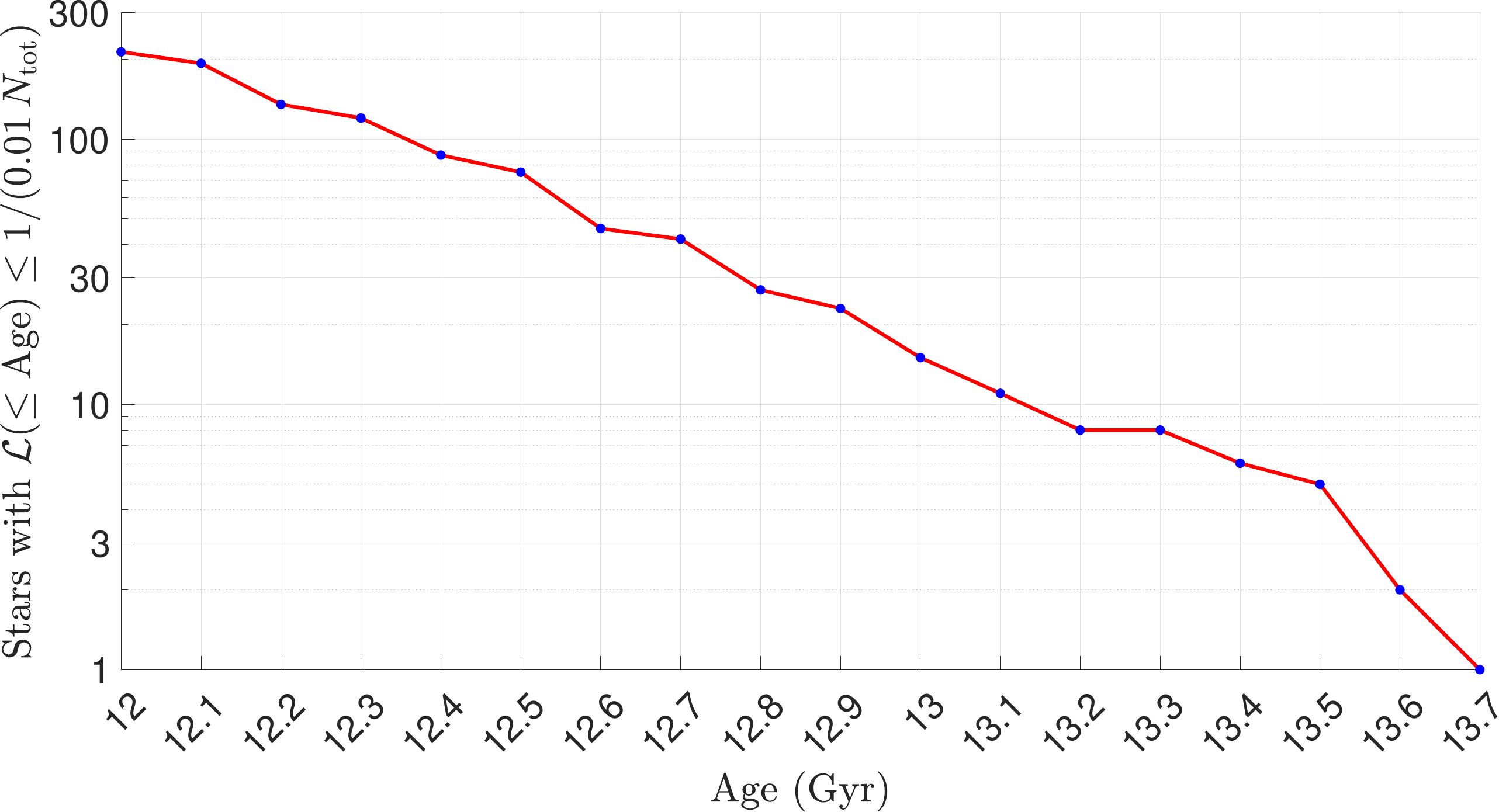} 
    \caption{The number of stars satisfying Equation~\ref{Ntot_inv_condition} for different possible values of $A_\star$. We expect that when there is only one such star, we have found the actual $A_\star$ (see the text).}
    \label{Ngta_nominal_result}
\end{figure}

Our results are shown in Figure~\ref{Ngta_nominal_result}. The threshold of one star satisfying Equation~\ref{Ntot_inv_condition} is reached at 13.7~Gyr, but there are no such stars at an assumed $A_\star \geq 13.8$~Gyr. The above argument therefore suggests that $A_\star \approx 13.7-13.8$~Gyr. Adding $\tf$, we estimate that $\AU \approx 13.9 - 14.0$~Gyr, which is remarkably close to $\AUCMB = 13.8$~Gyr \citep{SPT_2026}.

If we assume a lower $A_\star$ of 13~Gyr which is more in line with $\AstarETS$, we find that 15 stars satisfy Equation~\ref{Ntot_inv_condition}. This seems quite unlikely given the sample size. Moreover, we argued in Section~\ref{Introduction} that $\AstarETS \la 12.5$~Gyr with the latest low-redshift constraints on $H_0$ and $\OmegaM$. Adopting $A_\star = 12.5$~Gyr, we find that 75 stars satisfy Equation~\ref{Ntot_inv_condition}. It is exceedingly unlikely that 75 stars should be observed to have $A > A_\star$ at 99.93\% confidence, a statistically rare scenario that we expect to occur only once for our sample size and approximate latent age distribution.

\subsubsection{Accounting for the latent age distribution with a toy model}
\label{Toy_model}

\begin{figure}
    \centering
    \includegraphics[width=\linewidth]{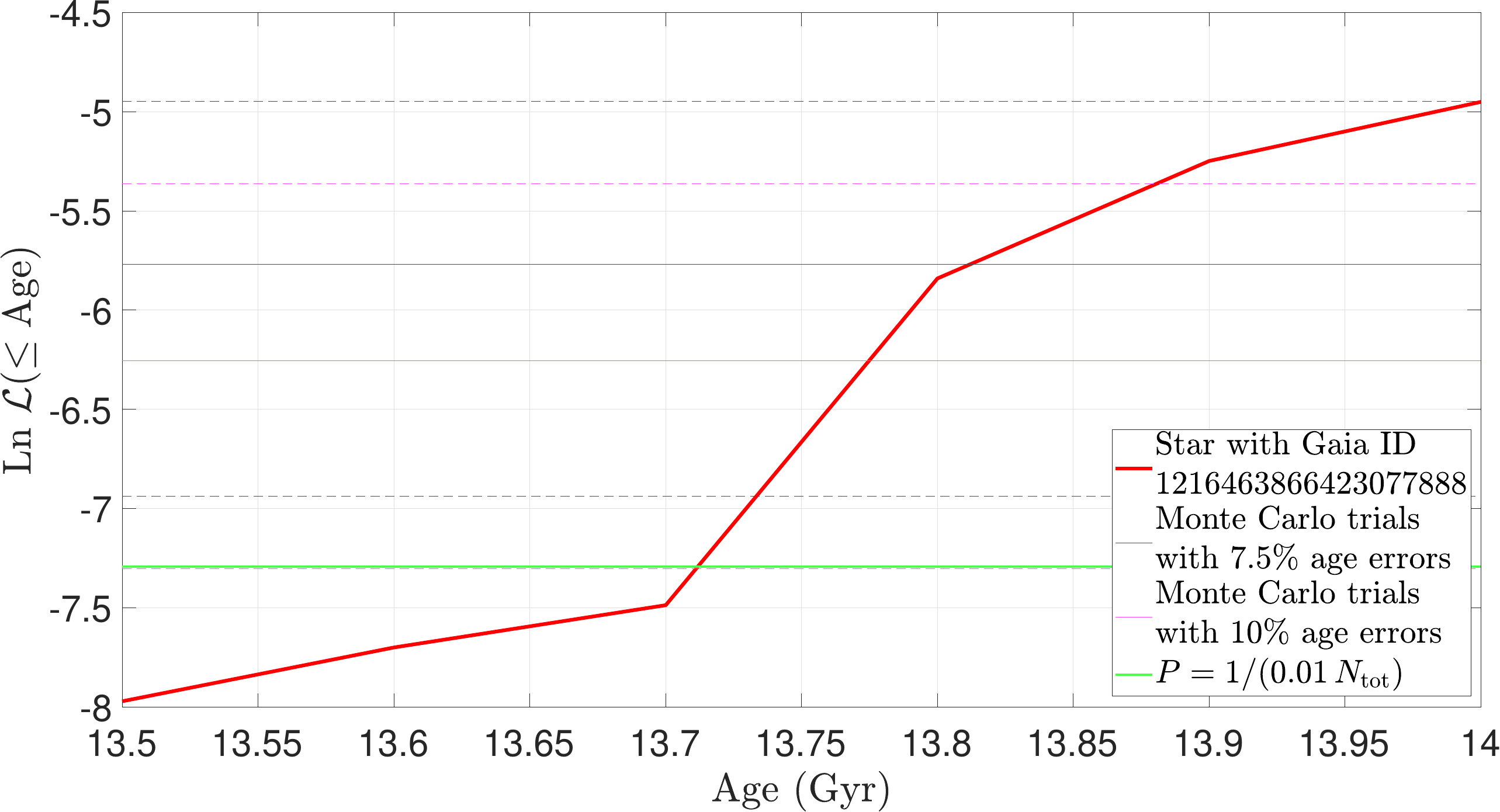} 
    \caption{The solid red line shows the cumulative age likelihood for our nominal sample's `oldest' star, which is the star at 13.7~Gyr in Figure~\ref{Ngta_nominal_result}. The other horizontal lines show different thresholds corresponding to different ways of viewing the problem, with the estimated $A_\star$ in each case being the intersection with the cumulative age likelihood (see the text). The solid green line shows a threshold of $1/(0.01 \, N_{\mathrm{tot}})$. The solid black line shows the result of Monte Carlo trials in our toy model, which give a $1\sigma$ uncertainty as indicated by the dashed black lines. This assumes $A_\star = 14$~Gyr and 7.5\% age uncertainties. If we instead assume 10\% age uncertainties, the thresholds decrease slightly to the solid and dashed magenta lines.}
    \label{Ln_Plta_oldest} 
\end{figure}

To obtain a more refined estimate of $A_\star$, we focus on the star in Figure~\ref{Ngta_nominal_result} which satisfies Equation~\ref{Ntot_inv_condition} for the largest assumed $A_\star$. This star is solely responsible for the single non-zero value plotted at 13.7~Gyr, beyond which no stars satisfy Equation~\ref{Ntot_inv_condition}. In some sense, it can be considered the `oldest' star in our sample.\footnote{Its \emph{Gaia} identifier is 1216463866423077888.} Figure~\ref{Ln_Plta_oldest} shows the cumulative age likelihood $\LX22(\leq A)$ for this star. The key question is what threshold $\LX22(\leq A)$ should reach at $A = A_\star$. Equation~\ref{Ntot_inv_condition} assumes that the answer is $1/(0.01 \, N_\mathrm{tot})$, which we indicate with a horizontal solid green line. We see that $\LX22(\leq A)$ crosses this threshold just after 13.7~Gyr, causing the star to satisfy Equation~\ref{Ntot_inv_condition} at 13.7~Gyr but not at 13.8~Gyr.

We can estimate the appropriate threshold for $\LX22(\leq A)$ more rigorously by directly sampling our toy model using Monte Carlo (MC) trials. We set these up by preparing a sample of size $N_\mathrm{tot} = 146,810$ in which the latent age of each star is drawn from the assumed distribution. The observed age is found by adding a Gaussian random measurement uncertainty of 7.5\%. Once we have set up a mock dataset in this way, we find $\LX22(\leq A_\star)$ for every star by integrating a Gaussian restricted to 0--20~Gyr, taking advantage of the fact that $A_\star$ is known to be 14~Gyr in our toy model. We then record the lowest value of $\LX22(\leq A_\star)$ across all the stars in the mock dataset. This serves as our estimate for the threshold $P_t$ we should use on the real $\LX22(\leq A)$ curve of the oldest star in our sample. Since this is a stochastic procedure, $P_t$ will differ each time we prepare a new mock dataset. We therefore prepare 20,000 mock datasets and find $P_t$ each time, building up its distribution in some detail. We then find the mode and $1\sigma$ confidence interval of $P_t$.

The results are shown in Figure~\ref{Ln_Plta_oldest} using a horizontal solid black line at the mode of the $P_t$ distribution, with the dashed black lines showing the $1\sigma$ uncertainty. The analogous versions of these results for 10\% age errors are shown similarly in magenta. Using these estimates of $P_t$, we find that $A_\star \approx 13.7$~Gyr. This matches our result using a threshold of $1/(0.01 \, N_\mathrm{tot})$ in Equation~\ref{Ntot_inv_condition}, accounting for the fraction of stars whose true age lies within $\sigma_A$ of $A_\star$ and thus might plausibly affect it.

\begin{figure} 
    \centering 
    \includegraphics[width=\linewidth]{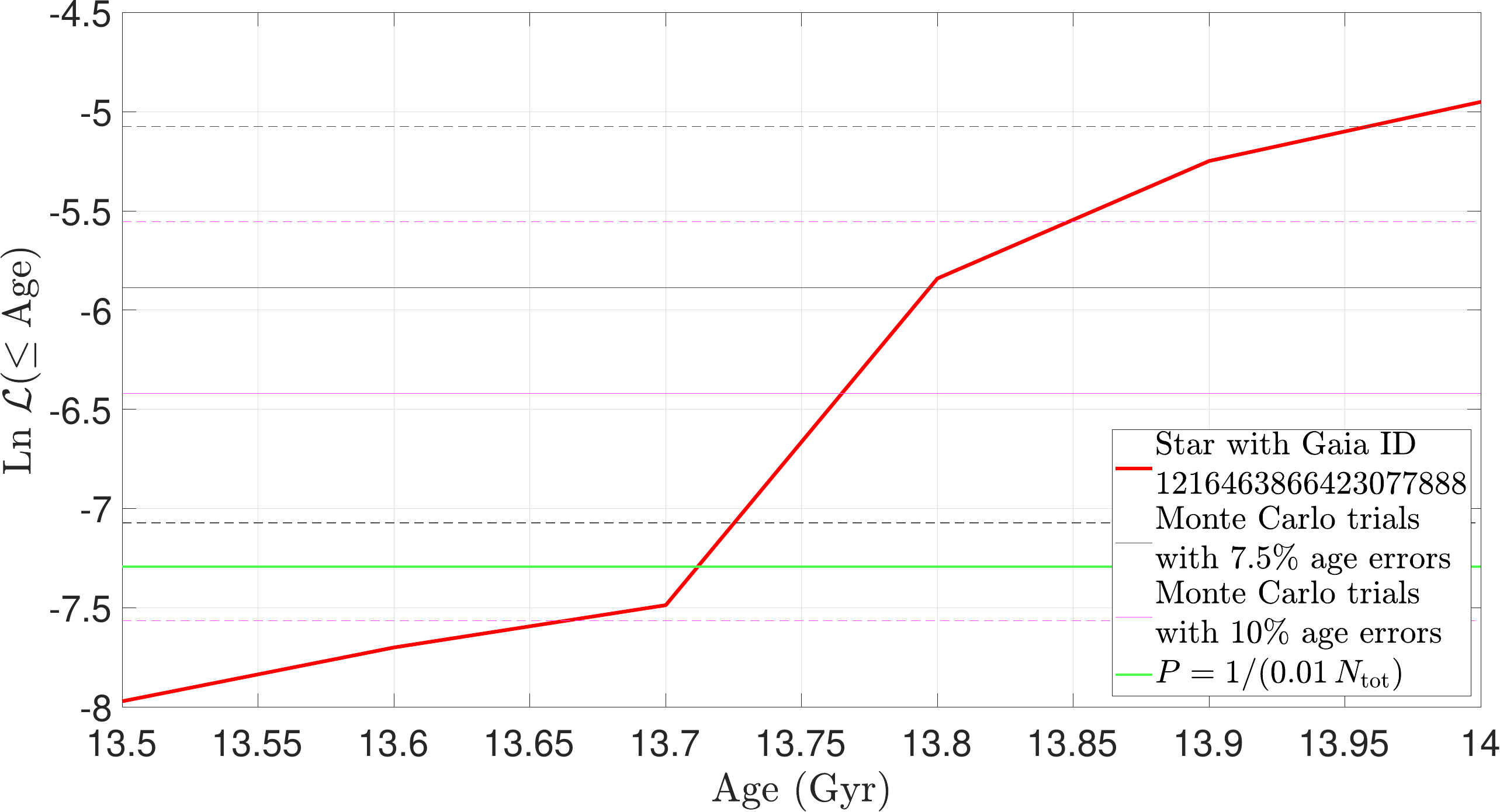} 
    \caption{Similar to Figure~\ref{Ln_Plta_oldest}, but using instead a toy model where $A_\star = 13$~Gyr, highlighting the negligible impact on our results.}
    \label{Ln_Plta_oldest_13Gyr}
\end{figure}

The above results use a toy model which has $A_\star = 14$~Gyr. One might worry that this affects our $P_t$ estimate in such a way that $A_\star$ inferred from the real sample is pushed towards 14~Gyr. To check this, we revise our toy model so that the latent age distribution tapers linearly to zero over the range $10-13$~Gyr, thereby reducing the assumed $A_\star$ to 13~Gyr. Figure~\ref{Ln_Plta_oldest_13Gyr} shows that this change has only a negligible impact on the estimated $P_t$ and thus the inferred $A_\star$. The small impact is no doubt related to the similar fraction of stars with latent $A > A_\star - \sigma_A$ in both toy models. It is therefore clear that if we focus on the oldest star in our sample bearing in mind its size and approximate age distribution, the inferred $A_\star$ is very similar to $A_\star^{\mathrm{CMB}}$ and far above $A_\star^{\mathrm{ETS}}$.

\subsection{Combined age likelihoods}
\label{Combined_age_likelihoods}

Our results in the preceding section appear compelling, but they rely on only a small number of stars. While this is inevitable to some extent when discussing the end point of a discretely sampled statistical distribution, it would be helpful to estimate $A_\star$ in a manner that gives at least some weight to a larger number of stars. Since much of our focus has been on the cumulative age likelihood $\LX22(\leq A)$ for one star, an obvious generalisation is to consider it for all stars. We multiply the values together by defining the statistic
\begin{eqnarray}
    S \left( A_\star \right) ~\equiv~ \Sigma_i \ln \LX22_i \left( \leq A_\star \right),
    \label{Sum_Ln_Plta}
\end{eqnarray}
where the stars are labelled using $i$. In principle, all stars should contribute to some extent -- even those stars much younger than $A_\star$.

To choose an appropriate threshold $S_t$ that we can identify with the true $A_\star$, we use the toy model developed in Section~\ref{Toy_model}. We prepare the mock datasets similarly, but instead of extracting information about only one star, we use all stars to calculate $S$ (Equation~\ref{Sum_Ln_Plta}). The idea is that only for one assumed value of $A_\star$ will $S$ calculated for the real sample match $S_t$, the value of $S$ in the toy model at its known $A_\star$.

\begin{figure}
    \centering
    \includegraphics[width=\linewidth]{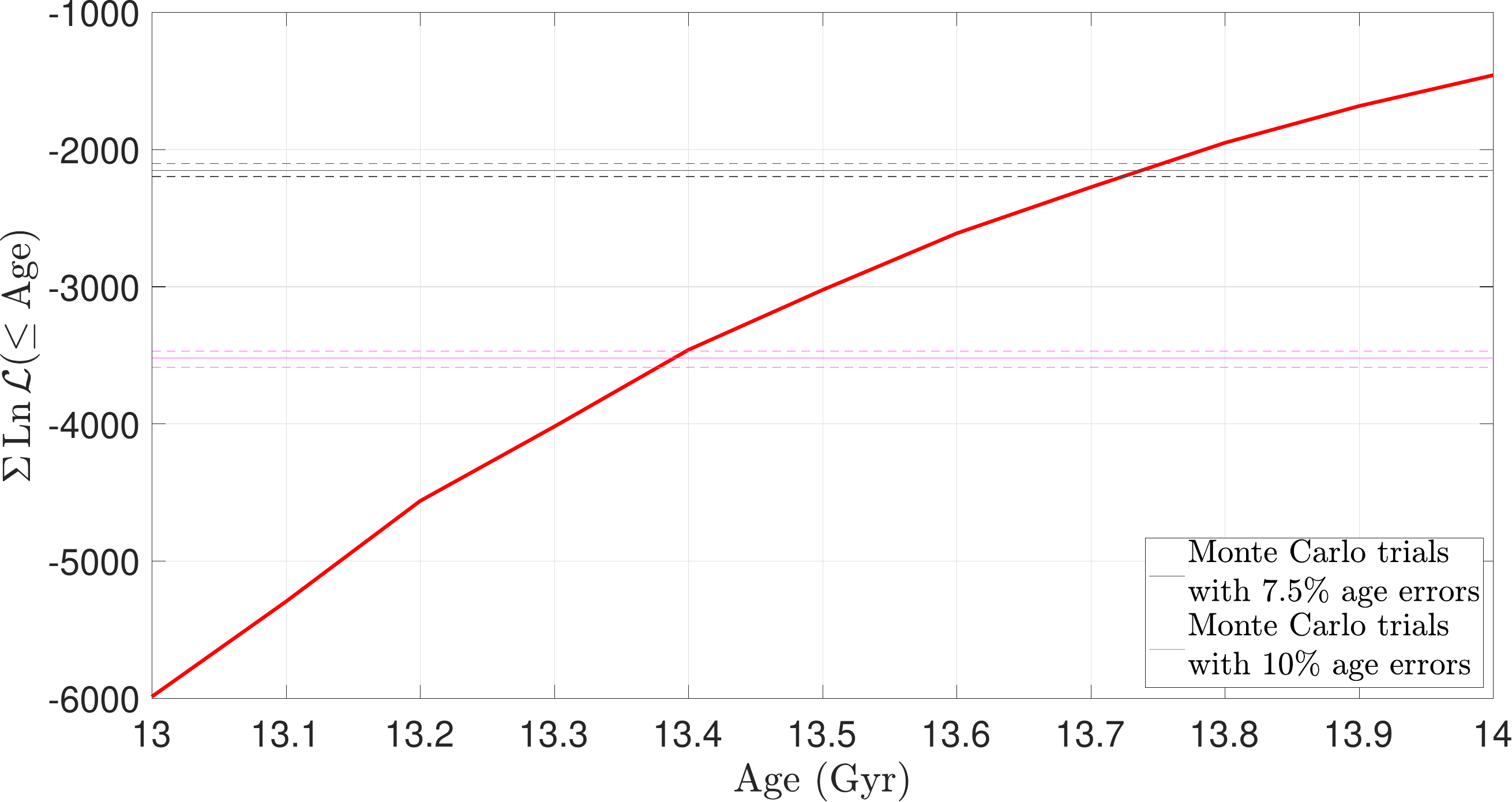} 
    \caption{The solid red line shows the sum of logarithmic cumulative age likelihoods for all stars in our nominal sample as a function of age (Equation~\ref{Sum_Ln_Plta}). The solid horizontal lines correspond to the most likely result of our toy model (Section~\ref{Toy_model}) with $A_\star = 14$~Gyr, while the dashed horizontal lines in the same colour show the $1\sigma$ uncertainty. Black (magenta) lines assume 7.5\% (10\%) age errors, the former being more representative (Figure~\ref{Age_error_turnover_nominal}).}
    \label{Ln_P_obs_values}
\end{figure}

Figure~\ref{Ln_P_obs_values} shows the observationally determined $S(A_\star)$ (Equation~\ref{Sum_Ln_Plta}). The horizontal black lines show the most likely $S_t$ and its $1\sigma$ uncertainty inferred from 20,000 realisations of our toy model assuming 7.5\% age errors, which matches the median error stated by \citetalias{Xiang_2022}. The resulting $A_\star$ is slightly over 13.7~Gyr. This is very similar to the result in Figure~\ref{Ln_Plta_oldest} based on the observed age likelihood of the oldest star. However, an important difference in Figure~\ref{Ln_P_obs_values} is that if we instead assume 10\% age errors and therefore need to find the intersection of $S(A_\star)$ with the horizontal magenta line, the inferred $A_\star$ drops to around 13.4~Gyr. It is unrealistic to assume a much larger $\widetilde{\sigma}_A$ given its distribution \citepalias[Figure~\ref{Age_error_turnover_nominal}; see also figure~1b of][]{Xiang_2022}. Even an $A_\star$ as low as 13.4~Gyr is still far above $A_\star^{\mathrm{ETS}}$, but it would leave much more room for alternative cosmologies which predict an $\AU$ only slightly below $\AUCMB$ \citep{Najera_2026}. This makes it important to consider individual age likelihoods in more detail, rather than assuming that all stars have the same Gaussian $\widetilde{\sigma}_A$.

\section{The reconstructed latent age distribution}
\label{Reconstructed_age_distribution}

One of the main challenges in accurately estimating $A_\star$ is that some stars will inevitably have an observed $A > A_\star$ due to measurement uncertainties. However, this can only occur for the small fraction of stars whose latent age is within striking distance of $A_\star$. To accurately estimate this fraction, we need to know the population latent age distribution $P(A)$. This must be reconstructed from the age likelihoods $\LX22_i(A)$ for each star $i$.

We reconstruct $P(A)$ using two approaches, both of which are non-parametric given the large sample size should allow a rather detailed reconstruction. Our first approach is iterative and designed to give a best-guess reconstruction without uncertainties (Section~\ref{Population_prior_protocol}). Our second approach is an MCMC reconstruction (Section~\ref{MCMC_reconstruction}). This serves as the basis for our $A_\star$ inference, which is our main result (Section~\ref{Results}).

\subsection{Iterative reconstruction with the population prior method}
\label{Population_prior_protocol}

Our iterative approach relies on the fact that $\LX22_i(A)$ is determined based only on observations of star $i$. In the absence of any other information, our posterior inference of $A_i$ would just be $\LX22_i(A)$. However, we have observations of a large number of stars. We can use this to create a population distribution by summing the individual $\LX22_i(A)$. This provides an initial guess for $P(A)$, but it is not very accurate because it represents a convolution between observational uncertainties and the latent $P(A)$. Even so, an initial guess for $P(A)$ provides a non-trivial prior. We can use this to get a more refined posterior inference on the age of star $i$ beyond merely using $\LX22_i(A)$. For instance, if a star is observed to have $A_i = 12 \pm 2$~Gyr but the vast majority of stars have $A \la 10$~Gyr, it is far more likely that $A_i$ is slightly overestimated from a true age of 10~Gyr, compared to being underestimated from a true age of 14~Gyr. We would simply not know this without observations of other stars.

Once the population $P(A)$ is used to refine the age posterior of every star, we can sum up these revised posteriors to obtain another guess for $P(A)$. Mathematically, this corresponds to the following:
\begin{eqnarray}
    P \left( A \right) ~=~ \sum_i P \left( A \right) \LX22_i(A) \, .
    \label{Population_prior_equation}
\end{eqnarray}
Thus, our reconstructed $P(A)$ at any stage serves as a prior. We multiply this with $\LX22_i(A)$ to get a posterior inference on the age of star $i$. We then stack these posteriors to obtain a revised $P(A)$ for the next stage. This process must be repeated until convergence is achieved. The final result for $P(A)$ is such that if it is used as a prior on the age of each star, then the resulting posteriors when summed lead to the original $P(A)$. We note that this $P(A)$ is specific to our sample and not necessarily representative of the Galaxy as a whole. Since older stars need to have a lower mass and are therefore generally fainter, our analysis may underestimate the prevalence of very old stars. However, our sample has stars with a wide range of distances $\leq 5$~kpc (Figure~\ref{d_distribution}), so we do not expect selection effects to cause a sharp cutoff in the reconstructed $P(A)$ at any particular age.

This procedure requires a large sample size to ensure negligible correlation between $P(A)$ and any single $\LX22_i(A)$, which we have implicitly assumed. A large sample also makes it much easier to interpolate $P(A)$ to any desired $A$, since otherwise the recovered distribution could become noisy. In fact, we expect this to some extent because measurement errors smooth out any sharp features in the latent $P(A)$. If a deconvolution algorithm such as Equation~\ref{Population_prior_equation} operates correctly, it would recover the original less smooth distribution.

\begin{figure}
    \centering
    \includegraphics[width=\linewidth]{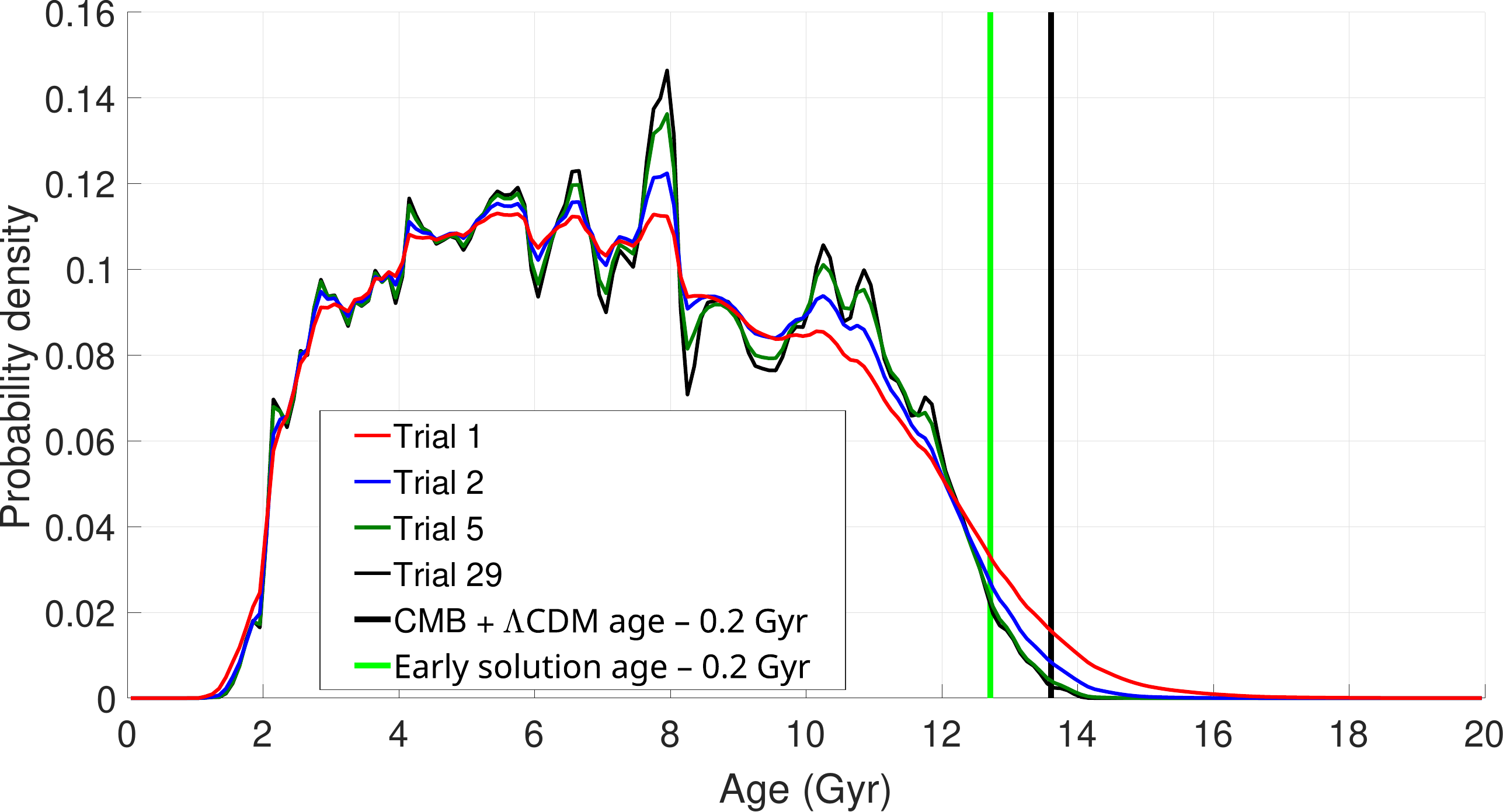}
    \caption{Our iterative reconstruction of $P(A)$ in 200 bins using Equation~\ref{Population_prior_equation} after 1, 2, 5, and 29 stages (red, blue, green, and black lines, respectively). Notice how the tail is suppressed as the algorithm progresses. We consider the result converged after 29 stages. The vertical green (grey) line shows $\AstarETS$ $\left( \AstarCMB \right)$, assuming $\tf = 0.2$~Gyr in both cases (Equation~\ref{AU_analytic}).}
    \label{PP_reconstruction_HR_nominal_stages}
\end{figure}

Our reconstruction is shown in Figure~\ref{PP_reconstruction_HR_nominal_stages} at stages 1, 2, 5, and 29, when we consider the results converged. The initial result follows simply from stacking $\LX22_i(A)$. The distribution narrows significantly at the next stage. Beyond stage 5, changes in the reconstructed $P(A)$ start to slow down. After 29 stages, changes are barely discernible.

Our final reconstruction of $P(A)$ is shown in Figure~\ref{PP_reconstruction_HR_nominal_stages} as the black line. It extends well beyond $\AstarETS$, suggesting that $A_\star$ is larger. However, $P(A)$ does not extend much beyond $\AstarCMB$. Since our procedure always gives some probability to each age bin, we cannot expect the recovered $P(A)$ to drop completely to zero at $A_\star$. It is therefore difficult to conclude that $A_\star > \AstarCMB$ on the basis of these results, especially since our iterative reconstruction does not provide any estimate of uncertainty.

\subsubsection{The old tail}
\label{Age_limit_PP}

\begin{figure}
    \centering
    \includegraphics[width=\linewidth]{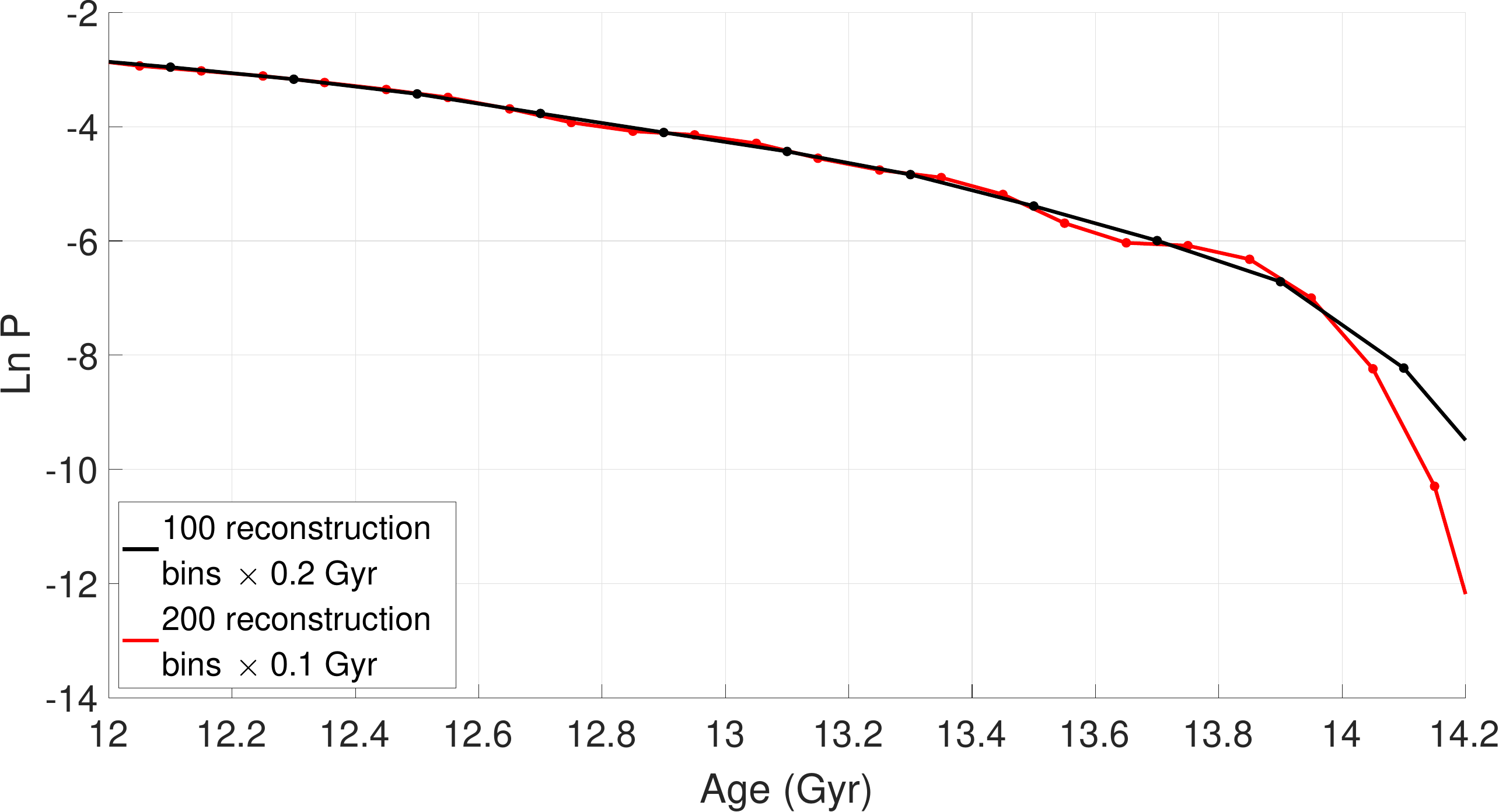} 
    \includegraphics[width=\linewidth]{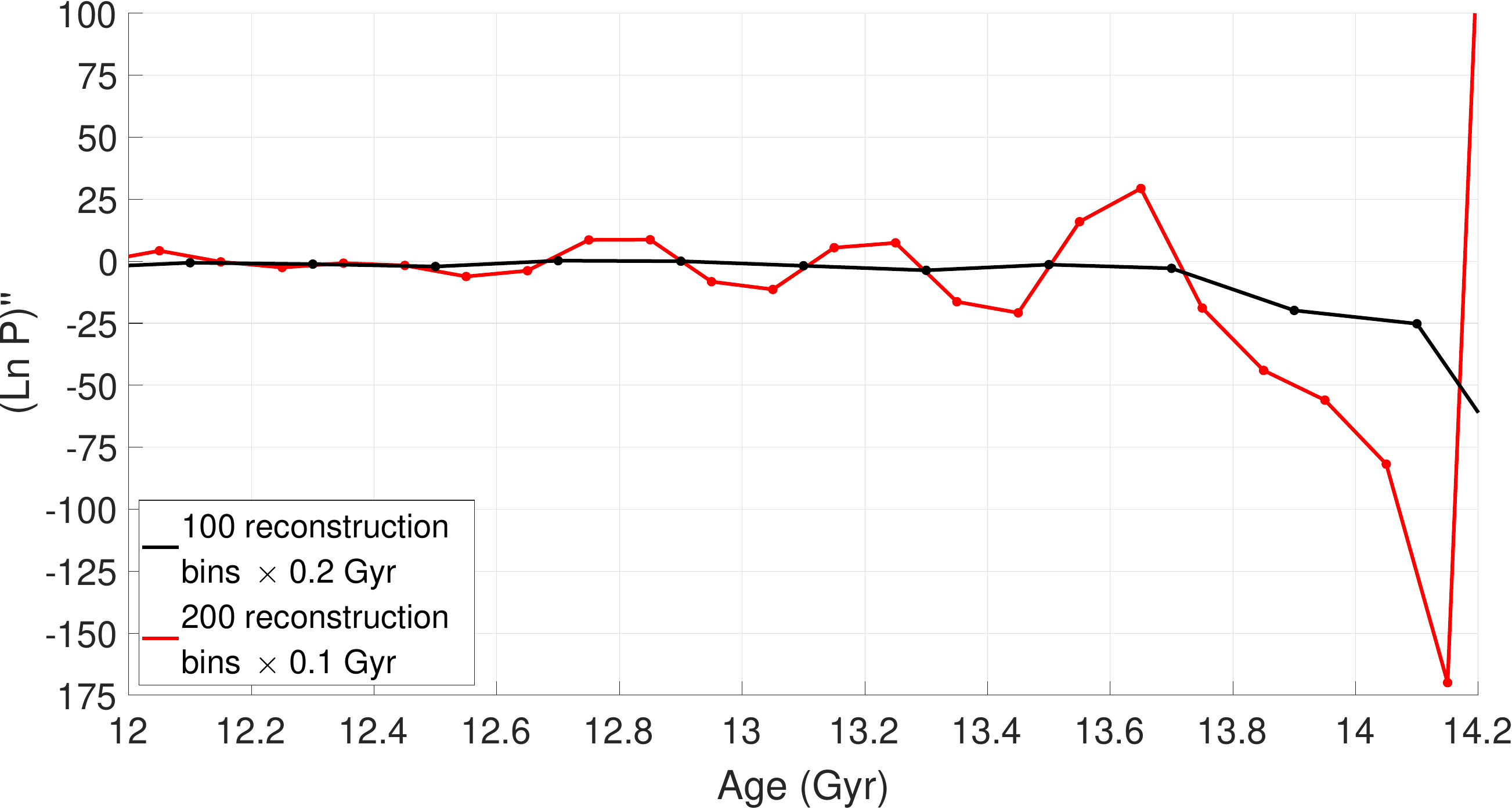}
    \caption{\emph{Top}: The red line shows our iterative reconstruction of the latent age distribution (black line on Figure~\ref{PP_reconstruction_HR_nominal_stages}). The black line shows a lower resolution 100-bin reconstruction. In both cases, the reconstructed $P(A)$ declines slightly faster than exponential, which we highlight by plotting $\ln P(A)$. \emph{Bottom}: We differentiate $\ln P(A)$ twice to help identify a sharp cutoff point. Beyond 13.7~Gyr, both reconstructions noticeably deviate from the behaviour at lower ages.}
    \label{Ln_P_obs_pp_values}
\end{figure}

To quantify $A_\star$ more rigorously based on our iteratively reconstructed $P(A$), we need to examine its tail region more closely. We do this in Figure~\ref{Ln_P_obs_pp_values}, where we show results using a 100-bin reconstruction in black and a higher resolution 200-bin reconstruction in red. In both cases, the bins uniformly cover ages of $0-20$~Gyr. The top panel shows the $\ln P(A)$ reconstructions, revealing an approximately exponential decline. Although the decline is steeper towards higher ages, there is no obvious end point feature that we can identify with $A_\star$.

To try and find such a feature given the roughly parabolic dependence of $\ln P(A)$ on $A$, we use the bottom panel of Figure~\ref{Ln_P_obs_pp_values} to show $(\ln P)''$, where each prime denotes differentiation. Our higher resolution reconstruction becomes somewhat noisy when explored in this level of detail, but our lower resolution reconstruction remains quite stable. $(\ln P)''$ maintains a nearly constant slightly negative value until an age of 13.7~Gyr. It then drops sharply in a manner unlike anything seen at lower ages. A similar feature is also evident in our higher resolution reconstruction, where there is a clear break from the oscillatory pattern evident at lower ages. Since a second derivative requires values from adjacent bins, this suggests that $A_\star \approx 13.7 - 13.8$~Gyr. The erratic behaviour at $A > 14$~Gyr is most likely an indication of numerical noise at $A > A_\star$. Our results therefore suggest that $A_\star \approx \AstarCMB$, though $A_\star$ could be slightly higher. There is no sign of a cutoff in the latent age distribution at $\AstarETS$. However, we can only quantify how strongly $\AstarETS$ is excluded using a reconstruction method that provides uncertainties.

\subsection{MCMC reconstruction}
\label{MCMC_reconstruction}

Instead of iteratively converging towards the true $P(A)$, we can infer its value in each age bin using MCMC. This requires us to quantify $\ln P$, the log-posterior of any particular $P(A)$ given the individual $\LX22_i(A)$, the observed age likelihood of star $i$ across the $N$ different age bins, which we index using $j$.
\begin{eqnarray}
    \ln P = \sum_i \ln \left[ \sum_{j = 1}^N P_j \LX22_i \left( A_j \right) \right] - \overbrace{\sum_{j = 1}^{N - 1} \left[ \ln \sigma + \frac{\left( \ln P_{j + 1} - \ln P_j \right)^2}{2 \sigma^2} \right]}^\mathrm{Gradient \, penalty} \, ,
    \label{Ln_P_MCMC}
\end{eqnarray}
where $\sigma$ is a model parameter discussed below and $P_j \equiv P(A_j)$ is the likelihood that the latent age of a star in our sample lies within bin $j$, which is centred on $A_j$. The basic idea behind the first term is that the contribution to $\ln P$ from star $i$ is the `overlap integral' $\ln P_i$, which is defined as follows:
\begin{eqnarray}
    \ln P_i ~=~ \int P \left( A \right) \LX22_i \left( A \right) \, \mathrm{d} A.
    \label{Ln_P_i_MCMC}
\end{eqnarray}

To speed up the computations, we reconstruct $P(A)$ using 100 bins uniformly covering the range $0.1 - 20.1$~Gyr. We integrate the 140-knot $\LX22_i(A)$ from \citetalias{Xiang_2022} across each of our reconstruction bins, assuming that it varies linearly between the knots where its value is known. Thus, $\LX22_i (A_j)$ in Equation~\ref{Ln_P_MCMC} refers to the integral of the original 140-knot $\LX22_i(A)$ across reconstruction bin $j$. To compute this integral in the oldest reconstruction bin, we extrapolate linearly from the last two knots to ages beyond 20~Gyr, imposing a floor of zero. If this is not reached, $\LX22_i(A)$ extends all the way to 20.1~Gyr, the upper limit of our reconstruction range. We found that we could get reliable results only with our particular choice of bin positions, presumably because the algorithm struggles to decide how much probability to assign to adjacent pixels when the available information is only really sufficient to constrain the total $P(A)$ across both bins. This leads to very severe oscillations if we use 100 bins uniformly covering $0 - 20$~Gyr, but this problem is solved if using instead $0.1 - 20.1$~Gyr, aligning each bin centre beyond 8~Gyr with a knot position in the original $\LX22_i(A)$.

We smooth the inevitably somewhat noisy reconstructed $P(A)$ by directly penalising differences between adjacent bins. We use the gradient penalty term in Equation~\ref{Ln_P_MCMC} for this, applying a standard $\chi^2$ penalty to every difference $\ln P_{j + 1} - \ln P_j$, with the typically expected such difference being $\sigma$. Since we do not know $\sigma$, we infer it alongside the $P_j$. We use a wide uniform prior on $\sigma$ because it refers to a difference in log-probability. However, it can also be argued that a scale parameter like $\sigma$ should have a scale-invariant Jeffreys prior of $1/\sigma$ due to its order of magnitude being unknown \citep{Jeffreys_1946}. We will see later that this has negligible impact on our results (Section~\ref{Results}). Even removing the gradient penalty term altogether does not much affect our inferred $A_\star$, though it does make the reconstructed $P(A)$ more noisy.

\begin{figure}
    \centering
    \includegraphics[width=\linewidth]{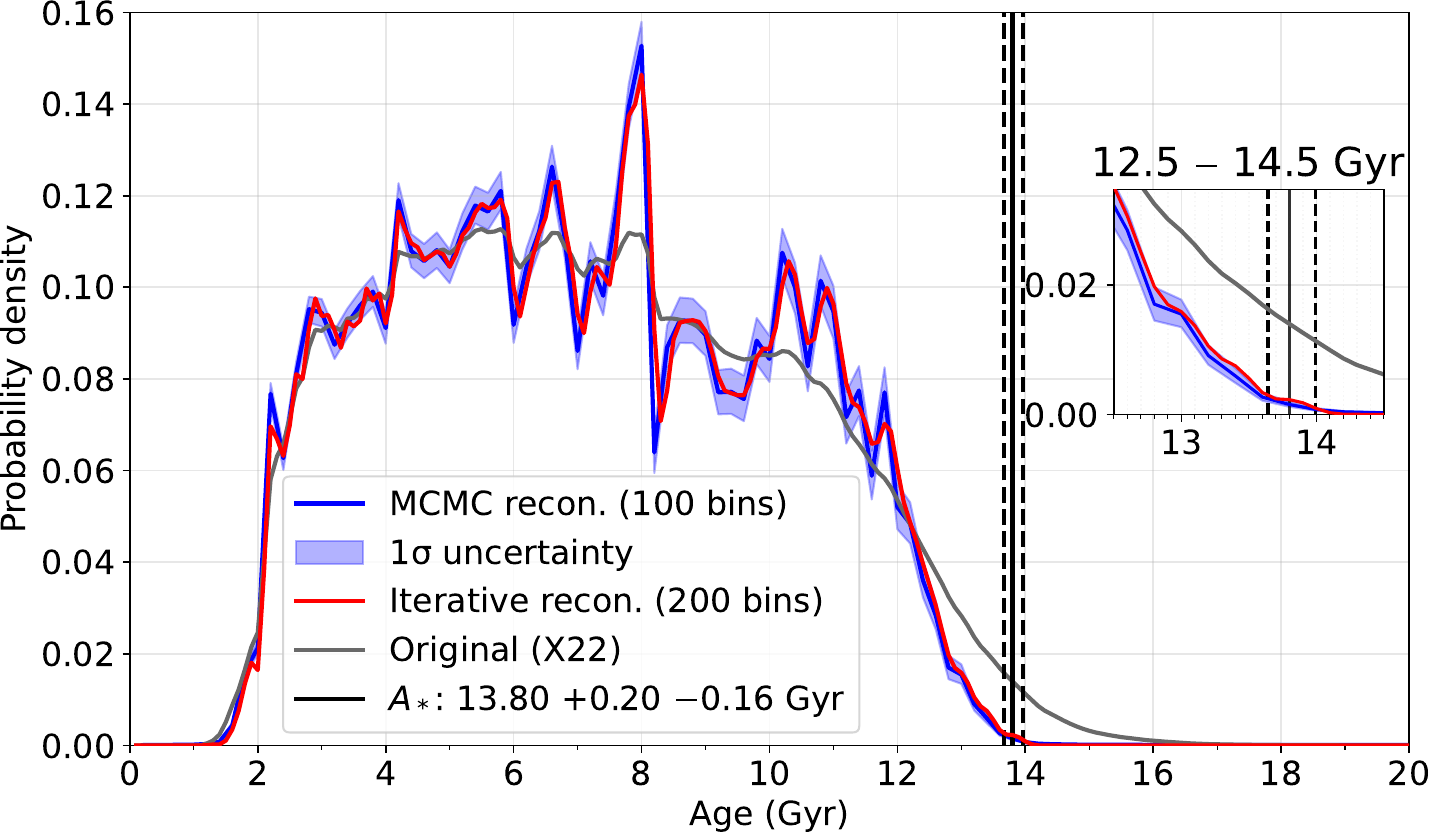} 
    \caption{The solid grey line shows a first estimate of $P(A)$ from stacking the \citetalias{Xiang_2022} age likelihoods in our nominal sample. Our MCMC reconstruction is shown in blue (Equation~\ref{Ln_P_MCMC}), with the solid line giving the mean and the shaded band the standard deviation, both using $3\sigma$ outlier rejection. The solid red line shows our higher resolution iterative reconstruction (Section~\ref{Population_prior_protocol}). The solid black vertical line marks the inferred $A_\star$, with dashed black lines indicating its $1\sigma$ uncertainty (Section~\ref{Age_limit_MCMC}). The inset zooms in on $12.5-14.5$~Gyr, highlighting the good agreement between the iterative and MCMC reconstructions of $P(A)$ in the tail region.}
    \label{MCMC_PP_recons}
\end{figure}

We reconstruct $P(A)$ with \textsc{numpyro} using 500 warm-up samples and 2500 main samples that enter further analysis. We find that this is more than sufficient to ensure our results are converged with respect to the Gelman-Rubin statistic, which is at most 1.01. Since we have 2500 possible values for each $P_j$, we find the mean $\overline{P}_j$ and standard deviation $\sigma_j$ using $3\sigma$ outlier rejection across the MCMC samples, guarding against samples that give an outlying $P_j$. Our mean MCMC reconstruction of $P(A)$ and its uncertainty are shown in blue on Figure~\ref{MCMC_PP_recons}. The red line here shows our higher resolution iterative reconstruction (Section~\ref{Population_prior_protocol}). Both reconstructions agree very well, especially in the critically important $12.5-14.5$~Gyr region shown in the inset. This is true despite the reconstructed $P(A)$ differing substantially from the input distribution, which is shown in grey. The main difference is that the MCMC reconstruction provides a measure of uncertainty, which is crucial in our case.\footnote{For other applications, the much faster and more robust iterative approach described in Section~\ref{Population_prior_protocol} may well be sufficient if the goal is simply to get the optimal reconstruction, e.g. if there are other much larger sources of uncertainty.}

\subsubsection{\texorpdfstring{$A_\star$}{A*} inference}
\label{Age_limit_MCMC}

\begin{figure}
    \centering
    \includegraphics[width=\linewidth]{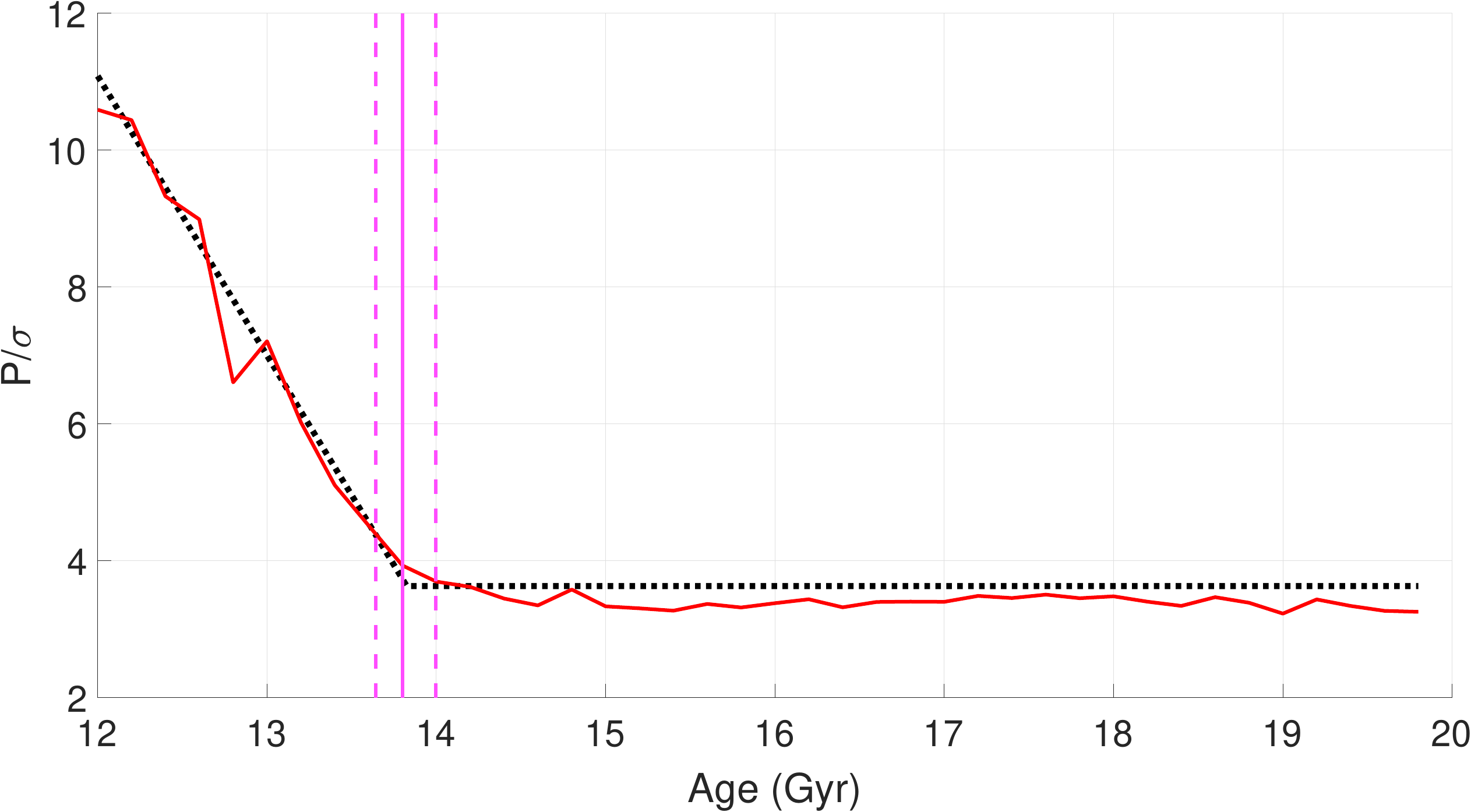} 
    \caption{The solid red line shows the ratio of the mean reconstructed $P(A)$ in each age bin to its standard deviation. This way of showing the MCMC reconstruction in Figure~\ref{MCMC_PP_recons} highlights an extended flat region at high ages, which we interpret as a bias floor due to our analysis requiring $P > 0$. The dotted black line shows our best fit to the reconstructed $P/\sigma$ with a linear~+~flat model. We identify the crossover point with $A_\star$, whose mode and $1\sigma$ uncertainty are shown using the solid and dashed vertical magenta lines, respectively.}
    \label{P_SN_ratio}
\end{figure}

Our reconstructed $P(A)$ in Figure~\ref{MCMC_PP_recons} lacks an obvious end point feature, but this is only true if we consider the $P(A)$ values on their own. The MCMC reconstruction also provides uncertainties. To exploit these, we divide each $\overline{P}_j$ by its uncertainty $\sigma_j$, providing a measure of how confident we should be that bin $j$ actually has any stars at all. The run of $\overline{P}_j/\sigma_j$ in Figure~\ref{P_SN_ratio} reveals an almost completely flat region at $A \ga 14$~Gyr, even though $P(A)$ continues to decline over this range. We interpret this as a minimum level of bias inherent to our approach, which cannot yield a negative probability. Since all 2500 MCMC samples must have $P_j > 0$ and the uncertainty is finite, the MCMC analysis inevitably infers that $\overline{P}_j/\sigma_j > 0$ to some extent in the bins beyond $A_\star$. In this regime, the distribution of $P_j$ is roughly Gaussian across the MCMC samples, with $\overline{P}_j$ decreasing towards higher ages. However, there is also a corresponding decrease in $\sigma_j$. This creates the observed flat trend in $\overline{P}_j/\sigma_j$. The level of this relative bias floor presumably depends on details of how we set up our MCMC analysis, for instance that we assume a uniform prior on $\ln P_j$ to obtain more numerically stable results.

At lower ages of $\approx 12 - 14$~Gyr, $\overline{P}/\sigma$ declines almost linearly with age. The intersection between this line and the relative bias floor at higher ages provides a very clear sharp feature that we identify with $A_\star$. We infer $A_\star$ using another MCMC analysis that fits a linear~+~flat model to the run of $\overline{P}/\sigma$. The model parameters are the relative bias floor, the slope, and the crossover point $A_\star$. Each model for $\overline{P}/\sigma$ is integrated across each bin to obtain a predicted value for that bin. We then quantify how well this model fits the actual values over the range $11.9-19.9$~Gyr, excluding the last bin centred on 20~Gyr to avoid edge effects. It is clear from Figure~\ref{P_SN_ratio} that going down to 11.9~Gyr is more than sufficient to accurately determine $A_\star$.

The uncertainty on $\overline{P}/\sigma$ is by definition unity $\forall j$. To obtain more accurate results, we also consider the covariance between adjacent bins. In each age bin $j$, we first find the difference $\delta P_j^s$ between the mean result $\overline{P}_j$ and each MCMC sample, which we index by $s$.
\begin{eqnarray}
    \delta P_j^s ~\equiv~ P_j^s - \overline{P}_j \, .
\end{eqnarray}
The covariance between $P_j$ and $P_k$ is the mean value of $\delta P_j^s \delta P_k^s$, which we find using $3\sigma$ outlier rejection. We then divide the covariance by $\sigma_j \sigma_k$ to compute the covariance between $P_j/\sigma_j$ and $P_k/\sigma_k$, with $\sigma$ in each bin also found using outlier rejection. In this way, we build up the covariance matrix between $P/\sigma$ in different bins. We then use this to calculate the $\chi^2$ between any linear~+~flat model and the MCMC reconstruction of $\overline{P}/\sigma$. As expected, correlations between different age bins are most pronounced for nearby bins. We find a weak alternating trend in the sign of the correlation coefficient: bin $j$ is usually negatively correlated with bin $j + 1$, while bins $j$ and $j + 2$ are generally positively correlated, etc. This is likely related to the reconstruction being able to more tightly constrain $P_j + P_{j + 1}$, which therefore become anti-correlated (Section~\ref{MCMC_reconstruction}). Our covariance matrix approach automatically accounts for such correlations.

For our secondary MCMC stage fitting a linear~+~flat model to $\overline{P}/\sigma$ (Figure~\ref{P_SN_ratio}), we again use \textsc{numpyro}. The very low computational cost permits us to take $10\times$ more samples despite the need to consider a $40 \times 40$ covariance matrix each time. We therefore use 5000 warm-up samples and 25000 main samples, the latter serving as the basis for our main result. The best-fitting linear~+~flat model has a low $\chi^2 = 1.8$ across the 40 age bins (rising to 10.6 when the gradient penalty is removed); such low values reflect strong correlations between adjacent bins and hence few independent degrees of freedom, suggesting that our quoted uncertainty on $A_\star$ is if anything conservative.

We also obtained similar results using \textsc{dynesty}, mainly so we could obtain a Bayesian evidence for our linear~+~flat model for $\overline{P}/\sigma$. We compared this to an alternative parabolic~+~flat model, but its Bayesian evidence is considerably lower due to the unnecessary extra complexity of the model. Moreover, the inferred value of the curvature term is fully consistent with zero. Taken together and given also the excellent linear~+~flat fit evident in Figure~\ref{P_SN_ratio}, it is clear that this model is already sufficient. We therefore do not consider a parabolic~+~flat model further.

\section{Results}
\label{Results}

\begin{table}
	\centering
	\begin{tabular}{lcccc}
        \hline
        Sample and & Sample & \multirow{2}{*}{$A_\star$ (Gyr)} & Relative & $P/\sigma$ slope \\
        analysis variant & size & & bias floor & per Gyr \\ \hline
        Nominal sample & \multirow{2}{*}{146,810} & \multirow{2}{*}{$13.80^{+0.20}_{-0.16}$} & \multirow{2}{*}{$3.63^{+0.25}_{-0.25}$} & \multirow{2}{*}{$-4.01^{+0.52}_{-0.60}$} \\ \relax
        and analysis & & & & \\ \relax
        Jeffreys prior on & \multirow{2}{*}{''} & \multirow{2}{*}{$13.75^{+0.19}_{-0.16}$} & \multirow{2}{*}{$3.62^{+0.25}_{-0.24}$} & \multirow{2}{*}{$-4.16^{+0.54}_{-0.60}$} \\ \relax
        gradient penalty $\sigma$ & & & & \\ \relax
        No gradient & \multirow{2}{*}{''} & \multirow{2}{*}{$13.58^{+0.17}_{-0.15}$} & \multirow{2}{*}{$2.40^{+0.17}_{-0.16}$} & \multirow{2}{*}{$-3.84^{+0.53}_{-0.69}$} \\ \relax
        penalty applied & & & & \\ \relax
        [Fe/H] cutoff & \multirow{2}{*}{149,730} & \multirow{2}{*}{$13.89^{+0.18}_{-0.16}$} & \multirow{2}{*}{$3.64^{+0.24}_{-0.25}$} & \multirow{2}{*}{$-4.05^{+0.51}_{-0.48}$} \\ \relax
        raised 0.1 dex & & & & \\ \relax
        [Fe/H] cutoff & \multirow{2}{*}{151,055} & \multirow{2}{*}{$14.04^{+0.20}_{-0.16}$} & \multirow{2}{*}{$3.63^{+0.26}_{-0.26}$} & \multirow{2}{*}{$-3.71^{+0.43}_{-0.46}$} \\ \relax
        raised 0.2 dex & & & & \\ \relax
        [Fe/H] cutoff & \multirow{2}{*}{141,748} & \multirow{2}{*}{$13.52^{+0.21}_{-0.16}$} & \multirow{2}{*}{$3.63^{+0.25}_{-0.24}$} & \multirow{2}{*}{$-4.40^{+0.68}_{-0.84}$} \\ \relax
        lowered 0.1 dex & & & & \\ \relax
        [Fe/H] cutoff & \multirow{2}{*}{134,166} & \multirow{2}{*}{$13.39^{+0.24}_{-0.20}$} & \multirow{2}{*}{$3.59^{+0.23}_{-0.24}$} & \multirow{2}{*}{$-3.84^{+0.83}_{-0.95}$} \\ \relax
        lowered 0.2 dex & & & & \\ \relax
        [$\alpha$/Fe] cutoff & \multirow{2}{*}{147,768} & \multirow{2}{*}{$13.85^{+0.22}_{-0.17}$} & \multirow{2}{*}{$3.66^{+0.27}_{-0.25}$} & \multirow{2}{*}{$-3.85^{+0.51}_{-0.60}$} \\ \relax
        lowered 0.05 dex & & & & \\ \relax
        No FLAME age & \multirow{2}{*}{180,391} & \multirow{2}{*}{$14.11^{+0.17}_{-0.14}$} & \multirow{2}{*}{$3.68^{+0.25}_{-0.24}$} & \multirow{2}{*}{$-4.18^{+0.42}_{-0.43}$} \\ \relax
        consistency test & & & & \\ \hline \relax
	\end{tabular}
	\caption{Parameter inferences for our nominal analysis and several variants, corresponding to the triangle plots shown in Figure~\ref{Triangles_Numpyro_linSNR}. The first three rows use our nominal sample, but with various changes to the gradient penalty term (Equation~\ref{Ln_P_MCMC}). The other rows fix the analysis method to that of our nominal result, but vary the sample. Changes to the age-dependent [Fe/H] ceiling are shown in Figures~\ref{age_feh_figure} and \ref{age_feh_intercept_offsets}, while changes to the [$\alpha$/Fe] floor are shown in Figure~\ref{age_alpha_figure}. The last row drops the Section~\ref{FLAME_comparison} consistency test between spectroscopic YY ages from \citetalias{Xiang_2022} and FLAME ages \citep{Creevey_2023, Creevey_2026} using only \emph{Gaia} data.}
	\label{Main_results_table}
\end{table}

\begin{figure*}
    \centering
    \includegraphics[width=0.49\linewidth]{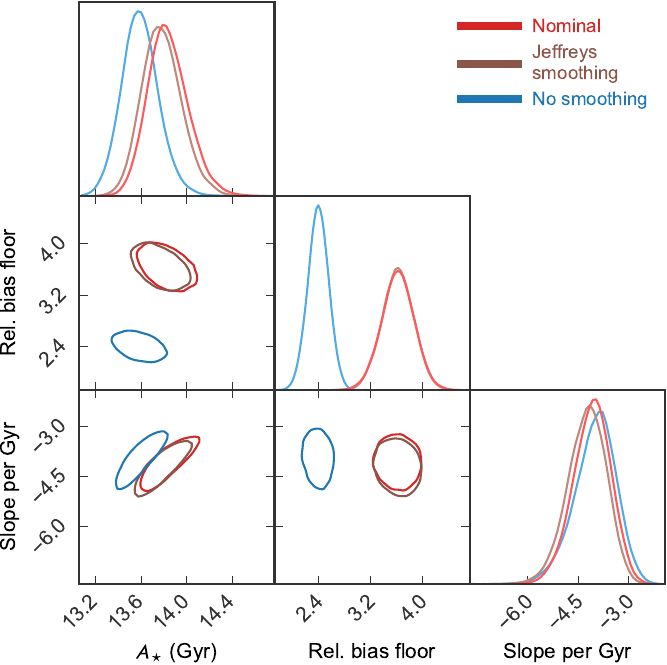} 
    \hfill
    \includegraphics[width=0.49\linewidth]{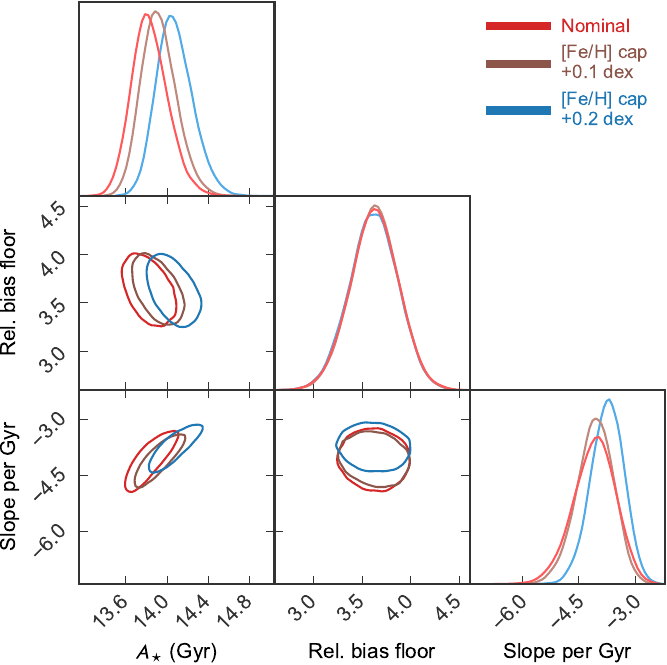}
    \includegraphics[width=0.49\linewidth]{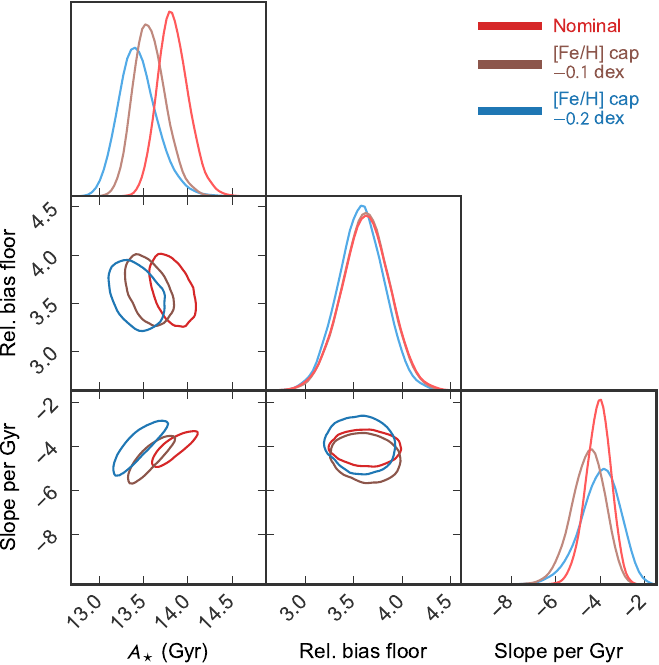}
    \hfill
    \includegraphics[width=0.49\linewidth]{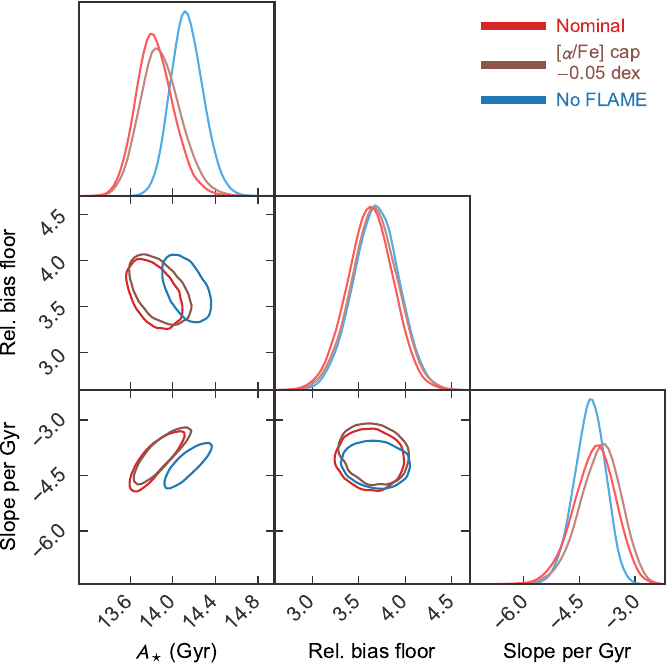}
    \caption{Triangle plots for our nominal analysis and several variants, with results summarised in Table~\ref{Main_results_table}. The solid red lines in all panels show our nominal result for reference. \emph{Top left}: Results with our nominal sample. The brown lines use a Jeffreys prior on $\sigma$ (Equation~\ref{Ln_P_MCMC}), while the blue lines skip its gradient penalty term. Other panels all show results using Equation~\ref{Ln_P_MCMC} with a wide uniform prior on $\sigma$. \emph{Top right}: The impact of raising the [Fe/H] floor at given age by 0.1~dex (brown) or 0.2~dex (blue), as shown by the higher dashed and dotted magenta lines, respectively, on Figures~\ref{age_feh_figure} and \ref{age_feh_intercept_offsets}. \emph{Bottom left}: The impact of reducing the [Fe/H] floor by 0.1~dex (brown) or 0.2~dex (blue), as shown by the the lower dashed and dotted magenta lines, respectively, on these figures. \emph{Bottom right}: The brown lines assume a 0.05~dex reduction in the [$\alpha$/Fe] floor (dashed magenta line on Figure~\ref{age_alpha_figure}). The blue lines drop the requirement for consistency between YY and FLAME ages (Section~\ref{FLAME_comparison}).}
    \label{Triangles_Numpyro_linSNR}
\end{figure*}

Our main results are shown in Table~\ref{Main_results_table}, which summarises the mode and $1\sigma$ confidence interval of each parameter. The corresponding posteriors are shown in Figure~\ref{Triangles_Numpyro_linSNR}, where different panels show analysis variants. Our nominal analysis is shown in solid red in all panels for comparison. Our main result is that $A_\star = 13.80^{+0.20}_{-0.16}$~Gyr, where the quoted uncertainty is statistical and assumes a fixed sample and analysis. A comparable common-mode stellar-model systematic uncertainty is discussed in Section~\ref{Estimated_systematic_uncertainties}, where we also estimate the total systematic uncertainty. Our inferred $A_\star$ is consistent with $\AstarCMB$ at just beyond the $1\sigma$ statistical uncertainty. Although inadequate quality cuts could inflate our $A_\star$ estimate, we consider this unlikely because the sample underpinning our main result lacks any discernible downturn in $\widetilde{\sigma}_A$ towards higher ages (Figure~\ref{Age_error_turnover_nominal}) and has a rather symmetric distribution of $\DJtz$, indicating no significant population of dynamically cold stars (Section~\ref{Jz_age_proxy}).

The top left panel of Figure~\ref{Triangles_Numpyro_linSNR} shows results using our nominal sample, but with changes to the gradient penalty term (Equation~\ref{Ln_P_MCMC}). The brown lines show the result of replacing the uniform prior on $\sigma$ with a Jeffreys prior. This slightly reduces $\sigma$ from $0.417^{+0.039}_{-0.033}$ to $0.409^{+0.042}_{-0.029}$ natural units, making the reconstructed $P(A)$ slightly smoother due to the larger gradient penalty. However, the inferred parameters are barely affected. The blue lines show the impact of removing the gradient penalty altogether. This leads to a less smooth reconstruction, with clear oscillations evident whereby adjacent pixels deviate from the moving average in opposite directions. This also feeds through to a much rougher $P/\sigma$ curve. Our analysis is able to deal with this and recover a precise $A_\star$, which moreover is very similar to our main result in red and our earlier estimates using extreme value statistics (Section~\ref{Simplified_Astar_estimates}) or an iteratively reconstructed $P(A)$ (Figure~\ref{Ln_P_obs_pp_values}). This provides confidence that for a given sample of stars, we can reliably determine $A_\star$. The $\approx 0.2$~Gyr spread in $A_\star$ across these analysis choices is comparable to our statistical uncertainty, providing an estimate of the systematic uncertainty associated with the reconstruction and fitting method.

The other panels in Figure~\ref{Triangles_Numpyro_linSNR} explore the impact of altering the quality cuts while applying a gradient penalty similarly to our nominal analysis. The top right panel shows the impact of raising the allowed [Fe/H], corresponding to the highest two magenta lines on Figure~\ref{age_feh_figure}, where the solid line shows our nominal ceiling. Mathematically, the intercept in Equation~\ref{Age_metallicity_cutoff_line} is raised by 0.1~dex for the brown lines or 0.2~dex for the blue lines in the top right panel of Figure~\ref{Triangles_Numpyro_linSNR}. Compared to our nominal result in red, this slightly increases the inferred $A_\star$, though only by $\approx 0.5\sigma$ in the former case and $1.2\sigma$ in the latter, creating moderate tension with $\AstarCMB$. However, even the latter result is still consistent with $\AUCMB$ at just over $2\sigma$ if we assume a slightly lower $\tf$ of perhaps 0.1~Gyr, at the low end of the range considered plausible in the literature \citep*{Cimatti_2023, Mazurenko_2025}. We note that relaxing the age--metallicity cutoff by 0.2~dex as described above appears somewhat problematic from the slight downturn in $\widetilde{\sigma}_A$ in the oldest age bin (see the middle right panel of Figure~\ref{Age_error_turnover_variants}). This behaviour is not apparent in our main sample (Figure~\ref{Age_error_turnover_nominal}), but is apparent prior to our quality cuts (Figure~\ref{Age_error_turnover_none}). While the downturn with the relaxed age--metallicity cutoff is nowhere near as severe, it is suggestive of stars with problematic age estimates creeping into our sample. This is also very strongly suggested by our investigation of the vertical action as an age proxy (Figure~\ref{jz_cold_tail_vs_FeHoffset}). We note that $A_\star$ changes most gradually with respect to the position of the age--metallicity cutoff when it is placed somewhere between its nominal position and 0.1~dex higher, suggesting that the cutoff line is placed almost optimally despite being chosen prior to our MCMC inference of $A_\star$.

Instead of relaxing the age--metallicity cutoff, we can tighten it, as shown in the bottom left panel of Figure~\ref{Triangles_Numpyro_linSNR}. These results correspond to the lowest two magenta lines on Figure~\ref{age_feh_figure}, which are drawn by reducing the intercept in Equation~\ref{Age_metallicity_cutoff_line} by 0.1~dex or 0.2~dex. The corresponding results are shown in the bottom left panel of Figure~\ref{Triangles_Numpyro_linSNR} using the brown and blue lines, respectively. The first reduction in the allowed [Fe/H] ceiling reduces $A_\star$ by almost $2\sigma$, while the second reduction further reduces $A_\star$ by a little over $1\sigma$ (Table~\ref{Main_results_table}). Even so, the slightly asymmetric nature of the error bars means the inferred $A_\star$ with a 0.2~dex reduction in threshold is still consistent with the mode inferred from our nominal analysis at $2\sigma$. Moreover, even this variant is still consistent with $\AstarCMB$ at just over $1\sigma$, which is also the case for our nominal analysis $-$ though the deviations are on opposite sides.

It is important to note that the results with a 0.2~dex reduction in the age-dependent [Fe/H] ceiling give the lowest $A_\star$ for any of our analysis variants. But even in this case, the inferred $A_\star = 13.39^{+0.24}_{-0.20}$~Gyr is in some tension with $\AstarETS$. Combining its $0.18$~Gyr uncertainty in quadrature with the low-side 0.20~Gyr uncertainty on our $A_\star$ inference yields a total uncertainty of 0.27~Gyr, making the 0.68~Gyr discrepancy with $\AstarETS$ a $2.5\sigma$ tension. This moderate tension would be further alleviated once we consider stellar-model and other systematic errors (Section~\ref{Estimated_systematic_uncertainties}). However, such a low ceiling on [Fe/H] is far too restrictive because the cutoff line now goes through or even below the high-density ridgeline evident towards the top right of Figure~\ref{age_feh_figure}. Since the aim of the cutoff line is to exclude stars in sparsely populated regions of the age--metallicity diagram, excluding everything above the lowest dashed line is clearly too restrictive.

To illustrate this more clearly, we consider in more detail the weighted number of stars in strips parallel to the nominal age--metallicity cutoff line, better quantifying the location of the high-density ridgeline apparent on Figure~\ref{age_feh_figure}. Each pixel here would ideally be assigned to a particular range of $\DFeH$ offsets from the nominal cutoff line. This is not possible in practice because the pixels cannot be arbitrarily small, so part of a pixel can easily be outside the considered range in offset. Instead of finding the fraction of the pixel area in between two parallel lines, we use the actual distribution of stars within each pixel. For this, we assign each star a weight that is $1/N_{\mathrm{p}}$ of the normalised count used to colour the age--metallicity pixel on Figure~\ref{age_feh_figure} containing that star, where $N_{\mathrm{p}}$ is the total number of stars in this pixel. The idea is that if a pixel here is coloured based on a normalised cell count of 0.8 and $N_{\mathrm{p}} = 8$, then each of the $N_{\mathrm{p}}$ stars can be assumed to contribute 0.1 to the colour for that pixel. Each of these 8 stars can then be uniquely assigned to the corresponding range in $\DFeH$. In this way, we can find the total weight in each $\DFeH$ bin, helping to assess how large an offset to our nominal cutoff line would still be a plausible alternative choice.

\begin{figure}
    \centering
    \includegraphics[width=\linewidth]{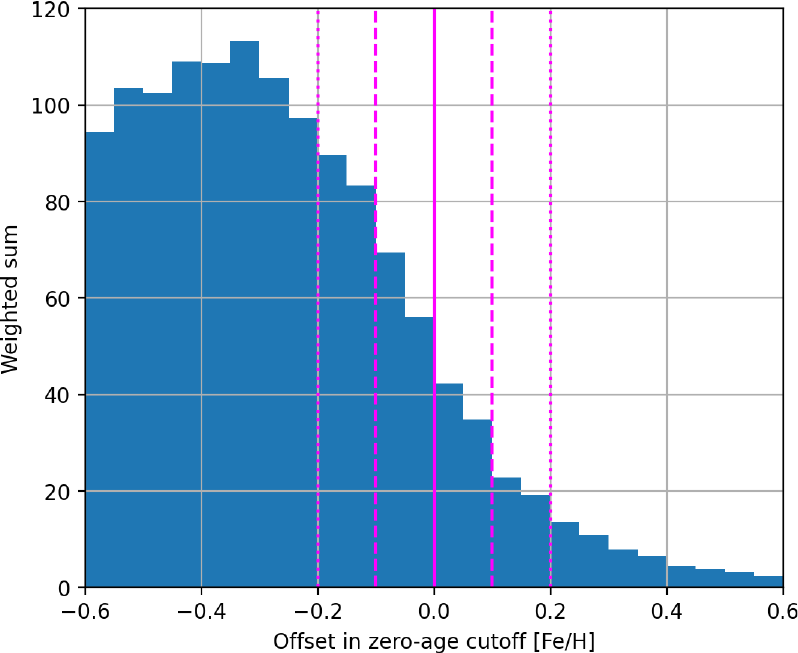}
    \caption{The solid blue bars show the total number of stars in between lines on the age--metallicity relation parallel to our nominal cutoff line (solid magenta line on Figure~\ref{age_feh_figure}). Each star is weighted so all the stars in each pixel on that figure have a total weight equal to the normalised count used to colour the corresponding pixel (see the text). Results are shown as a function of metallicity offset from the nominal position of the cutoff line, which therefore appears at 0, as illustrated with the solid vertical magenta line. The other vertical lines show the offsets used in analysis variants (Table~\ref{Main_results_table}). Since the idea of the cutoff line is to exclude outliers, it is clear that the age-dependent metallicity ceiling becomes too restrictive when we place it 0.2~dex below our nominal choice.}
    \label{age_feh_intercept_offsets}
\end{figure}

The results are shown in Figure~\ref{age_feh_intercept_offsets}, with the vertical magenta lines highlighting the offsets for which we run our detailed analysis. The solid vertical line shows the nominal cutoff lies well into the tail of the distribution, indicating a reasonable tradeoff between the desire to include more stars and a desire to avoid stars lying too far from typical scaling relations, which is suggestive of a problem with the stellar parameters. If the cutoff line is lowered by 0.1~dex from its nominal position, it is located only just outside the bulk of the distribution, suggesting the age-dependent metallicity ceiling is now too restrictive. This is especially clear if the cutoff line is placed 0.2~dex below nominal, moving it almost to the peak of the distribution. Clearly, such a restrictive cutoff line does not implement our idea of removing outliers from typical scaling relations between stellar properties. Instead, it would remove stars that are very much in line with the typical age--metallicity trend, almost certainly excluding a large number of stars with valid parameter values.

This is also apparent from the sample sizes shown in the top panel of Figure~\ref{Sample_size_vs_AC_offsets} for the old stars most relevant to finding $A_\star$. There is a steep loss of sample size from decreasing the nominal age--metallicity cutoff by 0.2~dex, which removes about 2/3 of the stars older than 13~Gyr. Even a 0.1~dex reduction removes over 1/3 of the stars at these ages. Moreover, Figure~\ref{Age_error_turnover_nominal} suggests that our nominal sample has been adequately cleaned, thus not requiring any further tightening of the quality cuts. In particular, Figure~\ref{jz_cold_tail_vs_FeHoffset} shows that the stars below but within 0.1~dex of our nominal age--metallicity cutoff have almost the same cold-tail asymmetry as stars more than 0.2~dex below it, based on a model for $\Jtz$ trained only on the latter. This provides little motivation to tighten our age--metallicity cutoff even by 0.1~dex. Doing so would likely underestimate $A_\star$ due to the reduced sample size, especially at the high ages most relevant to $A_\star$ (see the top panel of Figure~\ref{Sample_size_vs_AC_offsets}). This is because losing a significant fraction of predominantly old stars is likely to reduce the latent age of the oldest star, since the true $P(A)$ will no longer be sampled as well. Our $A_\star$ inference with the lowest considered age--metallicity cutoff line (0.2~dex below our nominal choice) should therefore not be considered a realistic estimate of $A_\star$. Instead, it is more of a very conservative lower limit. We also note that choosing the analysis variant which yields the lowest $A_\star$ is bound to improve the agreement with $\AstarETS$. Our results therefore show that there is indeed significant tension with $\AstarETS$ if we place the age--metallicity cutoff line somewhat above the typical trend rather than on top of it, as also suggested by the cold-tail asymmetry in $\DJtz$ relative to our $\Jtz$ model (Section~\ref{Jz_age_proxy}).

\begin{figure}
    \centering
    \includegraphics[width=\linewidth]{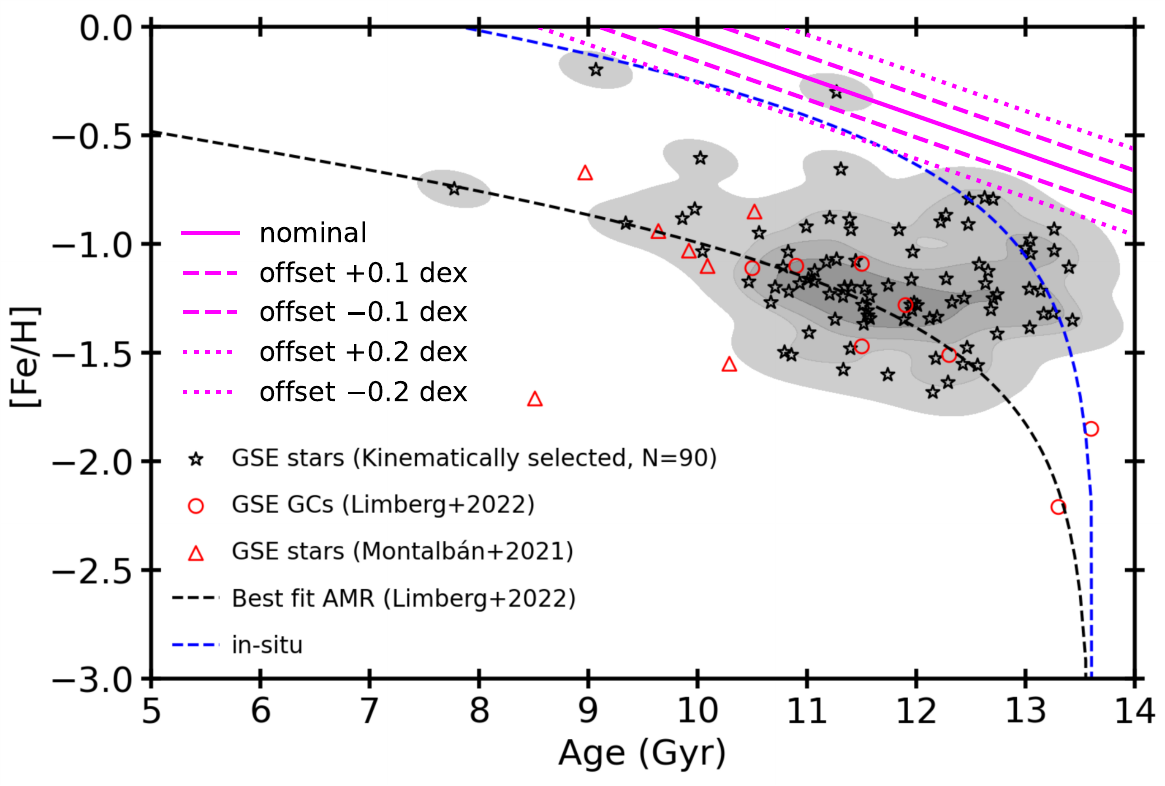}
    \caption{The age--metallicity relation for stars and GCs from the ancient GSE merger (open symbols). The black stars show 90 kinematically selected GSE stars with asteroseismic ages and uncertainties \citep{Montalban_2021}. The red triangles show GSE stars from their study, while the red circles show GSE GCs \citep{Limberg_2022}. The dashed black line shows their best-fitting GSE age--metallicity relation with an inferred $\AGC = 13.60 \pm 0.11$~Gyr, while the higher dashed blue line shows the relation for in-situ Galactic disc stars. The solid magenta line towards the top right shows our nominal age-dependent metallicity ceiling (Equation~\ref{Age_metallicity_cutoff_line}). The dashed (dotted) magenta lines move this vertically by $\pm 0.1$ ($\pm 0.2$) dex -- these are plotted similarly to Figure~\ref{age_feh_figure}. Adapted from figure~B1 of \citet{Nepal_2024}.}
    \label{GSE_age_Z_overlay}
\end{figure}

Another argument against our lowest considered age--metallicity cutoff line is provided by previous investigations of stars from the \emph{Gaia}-Sausage/Enceladus (GSE) minor merger early in Galactic history \citep{Belokurov_2018}. Figure~\ref{GSE_age_Z_overlay} shows the locations of several GSE stars and GCs, including 90 kinematically confirmed stars with asteroseismic ages \citep{Montalban_2021}. The dashed black line shows the age--metallicity relation for GSE, while the higher dashed blue line shows the relation for in-situ stars. The age-dependent metallicity ceilings we explore are shown using magenta lines towards the top right. This clearly shows that a threshold 0.2~dex below our nominal choice would remove stars even slightly more metal-rich than expected for their age, leaving no room for scatter about the expected in-situ age--metallicity relation. Indeed, the cutoff would be so restrictive that it would start to remove very metal-poor stars from the GSE merger. This is in good agreement with our previous finding that our most restrictive age--metallicity cutoff line goes through a high-density ridgeline on the diagram, making it too restrictive (see Figures~\ref{age_feh_figure} and \ref{age_feh_intercept_offsets}). Moreover, the lack of any downturn in $\widetilde{\sigma}_A$ towards high $A$ for our nominal sample (Figure~\ref{Age_error_turnover_nominal}) and the nearly flat trend of $\Act$ with $\DFeH$ below our nominal age--metallicity cutoff (Figure~\ref{jz_cold_tail_vs_FeHoffset}) suggest that the age estimates are already reliable. We therefore argue that there is no justification to lower our nominal [Fe/H] ceiling by a further 0.2~dex. The substantial loss of predominantly old stars with such an overly restrictive cut (top panel of Figure~\ref{Sample_size_vs_AC_offsets}) artificially reduces the inferred $A_\star$.

The bottom right panel of Figure~\ref{Triangles_Numpyro_linSNR} shows the remaining analysis variants that we consider. The brown curve reduces the [$\alpha$/Fe] cutoff line (Equation~\ref{Age_alpha_Fe_cutoff_line}) by 0.05~dex. This corresponds to using the dashed line on Figure~\ref{age_alpha_figure}, which we see is already well below the most densely populated part of the diagram. This is also evident from the sample size increasing by only 958 ($0.7\%$) from the 0.05~dex reduction in [$\alpha$/Fe] floor. The impact is proportionately somewhat larger on the number of old stars (see the bottom panel of Figure~\ref{Sample_size_vs_AC_offsets}). Even so, the extra stars have little impact on $A_\star$, increasing it by $\ll 1\sigma$ and leaving it consistent with $\AstarCMB$ at $\approx 1.5\sigma$.

The final analysis variant that we consider is to remove the requirement for consistency between YY and FLAME ages (Section~\ref{FLAME_comparison}). The result of relaxing this quality cut is shown in blue in the bottom right panel of Figure~\ref{Triangles_Numpyro_linSNR}. The inferred $A_\star$ increases by about $1.5\sigma$ and can no longer be reconciled with $\AstarCMB$. However, the bottom right panel of Figure~\ref{Age_error_turnover_variants} shows that there is a downturn in fractional age uncertainty at high ages. Though this is less severe than for our initial sample (Figure~\ref{Age_error_turnover_none}), it is still suggestive of contamination by stars with overestimated ages and/or underestimated uncertainties. Such a downturn in $\widetilde{\sigma}_A$ at high ages does not arise in our nominal sample (Figure~\ref{Age_error_turnover_nominal}). Since some of the most important quality cuts that this is based on (including the AC cuts and FLAME consistency) were fixed prior to developing the technique described in Section~\ref{Age_limit_MCMC} and the resulting $A_\star$ inference, we suggest that our nominal result is a reliable measurement of $A_\star$ in the large \citetalias{Xiang_2022} sample.

\section{Stars with chemistry indicative of early formation}
\label{Early_chemistry_analysis}

The oldest stars are likely to be metal-poor and enhanced in $\alpha$-elements \citep{Tinsley_1979}. Indeed, Figure~\ref{age_feh_figure} shows a low-metallicity tail towards the bottom right, where the stars largely have observed ages of $\approx 12.5 - 15$~Gyr. Assuming that much of the spread here is due to measurement errors, we can average these limits and estimate that $A_\star \approx 13.75$~Gyr. This is very much in line with our previous $A_\star$ estimates.

The narrow age spread at low metallicities motivates us to do a more detailed analysis targeting stars whose chemistry suggests formation early in Galactic history. We first define how we select such a sample (Section~\ref{Early_chemistry_sample}). We then explain how we infer $A_\star$ from this sample, exploiting the narrow age spread to simplify our $P(A)$ model (Section~\ref{A_star_inference_EC}).

\subsection{The early chemistry sample}
\label{Early_chemistry_sample}

We already apply age-dependent limits on [Fe/H] and [$\alpha$/Fe] (see Figures~\ref{age_feh_figure} and \ref{age_alpha_figure}, respectively), but so far our focus has been on removing stars whose age estimates may be problematic. We have not been concerned with the fact that our sample has many younger and more metal-rich stars. We therefore add further cuts for our `early chemistry' analysis. We only extend the quality cuts in our nominal sample, since we still need to avoid stars with potentially problematic age estimates.

The main issue with this approach is the inevitable large reduction in sample size. We therefore need to find out how far we can push the ceiling on [Fe/H] and floor on [$\alpha$/Fe] before a significant number of younger stars enters into our sample. Once this starts to happen, the age distribution broadens, requiring a more complicated $P(A)$ and thereby undermining any advantages that might be gained over our main analysis.

\begin{figure}
    \centering
    \includegraphics[width=\linewidth]{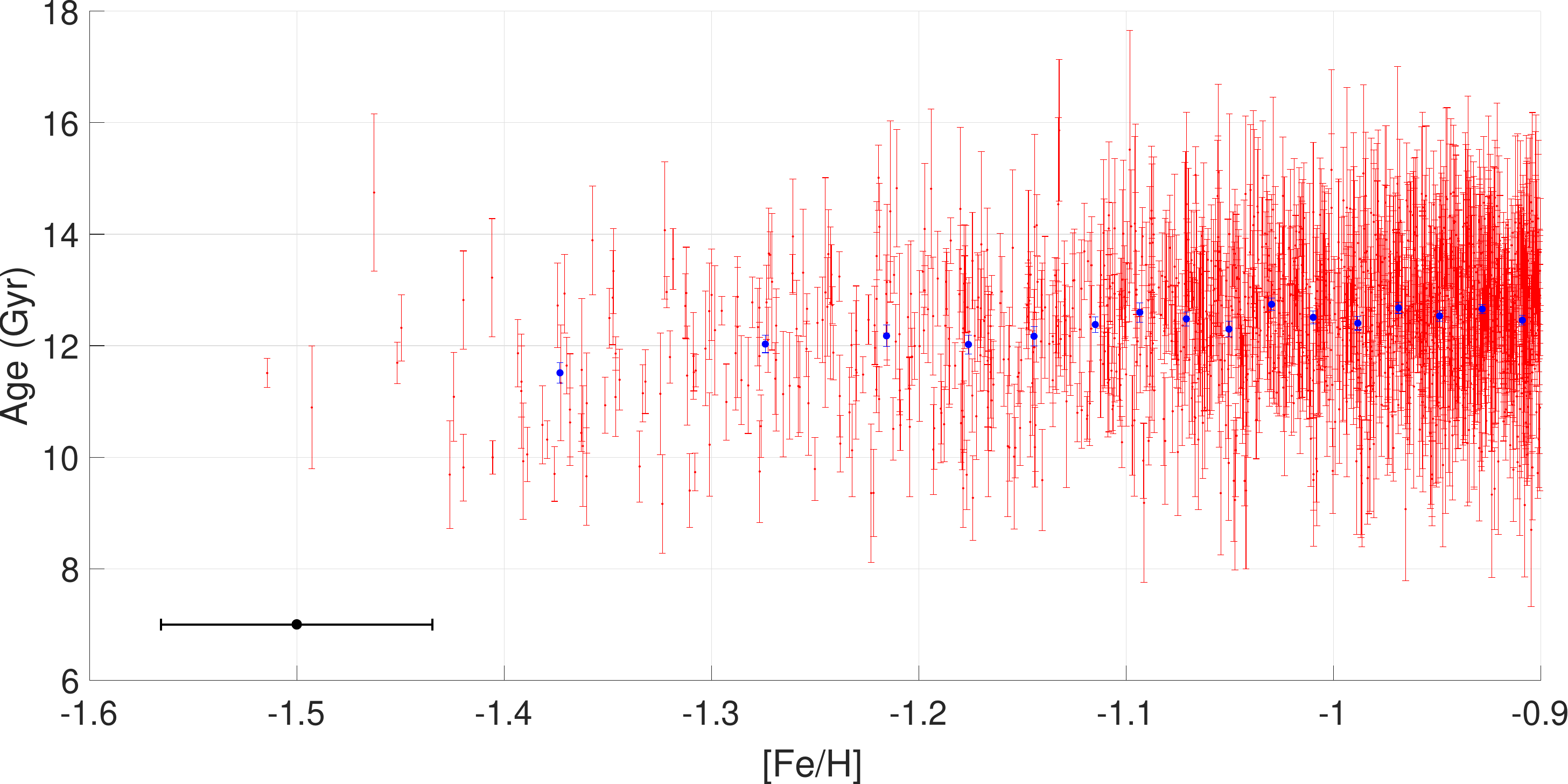} 
    \caption{The age--metallicity relation for stars in our early chemistry sample. Red points show individual stars, while the blue points with error bars show the average age and its uncertainty in each metallicity bin, for which we impose a minimum width and a minimum number of stars. The median [Fe/H] uncertainty of this sample is shown at the bottom left. For clarity, we omit [Fe/H] uncertainties, which are typically around 0.07~dex and usually $\la 0.1$~dex. Notice the lack of any age trend over the [Fe/H] range shown here.}
    \label{FEH_age_EC}
\end{figure}

Figure~\ref{FEH_age_EC} shows the age--metallicity distribution for our nominal sample restricted further to $[\mathrm{Fe}/\mathrm{H}] < -0.9$, also excluding a small number of stars with $A + \sigma_A < 10$~Gyr which would otherwise contaminate the sample. The red points show the observed ages of individual stars, while the blue points show the average observed age of stars in each [Fe/H] bin and the error on this average, assuming that intrinsic age dispersion dominates over measurement errors. Our results show no discernible trend towards lower ages at higher metallicity, suggesting that we can use the full range shown on Figure~\ref{FEH_age_EC}. However, stars do start to become younger at $[\mathrm{Fe}/\mathrm{H}] \ga -0.9$, preventing us from enlarging the sample by including even higher metallicity stars. We therefore limit our early chemistry sample to $[\mathrm{Fe}/\mathrm{H}] \leq -0.9$. This corresponds to the top of the low-metallicity tail evident towards the bottom right of Figure~\ref{age_feh_figure}. We include as much of this tail as possible while maintaining a narrow age distribution, in order to increase the sample size.

\begin{figure}
    \centering
    \includegraphics[width=\linewidth]{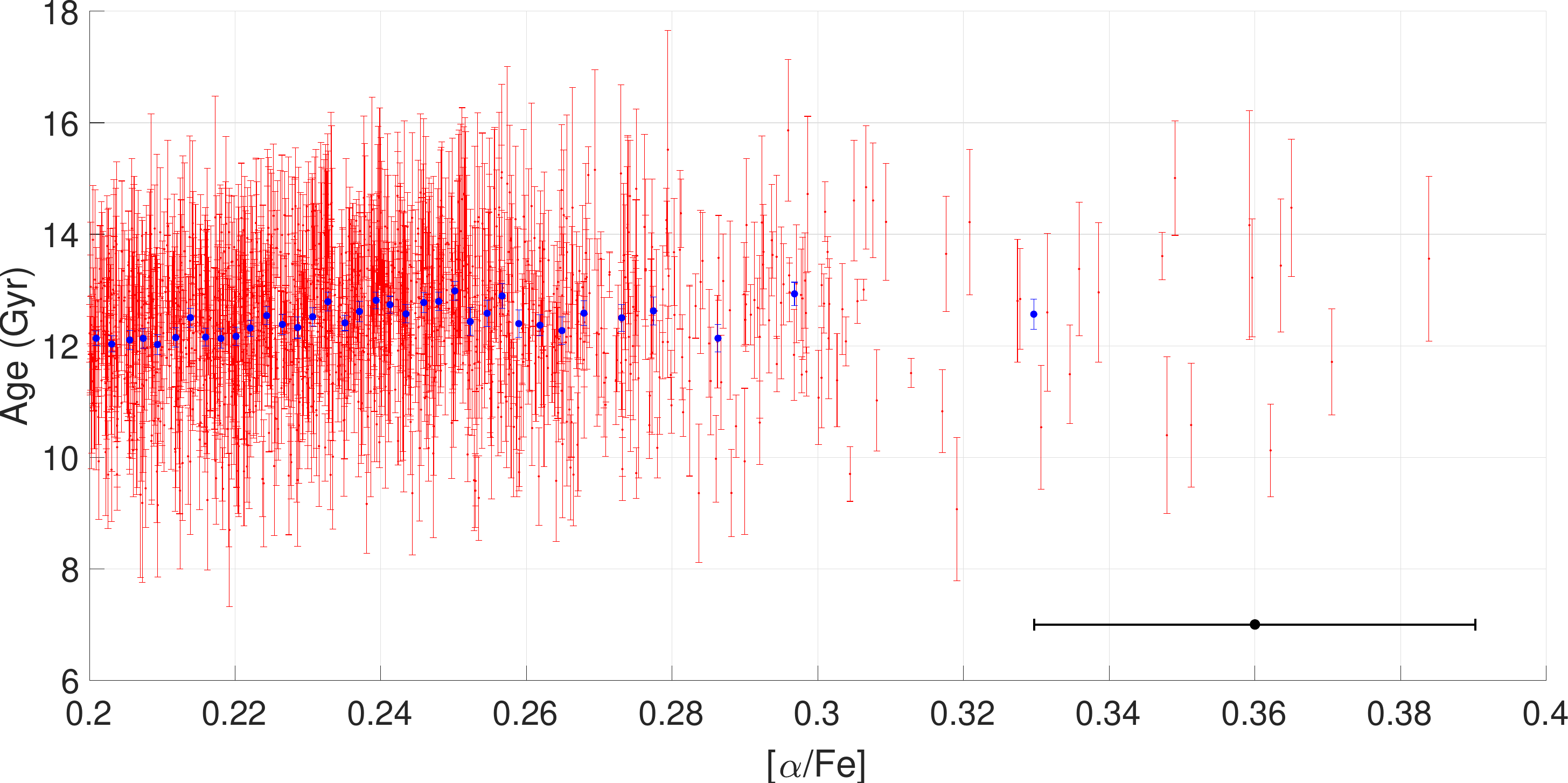} 
    \caption{Similar to Figure~\ref{FEH_age_EC}, but now showing the age--[$\alpha$/Fe] relation for our early chemistry sample. Its median [$\alpha$/Fe] uncertainty is shown at the bottom right. For clarity, we omit [$\alpha$/Fe] uncertainties, which are typically around 0.03~dex and usually $\la 0.05$~dex. These are significant compared to the range shown here, yet there is no appreciable trend in the mean age. This suggests the underlying relation between age and [$\alpha$/Fe] saturates at $[\alpha/\mathrm{Fe}] < 0.2$.}
    \label{ALPHA_FE_age_EC}
\end{figure}

We analyse the age--[$\alpha$/Fe] relation similarly in Figure~\ref{ALPHA_FE_age_EC}. There is no discernible trend in mean age down to $[\alpha/\mathrm{Fe}] = 0.2$, but stars with even lower [$\alpha$/Fe] do tend to be younger. We therefore limit our early chemistry sample to $[\alpha/\mathrm{Fe}] \geq 0.2$. This corresponds to the bottom of the $\alpha$-enriched spur evident towards the top right of Figure~\ref{age_alpha_figure}. We note that the [$\alpha$/Fe] errors are significant compared to the range shown here, so the lowest [$\alpha$/Fe] bins will include some stars at even lower [$\alpha$/Fe] than plotted. The lack of any trend in the mean age suggests that the underlying relation between age and [$\alpha$/Fe] saturates at $[\alpha/\mathrm{Fe}] < 0.2$, since otherwise an appreciable fraction of the stars with observed [$\alpha$/Fe] just above 0.2 would be younger, reducing the mean age for these bins.

Combining our metallicity and $\alpha$-enhancement cuts leads to a generally very old stellar population. Although a handful of much younger stars do slip through, these are obviously outliers, which we can easily remove by excluding stars with $A + \sigma_A < 10$~Gyr. After applying all these extra cuts on top of those in our nominal sample of 146,810 stars, we are still left with 1439 stars. This provides a chemistry-based confirmation of our assumption in Section~\ref{Individual_age_likelihoods} that about 1\% of the stars in our full sample might plausibly contribute to the inferred $A_\star$. While the remaining 99\% of the sample does in principle have some effect on the results discussed previously, this is not the case for our inference of $A_\star$ from our early chemistry sample, which we discuss next.

\subsection{Simplified \texorpdfstring{$A_\star$}{A*} inference}
\label{A_star_inference_EC}

\begin{figure}
    \centering
    \includegraphics[width=\linewidth]{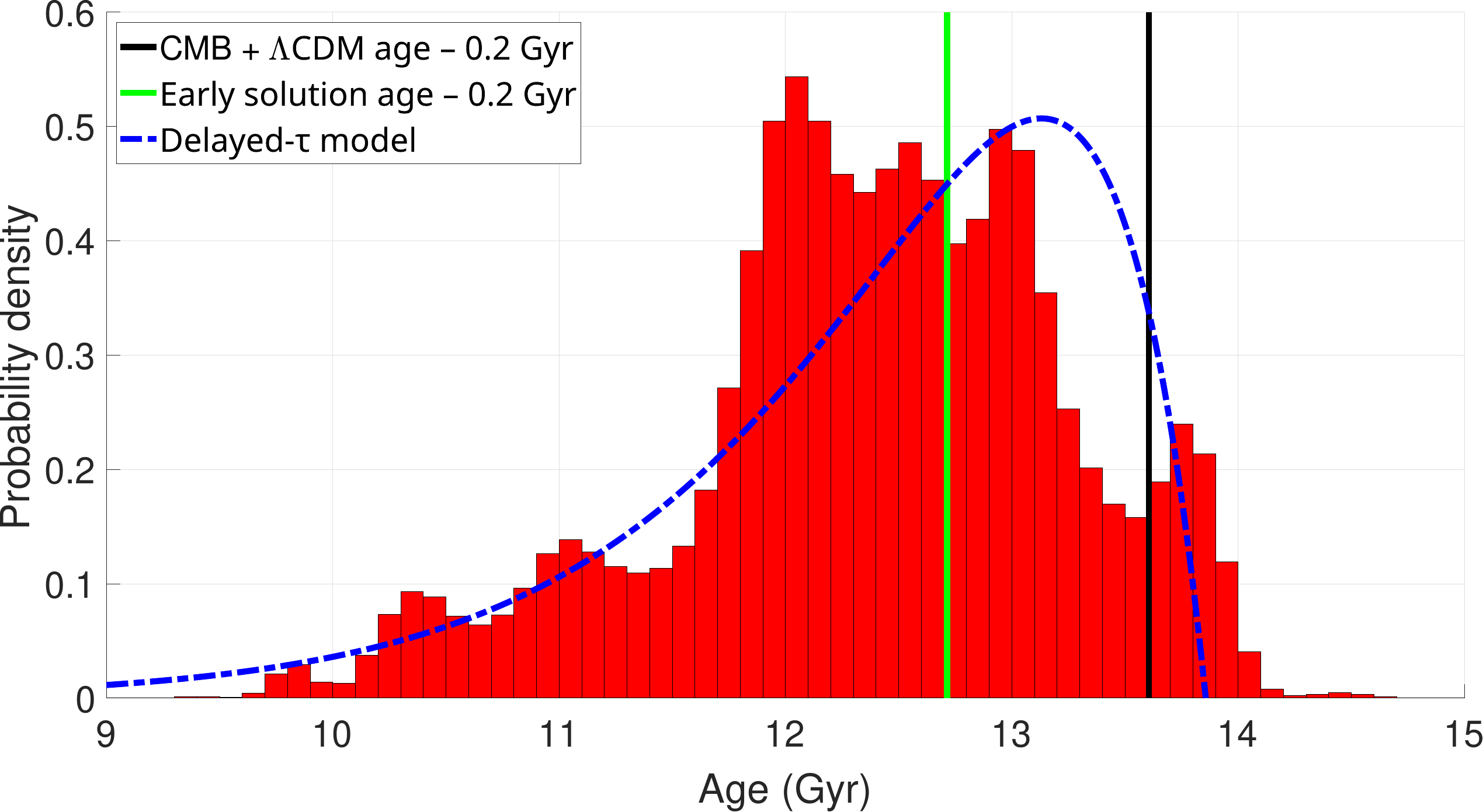} 
    \caption{Our 200-bin iterative population prior reconstruction of $P(A)$ for our early chemistry sample, using the method described in Section~\ref{Population_prior_protocol}. The solid green and black vertical lines show $\AstarETS$ and $\AstarCMB$, respectively. The dot-dashed blue line shows the best-fitting delayed-$\tau$ model truncated to $A > 7$~Gyr.}
    \label{PP_reconstruction_HR_EC_nominal_20}
\end{figure}

We expect that combining our [$\alpha$/Fe] and [Fe/H] cuts should provide an efficient way to select the oldest stars. The resulting narrow distribution of ages should then permit a simpler $P(A)$ model. To check this, we first run our iterative population prior reconstruction (Section~\ref{Population_prior_protocol}). Results converge very quickly, but due to the low computational cost, we run 20 stages. The final $P(A)$ reconstruction for our early chemistry sample is shown in Figure~\ref{PP_reconstruction_HR_EC_nominal_20}. This does indeed show a narrow distribution, which we model next. It also suggests that $A_\star$ slightly exceeds $\AstarCMB$ and significantly exceeds $\AstarETS$, in line with our main result (Table~\ref{Main_results_table}).

We model the latent age distribution of our early chemistry sample as a simple delayed-$\tau$ model \citep{Kroupa_2020} with timescale $A_{\rm s}$.
\begin{eqnarray}
    P \left( A \right) ~\propto~ x \exp \left( -x \right), \quad x \equiv \frac{A_\star - A}{A_{\rm s}} \, .
    \label{P_A_EC}
\end{eqnarray}
Since our sample lacks stars with $A + \sigma_A < 10$~Gyr, we truncate the distribution so that the latent $A \geq 7$~Gyr. We will see later that given the low inferred $A_{\rm s}$, truncating $P \left( A \right)$ to $A \geq 7$~Gyr has negligible impact on its overall normalisation and thus on our results.

Compared to our main analysis where we reconstruct $P(A)$ non-parametrically in 100 bins (Section~\ref{MCMC_reconstruction}), our early chemistry analysis is much simpler thanks to the narrower age distribution. The use of a parametrised $P(A)$ allows us to infer $A_\star$ directly as one of the model parameters, rather than in a secondary analysis of the reconstructed $P(A)$. These major advantages must be offset against the $102\times$ smaller sample size and inaccuracies arising from adopting the delayed-$\tau$ model, which Figure~\ref{PP_reconstruction_HR_EC_nominal_20} shows does not provide a good fit to the data. We expect that this would lead to a systematic uncertainty in $A_\star$ of up to the width of the $P(A)$ distribution, i.e., of order $A_{\rm s}$.

\begin{figure}
    \centering
    \includegraphics[width=\linewidth]{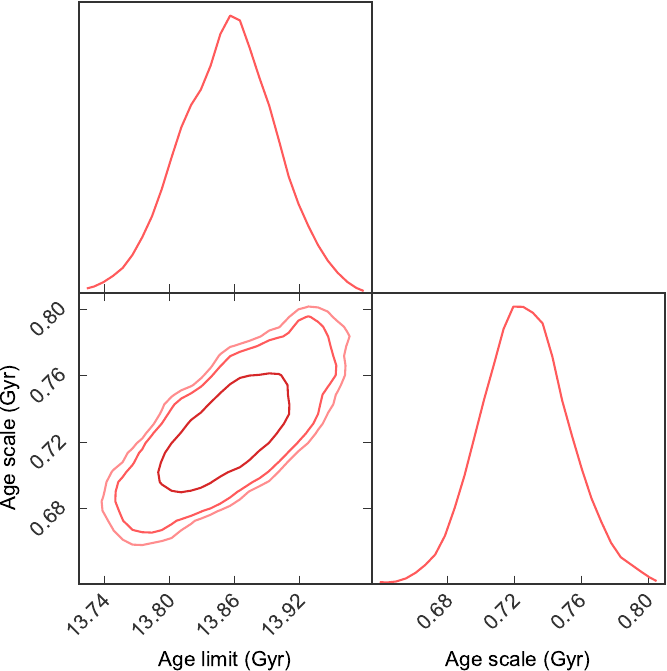} 
    \caption{Triangle plot showing the parameters inferred by our early chemistry analysis (Equation~\ref{P_A_EC}). The joint posterior shows the $1\sigma$, $2\sigma$, and $3\sigma$ contours.}
    \label{Triangle_EC}
\end{figure}

We infer the two model parameters in Equation~\ref{P_A_EC} using MCMC, again using \textsc{numpyro} with 500 warm-up samples and 2500 main samples. The results of the latter are shown in Figure~\ref{Triangle_EC}. The distribution has a quite narrow inferred `width' of $A_{\rm s} = 0.72^{+0.03}_{-0.02}$~Gyr and starts at $A_\star = 13.86^{+0.04}_{-0.05}$~Gyr. This error bar is statistical only and conditional on the assumed delayed-$\tau$ form (Equation~\ref{P_A_EC}), so it should not be taken at face value. A more realistic uncertainty estimate might be $A_{\rm s}$ given the poor fit of the delayed-$\tau$ model to the non-parametric population prior reconstruction of $P(A)$ (Figure~\ref{PP_reconstruction_HR_EC_nominal_20}). However, this figure suggests that $A_\star$ is quite close to our delayed-$\tau$ estimate.


Our result for $A_\star$ is very much in line with our simplified estimates (Section~\ref{Simplified_Astar_estimates}) and in excellent agreement with our main analysis (Table~\ref{Main_results_table}), both of which use our $102\times$ larger nominal sample. Note in particular that we find a similar $A_\star$ despite the use of much stricter cuts on the stellar chemistry. This shows that there is only limited scope to alter $A_\star$ by changing the AC cutoff lines used to define our main sample (Sections~\ref{Age_metallicity_relation} and \ref{Age_alpha_Fe_relation}). It is therefore clear that $A_\star$ is quite close to $\AstarCMB$ and possibly slightly higher, strongly disfavouring $\AstarETS$ on statistical and sample-selection grounds.

While the uncertainty on $A_\star$ in our early chemistry analysis is surely underestimated given it assumes the delayed-$\tau$ model (Equation~\ref{P_A_EC}), even reducing $A_\star$ by $A_{\rm s}$ would barely reconcile it with $\AstarETS$. Moreover, the population prior reconstruction for this sample's $P(A)$ peaks close to $\AstarETS$, making it rather unlikely that this is in fact the end point of the distribution (Figure~\ref{PP_reconstruction_HR_EC_nominal_20}). It would in any case be a remarkable coincidence if deficiencies with the assumed delayed-$\tau$ model cause it to overestimate $A_\star$ by precisely the amount needed to make $\AstarETS$ look like $\AstarCMB$. Given our early chemistry sample is selected to have a low metallicity, it is also unlikely that stellar-model systematics can shift the inferred age scale by a comparable amount (Section~\ref{Stellar_modelling_uncertainties}).

\section{Discussion}
\label{Discussion}

\subsection{Preparing a suitable sample}
\label{Sample_preparation}

Our initial sample of 247,103 subgiant stars comes from LAMOST~DR7 spectra \citep{Cui_2012, Zhao_2012} cross-matched with \emph{Gaia}~eDR3 astrometry \citep{Gaia_2021}. The ages of these stars were found by \citetalias{Xiang_2022} using YY isochrones up to 20~Gyr. Their reported median age uncertainty of 7.5\% appears reasonable based on comparing their reported ages for stars in wide binaries, which should be co-eval \citep*{Shariat_2026}. We nonetheless find an unexpected downturn in $\widetilde{\sigma}_A$ at $A \ga 16$~Gyr (Figure~\ref{Age_error_turnover_none}). Clearly, the quality cuts imposed by \citetalias{Xiang_2022} are not sufficient to allow a reliable inference of $A_\star$. While it is not clear why the \citetalias{Xiang_2022} sample has stars with very high ages and very low age uncertainties, this problematic behaviour is similar to the secondary peak in the age distribution at 19~Gyr reported by \citet{Tomasetti_2026}. The unusually high and precise ages (given the poorer precision at lower ages) might indicate observational difficulties in a small number of cases, which seems inevitable in a large survey like LAMOST. This is suggested by the problematic stars generally having a lower SNR (Section~\ref{Why_AC_cuts_help}). Another possibility is mass loss causing a star to appear older due to the tight age--mass relation (Figure~\ref{flame_lnage_lnmass}), or undetected binaries with a low-mass faint companion biasing the parameters of the brighter star \citep{Woody_2025}. We estimate that $\approx 0.1\%$ of the \citetalias{Xiang_2022} sample suffer from serious deficiencies with the inferred age based on this being incompatible with $\AstarCMB$, even at a reduced $\tf$ of just 0.1~Gyr.

To avoid cosmological biases, we cannot directly exclude these stars, or indeed stars which contribute to the unrealistic downturn in $\widetilde{\sigma}_A$ at $A \ga 16$~Gyr. We therefore develop a series of quality cuts that complement those applied by \citetalias{Xiang_2022}. In addition to basic cuts on the parallax (Section~\ref{Parallax_error_cut}) and how often each star appears resolved into multiple sources by \emph{Gaia} or has an elongated appearance (Section~\ref{Gaia_binary_cuts}), we apply a few additional quality cuts that are defined iteratively at the population level. These kinds of cuts are not normally considered in the literature because they require a large sample size. The basic idea is to reject outliers from the population distribution of some parameter or parameter pair. We apply this idea to the age--metallicity and age--[$\alpha$/Fe] relations, where we define a cutoff line manually prior to detailed inference of $A_\star$ (Figures~\ref{age_feh_figure} and \ref{age_alpha_figure}, respectively). The chemical idea behind this is that the oldest stars should be metal-poor and $\alpha$-enhanced \citep{Tinsley_1979}. Thus, any apparently very old star with high metallicity and/or low [$\alpha$/Fe] could indicate a problematic age measurement.\footnote{It is also possible the problem lies with the chemistry, but we reject such stars to be on the safe side.}

Our AC cuts are justified further in Section~\ref{AC_cuts_justification}, where we show that the AC-rejected stars contain a population with much lower vertical action than expected for their age, metallicity, and distance. We parametrise this using $\Act$ (Equation~\ref{cold_tail_definition}), which we consider a proxy for the level of contamination by stars with overestimated ages. We find that $\Act$ rises sharply upon crossing our adopted age--metallicity ceiling (Figure~\ref{jz_cold_tail_vs_FeHoffset}), with a less clear jump upon crossing our age--[$\alpha$/Fe] floor (Figure~\ref{jz_cold_tail_vs_alphaFeoffset}). Stars which fail both AC cuts have the highest $\Act$ compared to stars which fail only one of them (Table~\ref{jz_cold_tail_table}). The underlying reason for the age overestimates may be that the low spectral resolution ($R = 1800$) causes line blending issues that are exacerbated by the somewhat lower SNR of the AC-rejected stars compared to the nominal sample at a similar age and apparent magnitude (Figure~\ref{snr_ratio_vs_age}).

Both AC cuts are internal to the \citetalias{Xiang_2022} catalogue, but we can also benefit from comparing to a different catalogue. For this, we compare the \citetalias{Xiang_2022} age estimates with ages from the \emph{Gaia} FLAME package \citep{Pichon_2007, Creevey_2023, Creevey_2026}. Almost 80\% of the \citetalias{Xiang_2022} stars can be cross-matched in this way. Some of the stars with missing FLAME ages were apparently analysed by FLAME because other FLAME parameters are available -- except the age. This is due to the imposition of a cosmologically motivated age limit of 13.5~Gyr, beyond which the age has been deleted from the \emph{Gaia} catalogue -- but other parameters are still available. For stars with full FLAME data, we find the expected tight age--mass relationship for stars near the end of their life (Figure~\ref{flame_lnage_lnmass}). We use this relation (Equation~\ref{FLAME_age_mass_relation}) to find imputed FLAME ages for the stars with deleted FLAME ages. We then compare the imputed or published FLAME age to the YY age from \citetalias{Xiang_2022}. However, we cannot simply require consistent ages due to the much less rigorous nature of the FLAME ages, which use only \emph{Gaia} astrometric and photometric data, requiring simplifying assumptions about the stellar chemistry \citep{Fouesneau_2022}. We therefore use $3\sigma$ trendline outlier rejection between the FLAME and spectroscopic YY ages (Figure~\ref{yy_vs_flame_age_figure}).

After applying the quality cuts, we suggest three further checks that can be done on a sample to check its suitability for inferring $A_\star$:
\begin{enumerate}
    \item The fractional age uncertainty should rise up to the highest reported ages, as occurs for our nominal sample (Figure~\ref{Age_error_turnover_nominal}). The only quality cut variants to show a downturn in $\widetilde{\sigma}_A$ at the highest ages are those with a much higher metallicity ceiling and without a consistency check between YY and FLAME ages using trendline outlier rejection (see the middle and bottom right panel, respectively, of Figure~\ref{Age_error_turnover_variants}).
    \item Once $A_\star$ is determined for the sample, there should not be any star which challenges this $A_\star$ by more than is appropriate for the effective sample size. In our case, we estimate that the maximum plausible deviation with Gaussian errors is $3.4\sigma$ (Section~\ref{Individual_age_likelihoods}) -- though this would rise with a larger sample and/or one that is more targeted at the oldest stars. Our nominal sample does not contain stars with $A > A_\star + 3.4 \, \sigma_A$ (Figure~\ref{Age_error_turnover_nominal}). The only variant with such internally inconsistent stars does not use FLAME ages to cross-check the YY ages we use in this study (see the bottom right panel of Figure~\ref{Age_error_turnover_variants}).
    \item Using an even cleaner version of the sample to train a model for the vertical action, the $\DJtz$ residuals relative to the model should be similar between the cleaner training sample and the extra stars included in the proposed sample. In our case, we train the $\Jtz$ model (Equation~\ref{Jz_terms}) using stars with a metallicity offset relative to our nominal ceiling (Equation~\ref{Age_metallicity_cutoff_line}) of $\DFeH < -0.2$. The cold-tail asymmetry of these stars is consistent with that for the extra stars in our nominal sample with $\DFeH = (-0.2, 0)$, indicating that the latter stars can be safely included (Figure~\ref{jz_cold_tail_vs_FeHoffset}). A similar argument can be made for our floor on [$\alpha$/Fe] (Equation~\ref{Age_alpha_Fe_cutoff_line}), though in this case, it is also possible to justify relaxing the cut by a further 0.05~dex (Figure~\ref{jz_cold_tail_vs_alphaFeoffset}). The bottom left panel of Figure~\ref{Age_error_turnover_variants} shows that the resulting sample appears quite suitable for determining $A_\star$, making this one of the most plausible analysis variants among those we consider. However, it barely shifts the inferred $A_\star$ (Table~\ref{Main_results_table}).
\end{enumerate}

\subsection{Inference of \texorpdfstring{$A_\star$}{A*}}
\label{A_star_inference}

After applying our quality cuts, we are left with a nominal sample of 146,810 stars. $\widetilde{\sigma}_A$ now rises continuously with $A$, suggesting the quality cuts are sufficient for cosmological analysis (Figure~\ref{Age_error_turnover_nominal}). Naturally this is also true for the analysis variants we consider in which even stricter cuts are applied (top row of Figure~\ref{Age_error_turnover_variants}). However, its middle row shows that when the age--metallicity ceiling is relaxed, a mild downturn in $\widetilde{\sigma}_A$ is evident at the highest reported ages, suggesting there is little scope to safely relax our quality cuts. This is also apparent from the vertical action results in Figure~\ref{jz_cold_tail_vs_FeHoffset}. Since the most important quality cuts (including AC and FLAME) were set prior to developing the $P/\sigma$ technique (Section~\ref{Age_limit_MCMC}) underpinning our detailed estimates of $A_\star$, our nominal sample and analysis should give a reliable result for $A_\star$ -- even though this appears not to be the case for some of the considered variants.

Having defined our nominal sample, we consider a variety of ways to estimate $A_\star$, gradually building up the rigour and complexity of the analysis. The results consistently show that $A_\star \approx 13.7$~Gyr (or $\approx 13.4$~Gyr if larger age uncertainties are assumed) using the full 140-knot age likelihood $\mathcal{L}(A)$ of each star. Using our 100-bin MCMC reconstruction of the latent $P(A)$ and fitting the dependence of $\overline{P}/\sigma$ on age using a linear~+~flat model (Figure~\ref{P_SN_ratio}), we find that $A_\star = 13.80^{+0.20}_{-0.16}$~Gyr. This represents a much more sophisticated statistical analysis compared to techniques typically used in the literature to obtain $A_\star$. Varying our analysis methodology and especially our sample selection does change $A_\star$ somewhat, but the changes are broadly in line with our inferred uncertainty for a fixed sample and analysis (Table~\ref{Main_results_table}). The largest departures arise from the most extreme shifts to our quality cuts, which we argue are unreasonable on other grounds. In particular, a much stricter age--metallicity cutoff would exclude stars that cannot be considered outliers (Figures~\ref{age_feh_intercept_offsets} and \ref{GSE_age_Z_overlay}). This would artificially remove primarily old stars (see the top panel of Figure~\ref{Sample_size_vs_AC_offsets}), thereby reducing $A_\star$. Even so, our lowest result of $A_\star = 13.39^{+0.24}_{-0.20}$~Gyr still lies $2.5\sigma$ above $\AstarETS = 12.71 \pm 0.18$~Gyr (Equation~\ref{AU_analytic}). We stress that the 13.39~Gyr inference follows from placing the age--metallicity cutoff almost on top of the mean in-situ Galactic relation (Figure~\ref{GSE_age_Z_overlay}), so the resulting $A_\star$ should be treated as a very conservative lower limit rather than the result of plausible quality control choices. Our consideration of the vertical action as an age proxy also provides little motivation to lower the metallicity ceiling by even 0.1~dex (Figure~\ref{jz_cold_tail_vs_FeHoffset}). It is therefore clear that our analysis prefers $A_\star > \AstarETS$.

On the other hand, our $A_\star$ estimates are reasonably consistent with $\AstarCMB$ for our nominal $\tf$ in all cases apart from the variant where we remove the requirement for consistency with FLAME ages (Section~\ref{FLAME_comparison}). This leads to a downturn in fractional age uncertainty at the oldest ages and creates an internal inconsistency in that some stars have sufficiently high and precise ages to challenge $A_\star$ determined from the same sample at $3.4\sigma$, the maximum expected random scatter for Gaussian errors given the effective sample size (see the bottom right panel of Figure~\ref{Age_error_turnover_variants}). Keeping the FLAME quality cut, the highest result we obtain is $A_\star = 14.04^{+0.20}_{-0.16}$~Gyr, which corresponds to raising the nominal age--metallicity cutoff by 0.2~dex. This leads to a noticeable downturn in $\widetilde{\sigma}_A$ at the highest ages (see the middle right panel of Figure~\ref{Age_error_turnover_variants}) and creates a noticeably higher cold-tail asymmetry (Figure~\ref{jz_cold_tail_vs_FeHoffset}). Even this analysis is still consistent with $\AUCMB$ at just over $2\sigma$ with a reduced $\tf$ of 0.1~Gyr, at the low end of the range generally considered plausible. Other analyses are in $\la 1.5\sigma$ tension with $\AstarCMB$.

As our final analysis variant, we further restrict our nominal sample to stars with a very low metallicity and high [$\alpha$/Fe], both of which should select for the oldest stars (Section~\ref{Early_chemistry_sample}). Fitting a delayed-$\tau$ model for $P(A)$ to our early chemistry sample gives $A_\star = 13.86^{+0.04}_{-0.05}$~Gyr, which is practically identical to our main result that $A_\star = 13.80^{+0.20}_{-0.16}$~Gyr. We suggest that deficiencies with the delayed-$\tau$ model could create systematic uncertainties of order the distribution timescale $A_{\rm s} = 0.72$~Gyr. However, our non-parametric population prior reconstruction of $P(A)$ peaks around $\AstarETS$, making it very unlikely that this is the end point of the latent age distribution (Figure~\ref{PP_reconstruction_HR_EC_nominal_20}). Instead, the reconstruction suggests that our delayed-$\tau$ inference is approximately correct, in which case the end point is close to $\AstarCMB$. We stress that the good overall agreement with $\AstarCMB$ arises despite not assuming this at any stage, though we do assume that $A_\star < 20$~Gyr.

\subsection{Comparison to prior studies}
\label{Comparison_prior_studies}

Our result is very much in line with previous studies, which generally indicate that $A_\star \approx 14$~Gyr. For instance, the inverse variance weighted mean age of the 11 samples considered by \citet{Cimatti_2023} is $14.05 \pm 0.25$~Gyr \citep{Banik_2025_cosmology}. Using instead GCs shows that the oldest one has an age of $\AGC = 13.39 \pm 0.25$~Gyr \citep{Valcin_2021, Valcin_2025}, combining statistical and systematic uncertainties. The lower estimate could indicate that $\AGC < A_\star$, which is possible as GCs may well take longer to form than individual stars. There are also far fewer GCs, making it harder to statistically estimate $\AGC$ due to the larger sampling gap we expect between the true age of the oldest GC in the sample compared to in the Galaxy as a whole. If we neglect these issues and treat both studies on an equal footing given their similar uncertainties, their average $A_\star = 13.72$~Gyr, which closely matches our result. Importantly, both studies indicate that $A_\star > 13$~Gyr.

This conclusion is supported at high confidence by more recent analyses \citep{Nepal_2024, Tomasetti_2026}. For instance, a recent reanalysis of GCs allowing multiple stellar populations found that $\AGC = 13.61 \pm 0.34$~Gyr if systematic and statistical errors are combined \citep{Valcin_2026}. This is in line with a previous estimate that the moderately metal-poor GCs in the Galactic bulge region started forming $13.6 \pm 0.2$~Gyr ago \citep{Souza_2024}, while analysis of GCs from the ancient GSE merger \citep{Belokurov_2018} give an estimate of $13.60 \pm 0.11$~Gyr \citep{Limberg_2022}. These results are consistent with the subgiant Methuselah star (HD~140283) having an asteroseismic age of $14.2 \pm 0.4$~Gyr \citep{Lundkvist_2025}. A similar age of $14.1^{+1.0}_{-1.5}$~Gyr has been reported for the candidate hypervelocity star DESI-HVS1 \citep{Deng_2026}. Their figure~4 shows that improved photometry would substantially reduce its age uncertainty, which at present is large enough to allow $\AstarETS$.

Compared to previous studies, a major advantage of our work is that the sample is large enough to reduce dependence on any single star, yet it does still involve analysing individual stars rather than the integrated spectrum of many stars blended together (as in the cosmic chronometer or CC technique). GCs offer their own advantages in that their stars have a narrow age spread and share a common distance, which is important given much of the age uncertainty has historically been due to that in the distance \citep{Ying_2023, Ying_2025}. However, this is less of a concern nowadays given the very large number of \emph{Gaia} parallaxes accurate to within 10\%. When using individual stars, there is always a look-elsewhere effect that one has to be careful of, given the large number of stars with some available information that could be used to estimate the age and target old stars for follow-up. This makes $3\sigma$ or even $4\sigma$ outliers inevitable, but such effects can be taken care of in a statistical analysis, which we do in some detail.

Another advantage of our work is that we carefully avoid cosmological biases, apart from the weak prior that $\AU < 20$~Gyr \citepalias{Xiang_2022}. While this is mostly thanks to those authors, we also took precautions to mitigate the artificial removal of FLAME ages $>13.5$~Gyr based on other data columns still being available for these stars. Having access to a large sample enabled us to apply innovative population-level quality cuts, without which we cannot obtain reliable results (Figure~\ref{Age_error_turnover_none}). These cuts are based on the need for ages to be consistent between catalogues and for old stars to be metal-poor and $\alpha$-enriched (the AC cuts are further justified in Section~\ref{AC_cuts_justification} without assuming $A_\star$).

Due to the above considerations, our study is an important advance on previous estimates of $A_\star$, but our result is numerically similar. It also remains stable despite considering several techniques to obtain $A_\star$ from our nominal sample and variations to the quality cuts that define it (Table~\ref{Main_results_table}).

\subsection{Estimated systematic uncertainties}
\label{Estimated_systematic_uncertainties}

Our nominal result that $A_\star = 13.80^{+0.20}_{-0.16}$~Gyr is subject to various systematic uncertainties, which we consider in more detail here. We provide a combined systematic error budget in Section~\ref{Combining_systematic_errors}.

\subsubsection{Choice of quality cuts}
\label{Quality_cuts}

We estimate that our quality cut choices have a $1\sigma$ impact of 0.2~Gyr because most analysis variants lie within this margin of our nominal result (Table~\ref{Main_results_table}). The exceptions on the high side arise upon not using FLAME ages as a cross-check and upon relaxing the age--metallicity ceiling by 0.2~dex, but these samples suffer from a problematic downturn in fractional age uncertainties at high ages (see the middle and bottom right panel, respectively, of Figure~\ref{Age_error_turnover_variants}). The variant without FLAME ages gives the highest $A_\star$ that lies 0.31~Gyr above our nominal result, but this suffers from an additional internal consistency problem in that some stars have sufficiently high and precise ages to challenge the $A_\star$ estimated from this sample.

On the low side, we can reduce $A_\star$ by up to 0.41~Gyr if we tighten the age--metallicity ceiling by 0.2~dex, but we argue in Section~\ref{Results} that this is far too restrictive. None of our sample quality diagnostics indicate a problem with the nominal position of the age--metallicity ceiling, which appears to be at the optimal location based on our consideration of the vertical action (Figure~\ref{jz_cold_tail_vs_FeHoffset}). Lowering the ceiling by 0.1~dex is more plausible and reduces $A_\star$ by 0.28~Gyr, but this could be driven by the reduced sample size (see the top panel of Figure~\ref{Sample_size_vs_AC_offsets}) rather than improved quality, given that our nominal sample lacks any downturn in $\widetilde{\sigma}_A$ at high ages (Figure~\ref{Age_error_turnover_nominal}). Moreover, the stars older than 12.5~Gyr with $\DFeH = (-0.1, 0)$ relative to our nominal ceiling have a quite low cold-tail asymmetry of $\Act = 2.48 \pm 0.79\%$, which is quite consistent with the $\Act = 1.77 \pm 0.60\%$ of our training sample with $\DFeH < -0.2$. Our other analysis variants with $A_\star$ below our nominal result all lie within 0.22~Gyr of it.

\begin{figure}
    \centering
    \includegraphics[width=\linewidth]{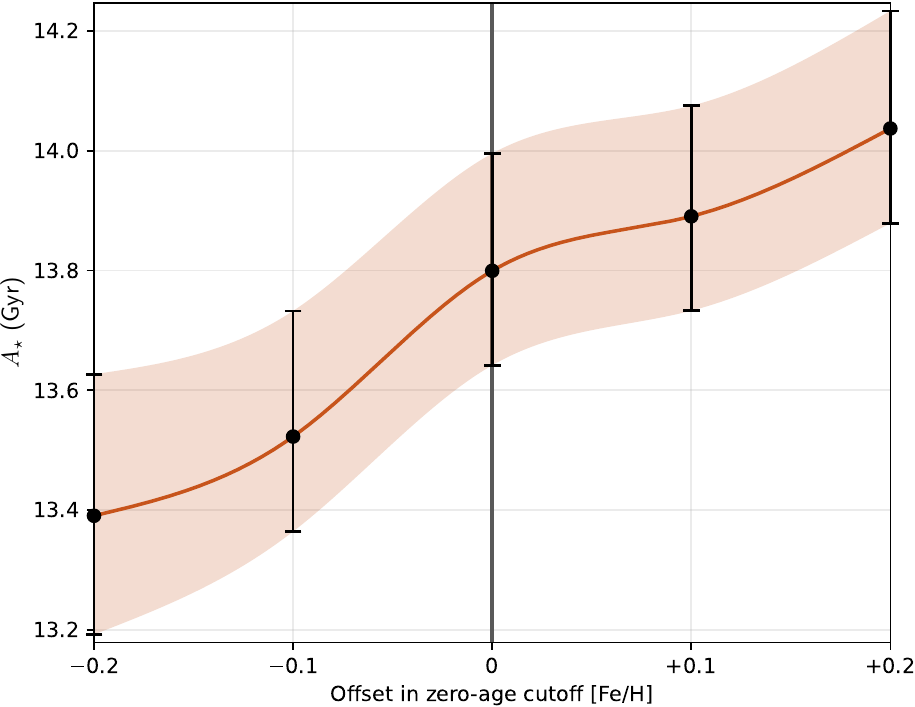} 
    \caption{Dependence of $A_\star$ on how far the age--metallicity cutoff line (Equation~\ref{Age_metallicity_cutoff_line}) is shifted relative to its nominal position, which is shown here as a vertical grey line. Notice the flattest region lies just to its right. The results shown here are from Table~\ref{Main_results_table}.}
    \label{Astar_vs_FeHoffset}
\end{figure}

These considerations leave little flexibility in how to implement the quality cuts which have the greatest impact on $A_\star$. As a result, the available flexibility does not introduce much systematic uncertainty to the $A_\star$ estimate. In particular, relaxing the age--metallicity cutoff by 0.1~dex raises the inferred $A_\star$ by only 0.09~Gyr, while relaxing the age--[$\alpha$/Fe] cutoff by 0.05~dex raises $A_\star$ by just 0.05~Gyr. The sensitivity of $A_\star$ to the position of the age--metallicity ceiling is minimised if the ceiling is placed close to its nominal position, as shown in Figure~\ref{Astar_vs_FeHoffset}. While some analysis variants do give a result for $A_\star$ that differs from our nominal result by $>0.2$~Gyr, this is to be expected given we consider $\gg 3$ analysis variants, 1/3 of which should differ from the nominal result by more than its $1\sigma$ systematic uncertainty assuming an idealised Gaussian distribution.

\subsubsection{LAMOST stellar parameters}
\label{LAMOST_stellar_parameters}

The ages used here are based on LAMOST spectroscopy combined with \emph{Gaia} astrometry from eDR3. Systematics with the parallax zero-point are by now fairly well understood for the fairly bright LAMOST stars at $D_\odot \la 2.5$~kpc relevant to our analysis (Figure~\ref{d_distribution}). The ages rely on stellar parameters derived by LAMOST, which appear to be very consistent with independent estimates using spectra with similar resolution obtained as part of the MaStar survey \citep{Yan_2019}. In particular, offsets in metallicity, surface gravity, and effective temperature are quite small, more so than for other surveys \citep{Chen_2020, Hill_2022}. This gives us little reason to suspect appreciable systematic errors with the stellar parameters derived from LAMOST~DR7 low-resolution spectra with SNR $\geq 20$ and \emph{Gaia}~eDR3 parallaxes precise to within 10\%. Our quality cuts can be expected to further reduce the potential for systematic errors in the parameters.

\subsubsection{Initial helium abundance}
\label{Initial_helium_abundance}

\citet{Tomasetti_2026} suggest in their appendix~B that one of the most important systematic uncertainties arises due to the uncertain initial helium abundance $Y_{\rm i}$, since increasing this by just 0.01 reduces the inferred age by $\approx 0.75$~Gyr due to the higher density impacting stellar evolution \citep*{Lebreton_2014_accuracy}. LAMOST spectra are helpful in estimating metallicity and [$\alpha$/Fe], but they cannot measure helium in old subgiant stars as $T_\mathrm{eff}$ is too low to excite helium lines. While the primordial helium abundance $Y_{\rm p}$ is known quite precisely independently of cosmology \citep{Aver_2026}, $Y_{\rm i}$ would be slightly higher due to enrichment by earlier generations of stars. This is correlated with metallicity enrichment, which therefore puts constraints on $\Delta Y \equiv Y_{\rm p} - Y_{\rm i}$, usually via the following assumption:
\begin{eqnarray}
    Y_{\rm i} ~=~ Y_{\rm p} + \alphaY Z \, ,
    \label{Yi_estimate}
\end{eqnarray}
where $Z$ is the absolute metallicity, or the mass fraction in nuclei heavier than helium. The slope $\alphaY \equiv \Delta Y/\Delta Z$ is constrained through diverse observational probes, ranging from Solar models to nebular emission lines in extragalactic H~II regions, where intense ionising radiation from massive stars excites observable helium transitions \citep{Nsamba_2021}. Their introduction shows that values roughly covering the range 1--2 have been reported in the literature. We therefore need to consider how much age uncertainties might be inflated once we consider plausible variations in $\alphaY$.

A crucial consideration here is that our age-dependent metallicity ceiling implies stars with $A \approx A_\star$ have [Fe/H] $\la -1$ (Figure~\ref{age_feh_figure}). Given that the stars most crucial to finding $A_\star$ are $\alpha$-enriched (Figure~\ref{age_alpha_figure}) and that the modern estimate of $Z_\odot = 1.39\%$ \citep{Asplund_2021}, we can estimate that these stars have $Z \la 0.2\%$. Even a conservative uncertainty in $\alphaY$ of $\pm 1$ implies an uncertainty in $Y_{\rm i}$ of $\la 0.002$, which then translates to $\la 0.15$~Gyr in the age \citep[see appendix~B of][]{Tomasetti_2026}. This is much lower than their estimate of 0.5~Gyr because those authors considered stars which nearly all have [Fe/H] $\ga -0.5$ (see the blue dots on their figure~1). Uncertainty in $Y_{\rm i}$ is expected to be particularly small for our early chemistry analysis given its explicit focus on metal-poor stars with $[\mathrm{Fe}/\mathrm{H}] < -0.9$ (Section~\ref{Early_chemistry_analysis}). This gives almost exactly the same $A_\star$ as our nominal analysis, where uncertainty in $Y_{\rm i}$ is presumably more important.

While helium spectral lines are not observable in the cool stars relevant to our analysis, $Y_{\rm i}$ impacts the density of a star and thus the frequencies of its normal modes, allowing asteroseismic observations to break the degeneracy between age and $Y_{\rm i}$ \citep*{Lebreton_2014_asteroseismology}. In this context, we note that \citet{Lundkvist_2025} report a high asteroseismic age of $14.2 \pm 0.4$~Gyr for the Methuselah star despite using a free $Y_{\rm i}$ and mixing length parameter $\alphaML$, which describes sub-grid convective processes \citep{Joyce_2023}. The good agreement with our inferred $A_\star = 13.80^{+0.20}_{-0.16}$~Gyr supports a value of $A_\star$ larger than $\AstarETS$, though HD~140283 carries comparable stellar-model systematics of its own \citep{Li_2020_age, Cunha_2021}. A cosmic age close to $\AUCMB$ is further supported by several GSE stars with asteroseismic ages $>13$~Gyr \citep{Montalban_2021}, as illustrated in Figure~\ref{GSE_age_Z_overlay}. These considerations reinforce our conclusion that uncertainty in $Y_{\rm i}$ is small for the metal-poor stars most relevant to our analysis because these stars formed out of material that was only slightly enriched by previous generations of stars, so we expect that $Y_{\rm p} \approx Y_{\rm i}$.

\subsubsection{Stellar modelling uncertainties}
\label{Stellar_modelling_uncertainties}

Inferring stellar ages ultimately requires a model of stellar evolution. This is calibrated using our detailed understanding of the Sun, but the metal-poor subgiants relevant to our study can potentially have non-Solar values for parameters relevant to the age, some of which are difficult to directly constrain. Since these systematic differences would be common to the stars in a wide binary, we cannot constrain them using reported age differences between the presumably co-eval stars in a wide binary \citep{Shariat_2026}. One uncertainty is $\alphaML$, but its impact on ages is likely $\ll 1$~Gyr for old metal-poor stars near the main sequence turnoff region, where $\alphaML$ mainly affects the inferred temperature rather than the age \citep[see appendix~B of][]{Tomasetti_2026}. The \citetalias{Xiang_2022} results use isochrones that assume $\alphaML = 1.743$ \citep[table~1 of][]{Yi_2001}, which is similar to a more modern estimate of 1.82 \citep[table~2 of][]{Choi_2016}. While these values are based on Solar observations, we can infer $\alphaML$ using asteroseismology for the Methuselah star, which is just the kind of metal-poor subgiant most relevant to finding $A_\star$ \citep{Lundkvist_2025}. Their results indicate that $\alphaML = 1.802^{+0.053}_{-0.065}$, which is quite similar to the Solar value despite not assuming this a priori (their table~3 indicates a prior range of 1.7--1.9).

\citet{Tomasetti_2026} also mention in their appendix~B that the impact of $\alphaML$ can be mitigated compared to the 1~Gyr estimated in \citet{Joyce_2023} using accurate measurements of the effective temperature, chemistry, and surface gravity. All of these are likely less accurate in their case than in ours because they focused on microlensed Galactic bulge stars \citep{Bensby_2017}, which moreover are too distant and faint to get percent-level \emph{Gaia} parallaxes. Given the \citetalias{Xiang_2022} stars underpinning our analysis are more comparable to the nearby field subgiants analysed by \citet{Tomasetti_2026} and those authors state in their appendix~B that uncertainty in $\alphaML$ induces age errors $\ll 1$~Gyr, it is unlikely to be an important issue in our case either -- especially given recent asteroseismic evidence that even a metal-poor subgiant has a nearly Solar $\alphaML$ \citep{Lundkvist_2025}. We therefore do not expect uncertainties in $\alphaML$ to play a significant role compared to the other uncertainties.

Another aspect of stars that is somewhat difficult to model is atomic diffusion, the process by which microscopic diffusion causes elements to move relative to one another. Gravitational settling generally depletes the surface of many heavy elements, which sink towards the core. This changes the core composition and stellar structure, altering the main-sequence lifetime and hence the inferred age for reasons similar to the impact of helium \citep{Dotter_2017}. There are strong empirical constraints on gravitational settling besides the known properties of the Sun. As a star evolves off the main sequence, its convective envelope deepens and mixes material that has been affected by diffusion back toward the surface. We can constrain this empirically by comparing stars in the same GC at different evolutionary stages assuming identical initial compositions, based on the fact that subgiants would have dredged up material while main sequence stars would not \citep{Gruyters_2013}. Another constraint comes from the Spite plateau, the approximately constant lithium abundance observed in warm, metal-poor stars over a broad range of [Fe/H] \citep{Spite_1982a, Spite_1982b}. The observed plateau lies well below the $\Lambda$CDM prediction for the primordial abundance, a discrepancy which has come to be known as the lithium problem \citep[for reviews, see][]{Fields_2011, Charbonnel_2026}. A leading interpretation is that at least part of this discrepancy is due to lithium depletion within stars \citep{Korn_2006, Gruyters_2013}, which is supported by a ``meltdown'' of the Spite plateau at very low [Fe/H] \citep{Mucciarelli_2022, Norris_2023}. Models can reproduce a lithium plateau substantially below the primordial abundance when atomic diffusion is moderated by additional transport processes. In particular, unmoderated diffusion would produce stronger and more strongly parameter-dependent depletion of surface lithium than is observed. This requires the effects of gravitational settling to be modified by processes like turbulent mixing and possibly radiative levitation. Different stellar evolution codes consequently adopt different prescriptions for diffusion and additional mixing, although these prescriptions are constrained by observations of abundance trends in GCs and the Spite plateau. The remaining uncertainties in the treatment of diffusion and mixing lead to systematic differences in stellar evolutionary tracks and therefore in ages inferred from them.

Our estimate of $A_\star$ relies on the YY isochrones used to obtain stellar ages from the available astrometric and spectroscopic data \citep{Yi_2001, Demarque_2004}. To estimate the impact of the isochrone model, we face a difficulty in that until recently, YY isochrones were one of the few to allow non-Solar [$\alpha$/Fe] ratios. \citetalias{Xiang_2022} showed in their extended data figure~4 that for the earliest stars with [$\alpha$/Fe] $\approx 0.2$, assuming a Solar value would have a much larger impact on the age than the choice of isochrone model, motivating them to use YY. In a direct comparison between a previous study and the YY ages used here, part of the discrepancy would be due to the latter using star-specific [$\alpha$/Fe] ratios, which would lead to an overestimate of the systematic uncertainty due to the isochrone model. However, YY can be forced to assume Solar-scaled elemental abundance ratios ($[\alpha/\mathrm{Fe}] = 0$) to compare with other models like MIST \citep{Choi_2016}. The results of this comparison are presented in extended data figure~4 of \citetalias{Xiang_2022}. At the ages most relevant to $A_\star$, the MIST isochrones yield ages that are consistently about 0.5~Gyr higher than YY. Assuming that this difference persists when star-specific $\alpha$-enhancements are used in the MIST~II isochrones \citep{Dotter_2026}, our results would be even further from $\AstarETS$, strengthening our main claim that $A_\star > \AstarETS$. This would require [$\alpha$/Fe] to have a similar impact on the age in both codes, which we assume for want of a more detailed investigation. We could then adopt the mean of the ages returned by each code, in which case the results would scatter by $\pm 0.25$~Gyr. We therefore conservatively estimate that uncertainties in the isochrone model have a systematic impact of 0.3~Gyr on the inferred $A_\star$, with the present evidence suggesting that it is more likely to increase with a more modern code. This estimate is in line with the 2.9\% theoretical uncertainty reported by \citet{Chaboyer_2002} for the age of a subgiant with known mass and metallicity, primarily due to uncertainties in the opacity.

A low systematic uncertainty is suggested by the wide range of studies in the literature which give an estimated $A_\star$ or $\AGC$ within about this range either side of $\AstarCMB$ (Section~\ref{A_star_inference}). Many of these studies use different isochrone models, with YY actually being one of the less commonly used. Only part of the scatter between different estimates is due to stellar physics modelling uncertainties like the treatment of atomic diffusion and $\alphaML$, both of which are subject to empirical constraints from the Sun and asteroseismic analyses \citep{Li_2020_age}. The high reported asteroseismic age of $14.2 \pm 0.4$~Gyr for the Methuselah star with free $Y_{\rm i}$ and $\alphaML$ \citep{Lundkvist_2025} and the fact that 90 kinematically selected GSE stars also have an asteroseismic age above 13~Gyr but not 14~Gyr \citep[Figure~\ref{GSE_age_Z_overlay}; see also][]{Montalban_2021} all suggest that our estimated $A_\star$ is not greatly susceptible to systematic uncertainties arising from stellar modelling. We note that there is a distinction between how much studies disagree on the age of any individual star versus how much they disagree on the upper limit to the latent age distribution. The latter is far more relevant to our discussion because our statistical analysis is not greatly reliant on any single star (Figure~\ref{Ngta_nominal_result}). Given the small differences in the age of the oldest star or GC returned by diverse approaches to stellar modelling, it is clear that after applying the relevant empirical constraints from the Sun and other stars, modelling uncertainties are $\ll 1$~Gyr and thus quite unlikely to reconcile our estimated $A_\star$ with $\AstarETS$.

\subsubsection{Combining the systematic uncertainties}
\label{Combining_systematic_errors}

\begin{table}
    \centering
    \begin{tabular}{lc}
        \hline
        Systematic uncertainty & Estimated impact on $A_\star$ \\ \hline
        Quality cut choices (Section~\ref{Quality_cuts}) & 200 Myr \\
        Initial helium abundance (Section~\ref{Initial_helium_abundance}) & 150 Myr \\
        Stellar isochrone model (Section~\ref{Stellar_modelling_uncertainties}) & 300 Myr \\ \hline
        \textbf{Total} & \textbf{390 Myr} \\ \hline
    \end{tabular}
    \caption{Our estimate of how various systematic uncertainties affect the inferred $A_\star$. The final row combines the contributions in quadrature.}
    \label{Astar_sys_budget}
\end{table}

Table~\ref{Astar_sys_budget} summarises the main systematic uncertainties that might appreciably impact our estimate of $A_\star$, as discussed in this section. Since these are qualitatively different sources of uncertainty, we combine them in quadrature to obtain a total systematic error budget of $\la 0.4$~Gyr. This is double the uncertainty of our nominal $A_\star$ inference, suggesting that for the time being, it is more important to focus on reducing systematic uncertainties than on increasing the sample size (Section~\ref{Further_improvements}).

\subsection{Cosmological implications}
\label{Cosmological_implications}

Our measurement of $A_\star$ is fully consistent with $\AstarCMB$ but lies well above $\AstarETS$. This substantial age tension casts doubt on mechanisms that attempt to resolve the Hubble tension by increasing $H(z)$ over a broad range of redshifts. This is because the cosmic time elapsed since a given redshift scales inversely with the expansion rate, causing a reduction in $\AU$ from enhancing $H(z)$ uniformly by 9\% above $\HCMBz$ back to the infancy of the universe. This compression of the cosmic timeline leaves an inadequate window for the first generation of long-lived stars to form, since a younger universe cannot accommodate our high inferred $A_\star$, a constraint that remains fundamentally independent of cosmology -- though it is dependent on stellar models and assumptions about the mixing length and unobserved helium abundance (Section~\ref{Estimated_systematic_uncertainties}). This result constrains the viable theoretical landscape of solutions to the Hubble tension, which should not be resolved by modifications to the expansion history that persist out to high redshift. In particular, it is unlikely that the tension can be solved solely through new physics prior to recombination, as also argued by several other studies using other arguments \citep{Vagnozzi_2022, Vagnozzi_2023, Poulin_2023, Banik_2025_cosmology, Calabrese_2025, Giovanetti_2026, Poulin_2026, SPT_2026}.

Apart from $A_\star$, various other lines of evidence also suggest a late Universe solution to the Hubble tension \citep*{Lin_2021_UCS, Perivolaropoulos_2024, Banik_2025_cosmology, Pantos_2026}. For instance, $\gamma$-ray attenuation due to scattering by extragalactic background light (EBL) yields an $H_0$ estimate of $62 \pm 4$~km/s/Mpc once $\OmegaM$ is fixed to a plausible value \citep{Dominguez_2019, Dominguez_2024}. We expect such an analysis to yield $H_0 \approx \H0CMB$ if the expansion rate only recently started deviating from $\HCMBz$. This is because long path lengths are required for significant attenuation of $\gamma$-rays, reducing their sensitivity to very low $z$. A similar argument can be applied to CCs, which require significant timespans to measure the gradient of the relation between redshift and lookback time \citep{Jimenez_2002, Moresco_2018, Moresco_2020, Moresco_2024}. CC datasets do generally line up very well with $\HCMBz$, not the 9\% higher value that might be expected in an early-time solution \citep{Cogato_2024, Guo_2025, Niu_2026}. Compared to the absolute ages of stars central to our work, CC measurements rely only on relative stellar ages, thereby providing an important consistency check. Recent work suggests that CC uncertainties have been overestimated \citep*{Kvint_2025}, strengthening the argument against early-time solutions.

Besides these late Universe arguments, the excellent performance of $\Lambda$CDM in fitting the CMB anisotropies by now limits the scope for new physics at early times \citep{Calabrese_2025, SPT_2026}. Looking back to even earlier times, the large-scale matter power spectrum has a characteristic turnover scale corresponding to modes that entered the cosmic horizon at matter--radiation equality \citep{Meszaros_1974}. The location of this feature also suggests that $H_0 = H_0^{\mathrm{CMB}}$ \citep{Zaborowski_2025}. While new physics prior to recombination could in principle bias this result, it seems unlikely that the new physics would have the same impact on the turnover scale as on $H_0$ inferred from the CMB, when the universe was $3\times$ larger and $7\times$ older than at matter--radiation equality \citep{Banik_2025_cosmology}. This is especially true in models which restrict the new physics to a narrow time window shortly prior to recombination, since the impact would then be minimal around the time of matter--radiation equality. Early-time solutions along these lines also face a more generic problem related to the early integrated Sachs-Wolfe effect \citep{Vagnozzi_2021} and primordial light element abundances \citep{Giovanetti_2026, Poulin_2026}.

Instead of modifying $a(t)$ at early times, we could do so at late times. This should have little impact on the CMB if the angular diameter distance to recombination is not modified. It should also have little impact on $\AU$. This was shown explicitly by \citet{Najera_2026} for their considered $f(Q)$ modified gravity models, where $\AU \ga 13.6$~Gyr despite solving the Hubble tension \citep[for further discussion of these models, see][]{Yang_2024, Kavya_2025}. Since late-time solutions by definition only affect $\HCMBz$ over a small portion of cosmic history, it is difficult to distinguish between them using presently available $A_\star$ measurements.

The above solutions all assume that $\HCMBz$ must be modified at some redshifts, but this is not necessarily the case. We could try to invoke non-standard physics at low redshifts that cause observers to underestimate distances. However, proposals involving a sharp change in the gravitational constant in the recent past are in severe conflict with terrestrial and Solar System constraints \citep*{Banik_2025_GSM}. Adding a fifth force that is differentially screened in the distance ladder is more promising \citep*{Desmond_2019, Desmond_2020}, although some such models have been shown not to offer a compelling solution \citep{Hogas_2026}.

If distances cannot be modified at low redshifts, the problem might lie with the redshifts themselves. Since these are spectroscopically determined, the observations themselves are not the issue. However, spectra do not directly tell us the relative contributions from peculiar velocities and from cosmic expansion over the light travel time. Significant outward peculiar velocities could potentially solve the Hubble tension, requiring us to be near the centre of a low-density region or void \citep{Haslbauer_2020}. We refer the reader to \citet{Banik_2026_void} for a review of this approach.

\subsection{Further improvements}
\label{Further_improvements}

While our $A_\star$ determination cannot presently distinguish a local void from background solutions with purely late-time modifications to $\HCMBz$, it does cast serious doubt on attempts to solve the Hubble tension entirely through new physics prior to recombination. Our results could in principle be made more accurate, but for now there are five comparable sources of uncertainty, so advances would be needed on all five fronts:
\begin{enumerate}
    \item With a fixed sample and analysis method, $A_\star$ still has an uncertainty of $\approx 0.2$~Gyr. This could be improved through larger samples. A relatively easy improvement would be to repeat the \citetalias{Xiang_2022} determinations of $\mathcal{L}(A)$ on a higher resolution grid. The present 140-knot system has a knot spacing of 0.2~Gyr for $A > 8$~Gyr, which limits the resolution of our $P(A)$ reconstruction. If a 200-knot system were used with a fixed 0.1~Gyr spacing throughout, that would simplify the analysis and perhaps yield tighter constraints.
    \item Varying the quality cuts also impacts our inferred $A_\star$ (Table~\ref{Main_results_table}). Improved understanding of what causes some age determinations to go wrong could lead to a better sample quality and reduced reliance on population-level quality cuts. For instance, binary stars could be detected by repeated observations of the same star, to check for RV drift \citep*{Manchanda_2023}. This is particularly promising in light of the much longer \emph{Gaia} and LAMOST observing baselines in subsequent data releases, which should make it far more likely that we can detect any astrometric or RV acceleration. Genuine binary motion is also easier to detect with improved understanding of \emph{Gaia} systematics and improved spectral resolution in subsequent LAMOST data releases, which already contain a significant number of medium-resolution spectra at $R = 7500$ \citep{Liu_2020, Wang_2022}. Higher spectral resolution makes it easier to notice the extra line broadening resulting from the two stars in a binary having a different RV. It also mitigates line blending issues, which we suspect caused some of the unusually high and precise ages in our initial sample given the poorer precision at lower ages (Figure~\ref{Age_error_turnover_none}). These improvements may well substantially reduce the incidence of stars with overestimated ages, making the inferred $A_\star$ less sensitive to population-level quality cuts. We note that since our sample mostly lies within 2~kpc (Figure~\ref{d_distribution}), it is sufficient to have accurate observations out to this distance, especially as higher quality data should increase the proportion of stars passing our quality cuts. Our consideration of systematic uncertainties also suggests that a smaller sample with less scope for systematic errors could quite plausibly yield a more robust $A_\star$ inference than what we obtain here (Table~\ref{Astar_sys_budget}).
    \item The \textsc{yrec} stellar models \citep{Pinsonneault_2026} that underpin our results require some parameters that are difficult to directly constrain, with uncertainty in $\alphaY$ perhaps contributing to a systematic age uncertainty of $\pm 0.15$~Gyr (Section~\ref{Initial_helium_abundance}). Improved understanding of metal and helium enrichment early in Galactic history could allow a generalisation of Equation~\ref{Yi_estimate} in which $Y_{\rm i}$ depends on both [Fe/H] and [$\alpha$/Fe]. The latter gives some idea of the relative importance of SNe~Ia. Given our understanding of element yields from different supernova types, it should then be possible to work out a more refined estimate of $Y_{\rm i}$ for each star. It is also possible to calibrate $Y_{\rm i}$ and $\alphaML$ using asteroseismic observations, especially for the metal-poor subgiants most relevant to determining $A_\star$, where Solar assumptions may not hold \citep[though see][]{Lundkvist_2025}. Many of the improvements in understanding stellar evolution over the last quarter century can be incorporated simply by switching from YY isochrones \citep{Yi_2001, Demarque_2004} to MIST~II \citep{Dotter_2026}. This has models available at [$\alpha$/Fe] values between $-0.2$ and $+0.6$ inclusive in steps of 0.2, removing the main motivation for using the older YY isochrones.
    \item Our estimates of $A_\star$ can only be related to $\AU$ by adding $\tf$ (Equation~\ref{tf_allowance}). While there is only very limited scope to reduce $\tf$ below our assumed 0.2~Gyr, a better understanding of star formation timescales in the early Universe would help to reduce the uncertainty in $\tf$.
    \item Even if $\AU$ is determined precisely, it is useful only once we compare it to the prediction of a cosmological model. Though there is negligible uncertainty in $\AUCMB$, this is not true for $\AUETS$, whose uncertainty is comparable to our uncertainty on $A_\star$ (Equation~\ref{AU_analytic}). Better observations in the low-redshift Universe are needed to better constrain the cosmological parameters without relying on assumptions about the physics prior to recombination. This is essential not only to refine $\AUETS$, but also to better assess the viability of late-time solutions to the Hubble tension.
\end{enumerate}
The above five uncertainties are all around 0.2~Gyr, making it difficult to substantially tighten our cosmology-independent constraints on $\AU$ with progress on only one of these fronts. It may therefore be some time before different late-Universe solutions can be distinguished using $A_\star$ alone. However, current measurements already create appreciable tension for purely early-time solutions, provided stellar models are as accurate as we have assumed. Indeed, the Methuselah star alone is already in some tension with $\AstarETS$ given its observed asteroseismic age of $14.2 \pm 0.4$~Gyr \citep{Lundkvist_2025}. Their detailed analysis of this ancient metal-poor subgiant only 60~pc away allows a free $Y_{\rm i}$ and $\alphaML$, making it difficult to reconcile with $\AUETS$ -- though of course asteroseismic ages have their own modelling uncertainties \citep{Li_2020_age, Cunha_2021}. Moreover, a recent reanalysis of GCs allowing multiple stellar populations found that $\AGC = 13.61 \pm 0.34$~Gyr \citep{Valcin_2026}. If these results continue to hold, the correct solution to the Hubble tension lies primarily or entirely in new physics that manifests at low redshifts. Better measurements of $A_\star$ and better understanding of $\tf$ would then be important to distinguish between proposals for what that new physics is \citep{Najera_2026}.

\section{Conclusions}
\label{Conclusions}

The age of the Universe is an important constraint on cosmological models if it can be obtained with minimal cosmological assumptions. We achieve this by finding the age of the oldest star, adding an appropriate allowance for the time needed to form the first long-lived stars (Equation~\ref{tf_allowance}). We find $A_\star$ using a large sample of 247,103 Milky Way subgiant stars with LAMOST~DR7 spectra and \emph{Gaia}~eDR3 astrometry, with age likelihoods determined by comparing to YY isochrones up to 20~Gyr \citepalias{Xiang_2022}. The estimated age uncertainties are $\widetilde{\sigma}_A \approx 5-10\%$ (Figure~\ref{Age_error_turnover_nominal}), which appears reasonable based on comparing the presumably co-eval stars in wide binaries \citep{Shariat_2026}. We complement the quality cuts applied by \citetalias{Xiang_2022} with a few additional cuts, our main motivation being to remove outliers from the population distributions of various quantities or pairs of quantities -- even if the star individually appears to have a reliable age likelihood.

The main population-level cuts we introduce are on the age--metallicity and age--[$\alpha$/Fe] relations, since the oldest stars should be metal-poor and $\alpha$-element enhanced (Figures~\ref{age_feh_figure} and \ref{age_alpha_figure}, respectively). While chemical evolution can be faster in the Galactic bulge, our sample does not extend that far (Figure~\ref{d_distribution}); LAMOST in any case focuses on rather high Galactic latitudes due to the northern location of the Xinglong station hosting it \citep{Liu_2020, Yan_2022}. Without the AC cuts, we are left with an unexpected downturn in fractional age uncertainties at the highest ages (Figure~\ref{Age_error_turnover_no_age_chemistry}). We defined the AC cuts internally to our sample prior to detailed statistical analysis. We further justify the AC cuts in Section~\ref{Jz_age_proxy} using the vertical action, identifying a kinematically cold population whose incidence is far higher than in our nominal sample for stars that fail one of our AC cuts and highest for stars that fail both (Table~\ref{jz_cold_tail_table}). Our proxy for a contaminating population of dynamically cold stars jumps precisely upon crossing our nominal age--metallicity ceiling (Figure~\ref{jz_cold_tail_vs_FeHoffset}), with a less clear jump apparent upon going below our age--{$\alpha$/Fe] floor (Figure~\ref{jz_cold_tail_vs_alphaFeoffset}). We find that the AC-rejected stars generally have a lower SNR than the retained stars, suggesting that the underlying reason for some of the age overestimates is simply line blending issues due to the low spectral resolution ($R = 1800$) compounded by a relatively low SNR, even though this is $\geq 20$ in all cases.

Our final quality cut is to cross-check the \citetalias{Xiang_2022} ages with those from FLAME \citep{Pichon_2007} using only \emph{Gaia} data. FLAME ages require various simplifying assumptions about the chemistry \citep{Creevey_2023, Creevey_2026} as \emph{Gaia} photometry barely constrains the metallicity \citep{Fouesneau_2022}. Another concern is that [$\alpha$/Fe] is fixed to the Solar value, even though it should be higher for the oldest stars. This is known to affect the inferred ages \citepalias[see extended data figure~4 of][]{Xiang_2022}, precluding a direct consistency check between their spectroscopic YY ages and FLAME ages. We instead perform $3\sigma$ iterative trendline outlier rejection between YY and FLAME ages, allowing for a growing discrepancy between the two for older stars (Figure~\ref{yy_vs_flame_age_figure}). Our final sample of 146,810 stars appears to be suitably clean given the average $\widetilde{\sigma}_A$ rises continuously with $A$ (Figure~\ref{Age_error_turnover_nominal}) and there is no single star with very high and precise $A$ that might dominate the inference of $A_\star$ (Figure~\ref{Ngta_nominal_result}). The distribution of $\DJtz$ relative to our model also lacks a substantial cold-tail asymmetry indicative of a population with overestimated ages (Section~\ref{Jz_age_proxy}).

We consider a variety of techniques to estimate $A_\star$ from our sample, gradually building up the rigour and complexity. These intuitive methods based on extreme value statistics consistently suggest that $A_\star \approx 13.7$~Gyr, or $\approx 13.4$~Gyr if larger age uncertainties are assumed (Figure~\ref{Ln_P_obs_values}). Our main results use the full age likelihood of each star to perform a non-parametric 100-bin MCMC reconstruction of the population latent age distribution (Figure~\ref{MCMC_PP_recons}). We find that dividing the probability in each age bin by the uncertainty across different MCMC samples leads to a clearly defined sharp feature, which we identify with $A_\star$ (Figure~\ref{P_SN_ratio}). We then run a secondary MCMC analysis to infer $A_\star$.

Our main result is that $A_\star = 13.80^{+0.20}_{-0.16}$~Gyr based on a sample of 146,810 carefully selected Galactic stars with low-resolution ($R = 1800$) LAMOST~DR7 spectra \citepalias{Xiang_2022} with SNR $\geq 20$ and \emph{Gaia} parallaxes $>0.2$~mas accurate to within 10\% \citep{Gaia_2021}. For this sample, $A_\star$ remains similar using somewhat different protocols for how the reconstructed $P(A)$ is smoothed (Table~\ref{Main_results_table}). We also consider alterations to the quality cuts here, but in all analysis variants, $A_\star$ is far closer to $\AstarCMB$ than to $\AstarETS$. This is true even in the extreme case of a 0.2~dex reduction in the age-dependent [Fe/H] ceiling, which considerably reduces the sample size at the oldest ages (see Appendix~\ref{Sample_size_revised_AC_cuts}) but offers no discernible improvement in quality (compare the bottom panel of Figure~\ref{Age_error_turnover_nominal} to the top row of Figure~\ref{Age_error_turnover_variants}). Moreover, the cutoff line now passes through a densely populated ridgeline on the age--metallicity diagram (lower dotted magenta line on Figure~\ref{age_feh_figure}). This is too restrictive given the idea is to exclude low-density parts of the diagram (Figure~\ref{age_feh_intercept_offsets}). Indeed, such a low metallicity ceiling starts to exclude stars from the metal-poor GSE merger and coincides with the mean relation for in-situ stars, leaving no room for stars to have a higher metallicity at given age (Figure~\ref{GSE_age_Z_overlay}). All of our analyses are in at least $2.5\sigma$ tension with $\AstarETS$ (Table~\ref{Main_results_table}). The lowest tension arises if we tighten our age--metallicity ceiling by 0.2~dex, but we argue that this analysis significantly underestimates $A_\star$. Even so, the result is still far above $\AstarETS$. Other analysis variants all yield a higher $A_\star$ with similar precision, leading to even greater tension with $\AstarETS$. This highlights the difficulty of reconciling our results with the low $\AstarETS$.

Our nominal analysis and all our variants to it are reasonably consistent with $\AstarCMB$, apart from the variant lacking a consistency check between YY and FLAME ages. This leads to a downturn in fractional age uncertainty towards higher $A$ (see the bottom right panel of Figure~\ref{Age_error_turnover_variants}), suggesting the sample now includes some stars with problematic age likelihoods. We suggest that this leads to an overestimated $A_\star$, but there is still an internal tension in that some stars have sufficiently high and precise ages to challenge the high estimated $A_\star$ (stars below the green lines in this panel). Leaving aside this variant, the analysis with the highest $A_\star$ arises from raising the [Fe/H] ceiling by 0.2~dex (higher dotted magenta line on Figure~\ref{age_feh_figure}). This result is still consistent with $\AstarCMB$ at $2.1\sigma$ if we adopt a lower $\tf$ of 0.1~Gyr, or at $2.8\sigma$ for our nominal $\tf = 0.2$~Gyr. This sample also has a downturn in $\widetilde{\sigma}_A$ at high $A$, making it likely that $A_\star$ is overestimated (see the middle right panel of Figure~\ref{Age_error_turnover_variants}). The other analysis variants shown here lack this problematic feature, but they are also in $\la 1.5\sigma$ tension with $\AstarCMB$, which is therefore quite plausible according to our analysis despite not being assumed.

Our measurement of $A_\star$ is in agreement with applying a delayed-$\tau$ model for $P(A)$ to our early chemistry sample (Section~\ref{Early_chemistry_analysis}). The $102\times$ smaller sample size compared to our nominal sample is offset by the much simpler parametric $P(A)$, yielding the precise constraint that $A_\star = 13.86^{+0.04}_{-0.05}$~Gyr. Although the uncertainties are almost certainly underestimated given the actual $P(A)$ may differ from the assumed delayed-$\tau$ form (Equation~\ref{P_A_EC}), our non-parametric reconstruction of the latent age distribution peaks near $\AstarETS$, making it quite unlikely that this is the end point of the distribution (Figure~\ref{PP_reconstruction_HR_EC_nominal_20}). The excellent agreement of this early chemistry result with our nominal result of $A_\star = 13.80^{+0.20}_{-0.16}$~Gyr and our other $A_\star$ estimates (Table~\ref{Main_results_table}) provides strong support for our main result. We also note that it is in good agreement with results from previous studies \citep{Montalban_2021, Valcin_2021, Limberg_2022, Cimatti_2023, Souza_2024, Lundkvist_2025, Valcin_2025, Tomasetti_2026, Valcin_2026}, some of which use asteroseismology to reduce dependence on modelling assumptions \citep{Li_2020_age}. The results are systematically higher than $\AstarETS$, yet they generally cluster within a few hundred Myr of $\AstarCMB$.

We find that $A_\star$ is plausibly consistent with $\AUCMB$, but in tension with $\AUETS$ (Equation~\ref{AU_analytic}) -- modulo stellar-model systematics like the initial helium abundance (Section~\ref{Initial_helium_abundance}). Its uncertainty is smaller in our early chemistry analysis due to the low metallicity, so it is reassuring that this sample gives almost the same $A_\star$. Since these ancient stars are less similar to the Sun than our nominal sample as a whole, it is possible that other aspects of the stellar model become less accurate. However, asteroseismic analysis of the Methuselah star indicates that its $\alphaML$ is similar to the Solar value \citep{Lundkvist_2025}. Relaxing the quality cuts or allowing a longer formation time after the Big Bang (Equation~\ref{tf_allowance}) would only exacerbate the tension with $\AUETS$, but there is very little scope to do the opposite. Combining the choice of quality cuts with other systematic uncertainties in our estimate of $A_\star$, we suggest that there is a 0.4~Gyr systematic uncertainty in the final result (Table~\ref{Astar_sys_budget}). This is larger than our formal uncertainty considering only statistical errors in our nominal sample, but too small to reconcile our inferred $A_\star$ with $\AstarETS$.

The high inferred $A_\star$ is very problematic for solutions to the Hubble tension through new physics prior to recombination, an approach which also faces other challenges (Section~\ref{Cosmological_implications}). Moreover, there is growing recognition of a distortion to the shape of $\HCMBz$ at $z \la 1$, not merely a uniform 9\% uplift out to $z \gg 1$ \citep{Jia_2023, Jia_2025a, Jia_2025b, Lopez_2025, Jia_2026}. This is especially apparent with baryon acoustic oscillations, which systematically depart from predictions using $\HCMBz$ at $z \la 1$ \citep{Banik_2025_BAO, DESI_2025}.

Taken together, these results suggest a late Universe solution to the Hubble tension \citep{Lin_2021_UCS, Vagnozzi_2022, Vagnozzi_2023, Banik_2025_cosmology}. If this is through an adjustment to the background expansion history, it would not much affect $\AU$ \citep{Najera_2026}. Another possibility is that the Hubble tension is due to a large local underdensity or void \citep{Haslbauer_2020, Banik_2025_BAO, Mazurenko_2025, Banik_2026_z0}. Outflows from the void would create additional non-cosmological contributions to the redshift, giving the appearance of a high $H_0^\mathrm{local}$ \citep[\citealt{Banik_2026_void}; see also section~4.5.1 of][]{Valentino_2025}. Progress on several fronts would be required for $A_\star$ measurements to distinguish between these scenarios (Section~\ref{Further_improvements}).

Cosmology usually involves looking out to great distances in order to reach lookback times that are a significant fraction of $\AU$. We can also achieve this by studying the oldest Galactic stars, which serve as ancient `fossils' telling us about the history of the Universe. This more closely resembles techniques used in other fields concerned with the past. As cosmologists seek to peer ever further into the depths of space, they would be well advised to constrain their models using the much more detailed observations that are possible in the Solar neighbourhood, taking advantage of the well understood physics of star formation and evolution.

\section*{Acknowledgements}

IB and HD are supported by Royal Society University Research Fellowship grant 211046. The contribution of TKK was enabled by an undergraduate research bursary from the Royal Astronomical Society. The authors are very grateful to Maosheng Xiang and Hans-Walter Rix for providing the age likelihoods used in this work and to Richard Stiskalek for helping develop the MCMC algorithm used to reconstruct the latent age distribution from these age likelihoods. We thank Claudia Maraston for helpful comments on the quality cuts and Elena Tomasetti for comments on the article which improved the discussion of systematic uncertainties. We are also very grateful to the referee for comments which substantially improved this manuscript. All the triangle plots in this contribution were prepared using \textsc{pygtc} \citep{Bocquet_2016}. 


\section*{Data Availability}

The \citetalias{Xiang_2022} catalogue is publicly available. Age likelihoods and a list of which of their stars are in our nominal sample will be made available upon reasonable request to the lead author.

\bibliographystyle{mnras}
\bibliography{AOU_bbl}

\begin{appendix}

\section{Excluding a duplicate star}
\label{Duplicate_star}

We found two entries in the \citetalias{Xiang_2022} catalogue corresponding to the same star. Both entries list the \emph{Gaia} identifier as 6898921549983327744 and have extremely similar sky coordinates. The star appears to have been observed twice, with derived stellar properties in the two entries consistent within uncertainties. For instance, $T_\mathrm{eff} = 5488 \pm 37$~K in one case and $5509 \pm 9$~K in the other. The large difference in uncertainty arises from the spectral SNR, which is 36.24 for the first entry and 130.19 for the second entry. The spectroscopic identifiers indicate the underlying spectra were taken on 28 September 2012 and 9 October 2012. Since the latter has a much higher SNR, we choose to keep this entry and discard the earlier lower SNR spectrum. This reduces our initial sample to 247,103 stars, which we sort in ascending order of the \emph{Gaia} identifier.

\section{The relation between age and its uncertainty in variants to our main sample}
\label{Age_error_appendix}

\begin{figure*} 
    \centering
    \setlength{\unitlength}{1cm}
    \begin{picture}(0,0)
        \put(2.5, 4.86){\makebox(0,0)[lt]{\textbf{[Fe/H] ceiling lowered 0.2 dex}}}
    \end{picture}
    \includegraphics[width=0.49\linewidth]{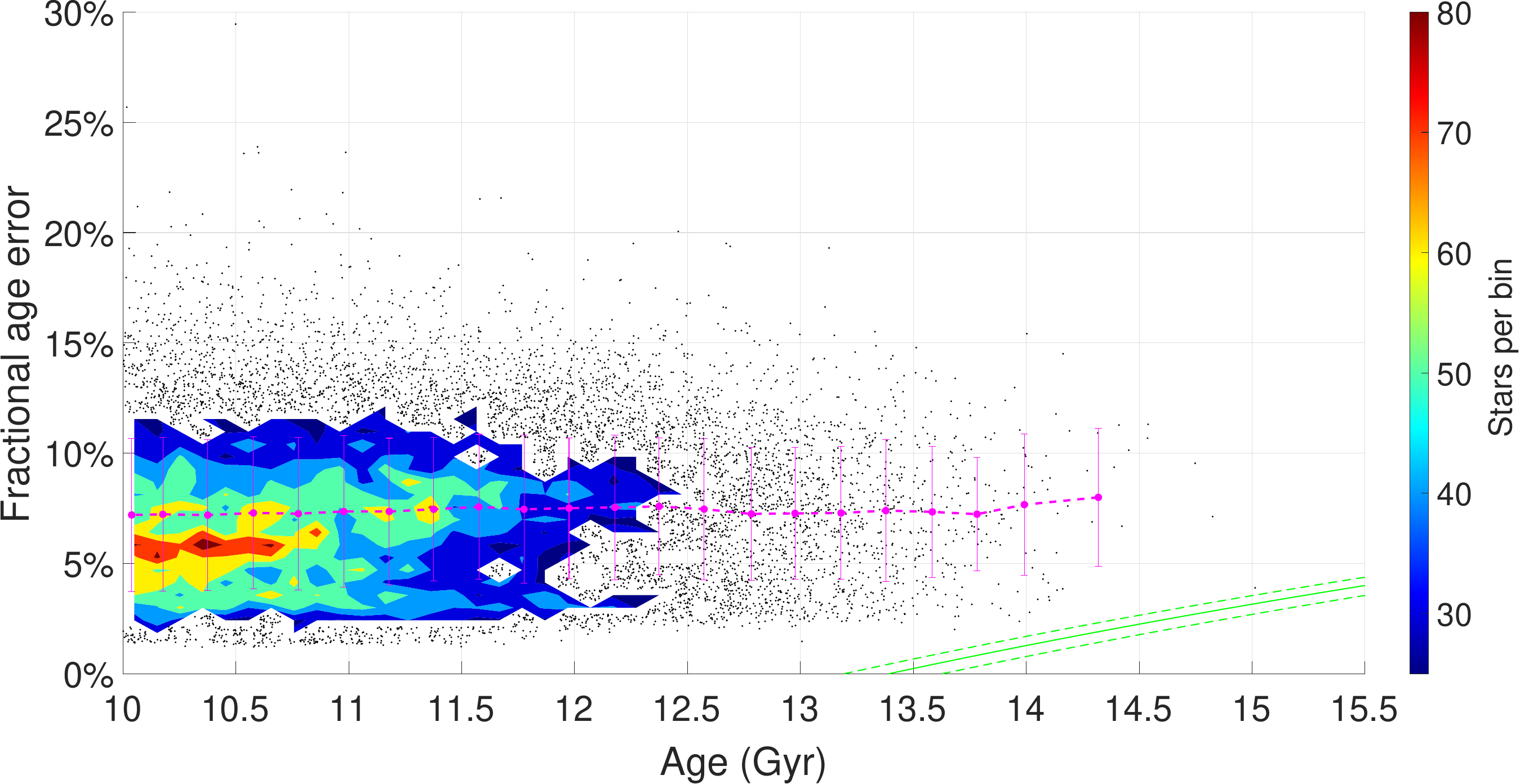} 
    \hfill
    \begin{picture}(0,0)
        \put(2.5, 4.83){\makebox(0,0)[lt]{\textbf{[Fe/H] ceiling lowered 0.1 dex}}}
    \end{picture}
    \includegraphics[width=0.49\linewidth]{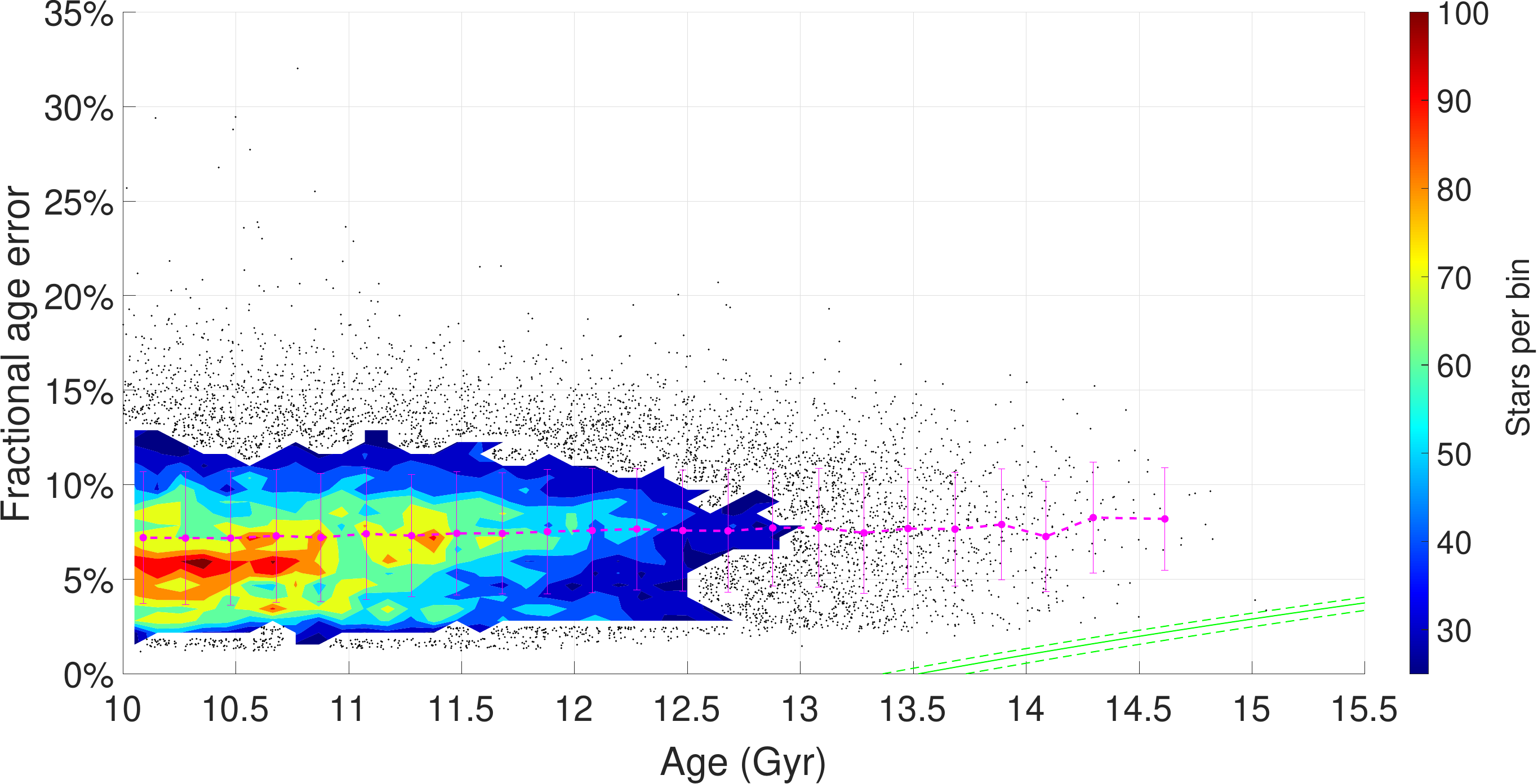}

    \par\vspace{5ex} 

    \begin{picture}(0,0)
        \put(2.7, 4.83){\makebox(0,0)[lt]{\textbf{[Fe/H] ceiling raised 0.1 dex}}}
    \end{picture}
    \includegraphics[width=0.49\linewidth]{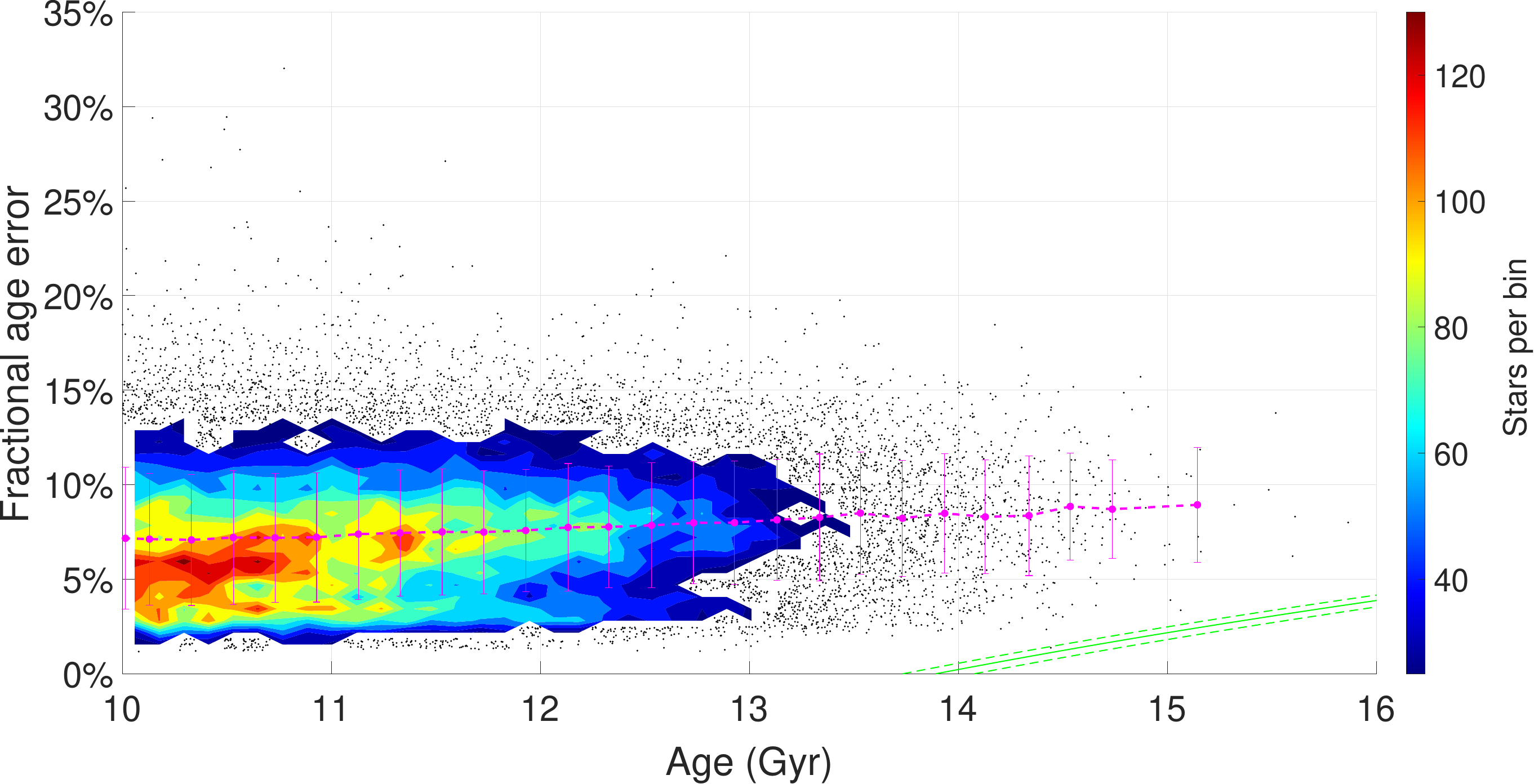}
    \hfill
    \begin{picture}(0,0)
        \put(2.7, 4.83){\makebox(0,0)[lt]{\textbf{[Fe/H] ceiling raised 0.2 dex}}}
    \end{picture}
    \includegraphics[width=0.49\linewidth]{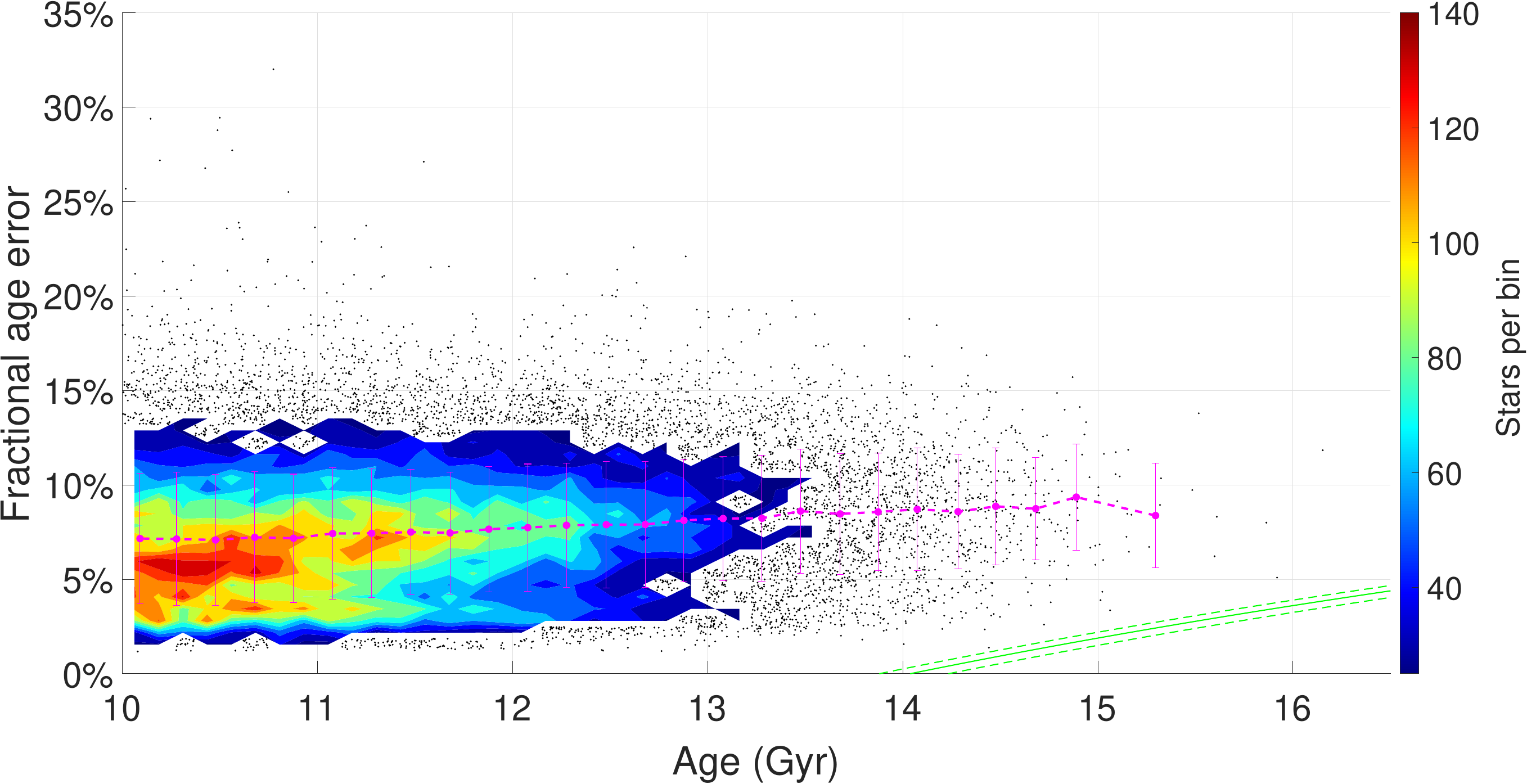}

    \par\vspace{5ex} 

    \begin{picture}(0,0)
        \put(2.6, 4.83){\makebox(0,0)[lt]{\textbf{[$\boldsymbol{\alpha}$/Fe] floor lowered 0.05 dex}}}
    \end{picture}
    \includegraphics[width=0.49\linewidth]{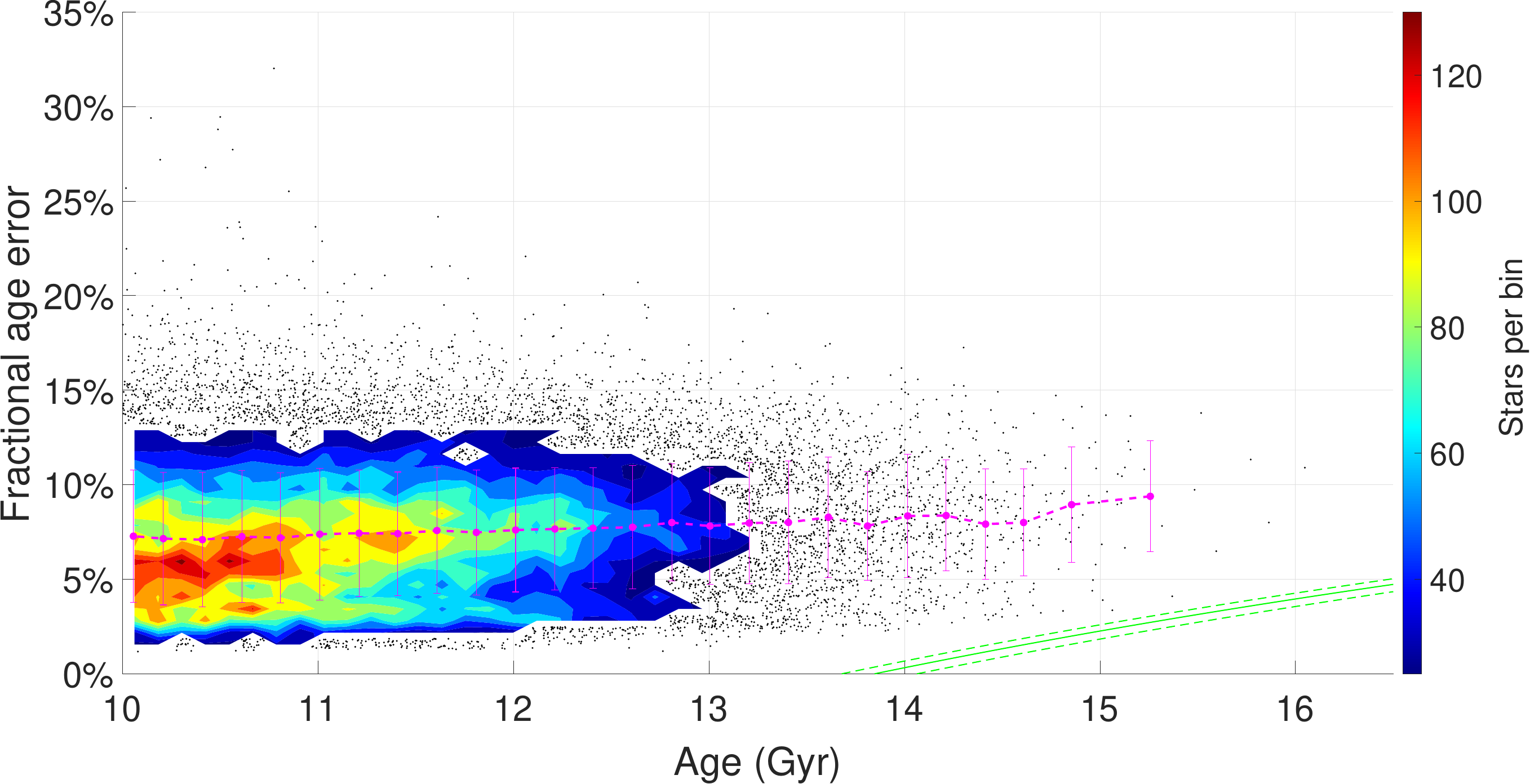}
    \hfill
    \begin{picture}(0,0)
        \put(2.4, 4.83){\makebox(0,0)[lt]{\textbf{No FLAME age consistency check}}}
    \end{picture}
    \includegraphics[width=0.49\linewidth]{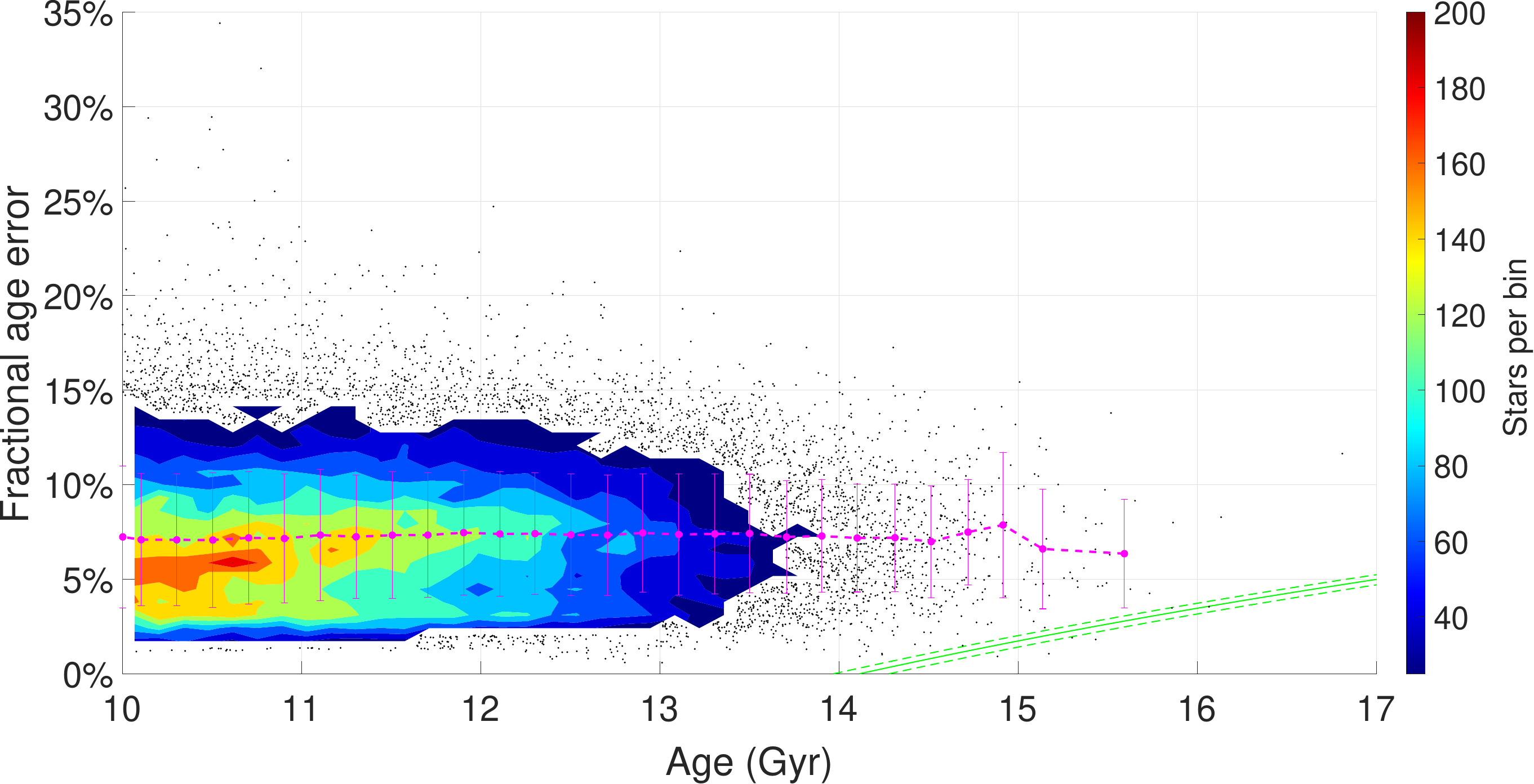}
    \caption{The fractional age uncertainty at different ages, shown similarly to the bottom panel of Figure~\ref{Age_error_turnover_nominal}. Each panel shows a variant to our nominal sample, as described in the title. Parameter inferences with these alternative choices of quality cut are summarised in Table~\ref{Main_results_table}, from which the $A_\star$ value and its $1\sigma$ range are used to show the minimum uncertainty required for consistency with $A_\star$ at $3.4\sigma$ (green lines). Notice that relaxing the age--metallicity quality cut by 0.2~dex and skipping the consistency check between YY and FLAME ages yield samples with a slight downturn in $\widetilde{\sigma}_A$ at the highest ages. This suggests the presence of stars with inaccurately estimated age likelihoods, which can inflate the estimated $A_\star$. The variant without FLAME ages also suffers a consistency problem in that some stars (below the green lines) have sufficiently high and precise ages to create tension with $A_\star$ estimated using the same sample.}
    \label{Age_error_turnover_variants}
\end{figure*}

The full \citetalias{Xiang_2022} sample of 247,103 stars cannot be directly used to find $A_\star$ because of the downturn in $\widetilde{\sigma}_A$ at high ages, which indicates a significant problem with some of the age estimates (Figure~\ref{Age_error_turnover_none}). This problem is absent in our nominal sample (Figure~\ref{Age_error_turnover_nominal}). To check the robustness of our inferred $A_\star$, we considered several variants to our nominal quality cuts and presented the results in Table~\ref{Main_results_table}. Figure~\ref{Age_error_turnover_variants} shows the age-dependence of $\widetilde{\sigma}_A$ in these variants. The nominal quality cuts are made tighter in the top row, while the middle and bottom rows relax them in various ways, as summarised in the title to each panel.

The top panels show that stricter versions of our nominal quality cuts yield a sample that behaves similarly to our nominal sample, with no clear downturn in $\widetilde{\sigma}_A$ at high ages. This is to be expected since our nominal sample appears to be adequately cleaned of stars with problematic age estimates, so this should also be the case with even stricter quality cuts.

There is little sign of a downturn in $\widetilde{\sigma}_A$ at high ages if we relax the nominal age--metallicity ceiling (Equation~\ref{Age_metallicity_cutoff_line}) by 0.1~dex (middle left panel). However, a downturn is apparent if we relax it by 0.2~dex (middle right panel). An even clearer downturn is evident in the bottom right panel, where we drop the requirement for consistency between YY and FLAME ages (Section~\ref{FLAME_comparison}). This panel also exhibits an internal inconsistency in that some stars lie below the green lines, indicating they have a sufficiently high and precise age to be in tension with $A_\star$ inferred from the same sample, despite allowing for up to $3.4\sigma$ random age overestimates given the sample size. These results show that there is little scope to safely relax the age--metallicity and FLAME quality cuts, all of which we specified prior to developing the $P/\sigma$ technique to infer $A_\star$ (Section~\ref{Age_limit_MCMC}).

On the other hand, the bottom left panel shows that $\widetilde{\sigma}_A$ maintains a healthy rise with age if we relax the nominal age--[$\alpha$/Fe] floor (Equation~\ref{Age_alpha_Fe_cutoff_line}) by 0.05~dex. This sample may well be as clean as our nominal sample (as also suggested by Figure~\ref{jz_cold_tail_vs_alphaFeoffset}), perhaps explaining why both yield a very similar $A_\star$ (Table~\ref{Main_results_table}).

\section{The sample size with revised AC cuts}
\label{Sample_size_revised_AC_cuts}

The AC cuts can have a significant impact on the number of stars in our sample, especially at the high ages most relevant to determining $A_\star$. The top panel of Figure~\ref{Sample_size_vs_AC_offsets} shows the sample size with variations to our nominal age-dependent metallicity ceiling, highlighting the severe loss of stars older than 14~Gyr if we lower the ceiling by 0.1 or especially 0.2~dex. The bottom panel of Figure~\ref{Sample_size_vs_AC_offsets} shows the sample size with an altered floor on [$\alpha$/Fe]. There is a very steep loss of stars older than 14~Gyr if we raise the floor by 0.025 or especially 0.05~dex, explaining why we do not consider these variants. However, the sample size is quite stable if we lower the [$\alpha$/Fe] floor, even for the $A > 14$~Gyr case where the sample size is most sensitive to the adopted AC cuts. This makes it more useful to explore the impact of relaxing the nominal [$\alpha$/Fe] floor, which we do in one analysis variant.

\begin{figure}
    \centering
    \includegraphics[width=\linewidth]{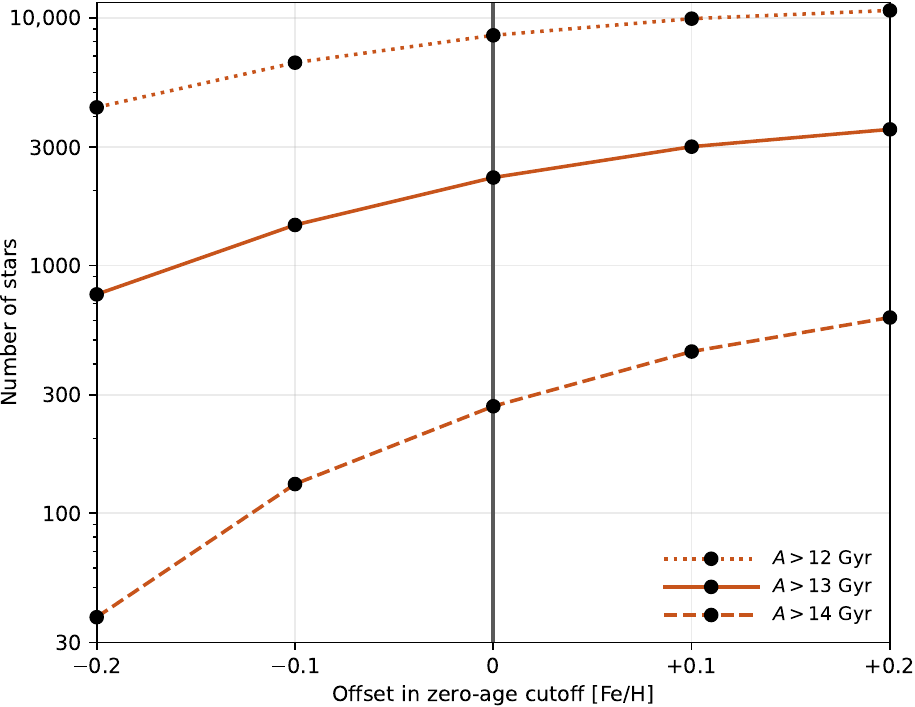} 
    \vspace{4pt}
    \includegraphics[width=\linewidth]{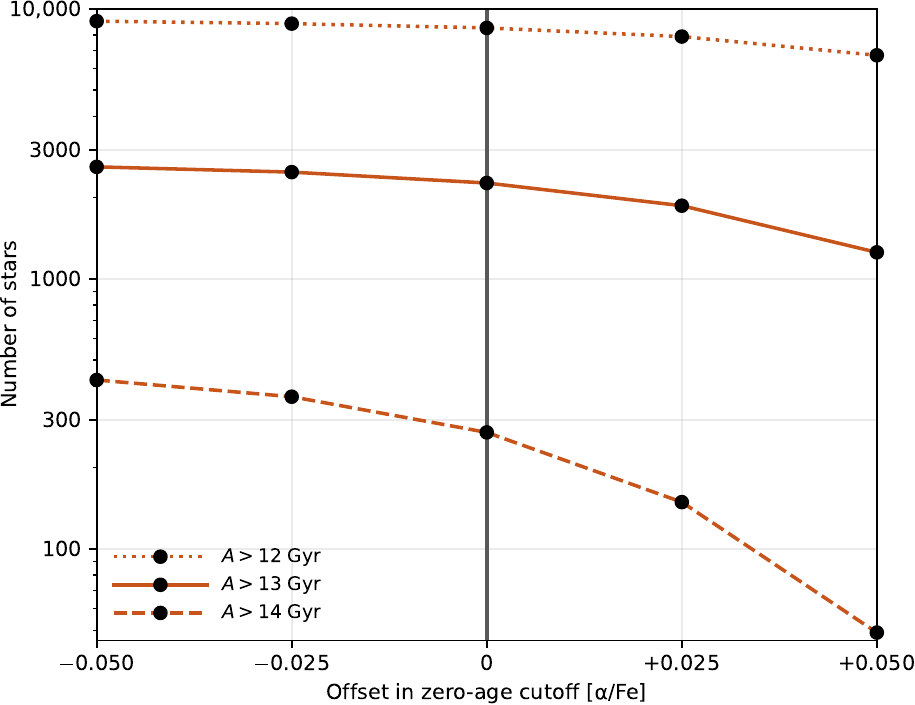}
    \caption{The sample size with a shifted age-dependent [Fe/H] ceiling (Equation~\ref{Age_metallicity_cutoff_line}; \emph{top}) and [$\alpha$/Fe] floor (Equation~\ref{Age_alpha_Fe_cutoff_line}; \emph{bottom}). The dotted, solid, and dashed lines show results for stars measured to be older than 12, 13, and 14~Gyr, respectively. In both panels, the nominal quality cut position is indicated using a thick grey vertical line. Notice the sharp drop in sample size at $A > 14$~Gyr if we substantially tighten either AC quality cut.}
    \label{Sample_size_vs_AC_offsets}
\end{figure}

\section{The vertical action model and its dimensionless residuals}
\label{Vertical_action_model}

In Section~\ref{Jz_age_proxy}, we used the vertical action $J_z$ as a proxy for the age of a star. We use this appendix to present additional related results.

\subsection{Vertical heating trends}
\label{Vertical_heating_trends}


Due to the inevitable differences in stellar parameters between the retained and AC-rejected samples, we fit a model for $J_z$ in terms of the most important parameters on which it depends (Equation~\ref{Jtz_relation}). Since these parameters are themselves correlated (e.g. older stars are typically more metal-poor), we had to jointly fit all three $F_i$, where $i$ labels a parameter relevant to $J_z$. To better illustrate each of the $F_i$, we make use of partial residuals $R_i$. These are defined as the residuals between the data and the model excluding the $F_i$ contribution. For instance, the age residuals $R_A$ are defined as follows:
\begin{eqnarray}
    R_A ~\equiv~ \Jtz - \Jtzo - F_{\mathcal{M}} \left( \mathcal{M} \right) - F_D \left( D_\odot \right),
    \label{partial_residual}
\end{eqnarray}
with $\Jtzo$, $F_{\mathcal{M}}$, and $F_D$ jointly fit to the data. The idea is that without including an explicit age dependence in $\Jtz$, the residuals relative to this restricted model will show strong age-dependent residuals. $R_{\mathcal{M}}$ and $R_D$ are defined analogously, with every parameter appearing exactly once in the corresponding versions of Equation~\ref{partial_residual}. Plotting $R_i$ against parameter $i$ shows the residual dependence that must be captured by $F_i$ in order to have a good model for $\Jtz$. We can then plot our assumed functional form and check that it provides a good fit to the partial residuals.

The results are shown in Figure~\ref{jz_partials}, where each panel shows the partial residuals for one parameter. Given the 0.37~dex scatter about the mean relation, we do not show individual stars for clarity. The points instead show a binned median and its standard error. In all cases, $F_i$ provides a good fit to the partial residuals in parameter $i$, with a reduced $\chi^2$ of 1.7, 2.0, and 1.6 for the distance, metallicity, and age residuals, respectively. This provides confidence that our model for $\Jtz$ (Equation~\ref{Jz_terms}) provides an adequate accounting of differences between the retained and AC-rejected samples due purely to the different stellar parameters.

\begin{figure}
    \centering
    \includegraphics[width=\linewidth]{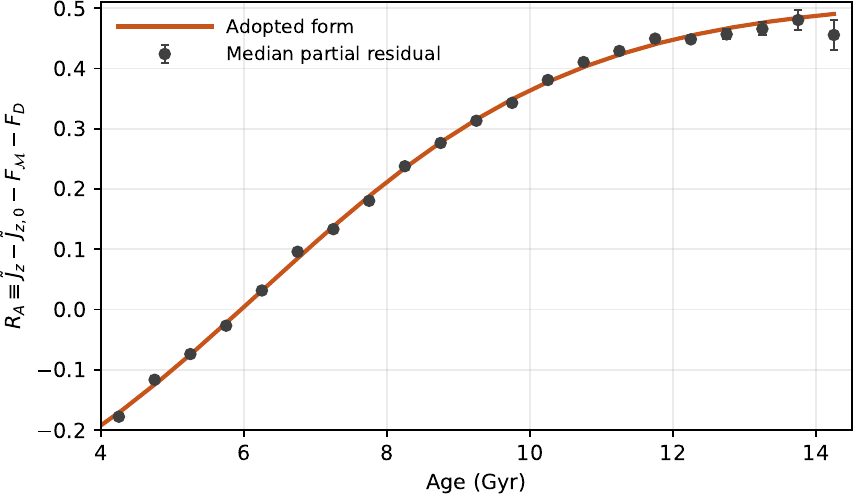} 
    \vspace{4pt}
    \includegraphics[width=\linewidth]{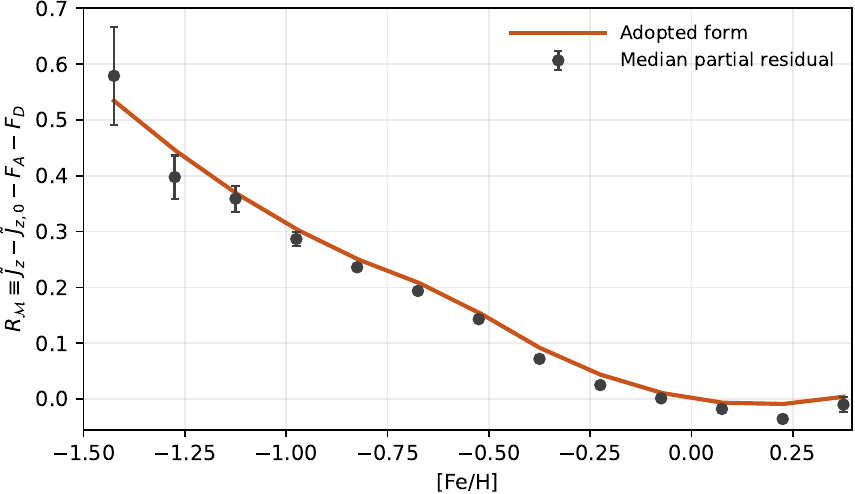}
    \vspace{4pt}
    \includegraphics[width=\linewidth]{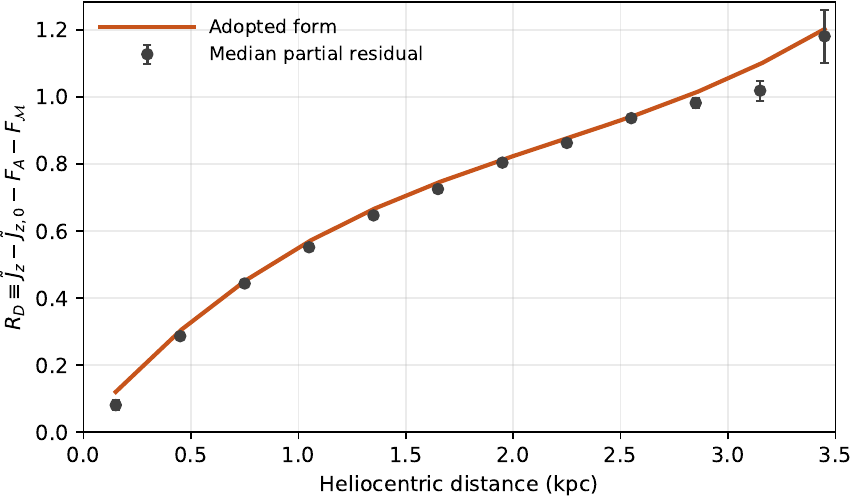}
    \caption{The solid black points show the partial residuals of $\Jtz$ relative to our model (Equation~\ref{partial_residual}) without the dependence on age (\emph{top}), metallicity (\emph{middle}), or distance (\emph{bottom}). The partial residuals are plotted against the removed parameter, providing a target for the term in Equation~\ref{Jtz_relation} that accounts for dependence on this parameter. The solid black points with uncertainties show the median value and its standard error. The solid red line shows our fitted dependence of $\Jtz$ on the plotted parameter. In all cases, we found a good fit to the partial residuals using a simple closed-form expression (Equation~\ref{Jz_terms}).}
    \label{jz_partials}
\end{figure}

The top panel of Figure~\ref{jz_partials} highlights the saturation of disc heating for old stars. This is presumably due to the smaller fraction of their orbits spent close to the Galactic disc, where heating processes take place \citep{Ting_2026}. The shallow gradient at high ages makes it difficult to estimate the age using $\Jtz$. The middle panel shows a knee at $\mathcal{M} \approx -0.6$, preventing us from using a single parabola over the full range. The break may indicate the metallicity separating the thin and thick Galactic discs \citep{Beraldo_2021}. The bottom panel shows that the distance partial residuals rise continuously with distance, though with a slight flattening beyond 1~kpc. The rising behaviour of $R_D$ is most likely a selection effect given the obviously anthropic definition of $D_\odot$. Since the Sun is located only $10 \pm 2$~pc from the Galactic disc mid-plane \citep{Griv_2021} and dynamically hotter stars are on average further from this plane, they are also on average further from the Sun. Our comparison between retained and AC-rejected stars is barely sensitive to this effect because both samples have almost the same distance distribution, so the most important correction is the one for metallicity (Section~\ref{Jz_age_proxy}).

\subsection{The dimensionless cold-tail statistic}
\label{Dimensionless_cold_tail}

In Equation~\ref{cold_tail_definition}, we defined the cold-tail statistic $\Act$ based on the proportion of stars where $\left| \DJtz \right|$ exceeds a certain number of dex. It is possible that the $\Jtz$ model is less accurate at the parameters typical of the AC-rejected stars, or that the intrinsic dispersion and/or measurement error are larger here. To account for these possibilities, we can subtract the mean value $\mu$ and divide by the standard deviation $\sigma$ of the $\DJtz$ distribution. This gives a `dimensionless' version of the cold-tail statistic in which a threshold in dex must be replaced by a threshold number of standard deviations.

To limit the impact of outliers, we use iterative $3\sigma$ outlier rejection as used elsewhere in this work when calculating $\mu$ and $\sigma$ of the $\DJtz$ distribution, running the process until convergence. We then recentre and rescale the $\DJtz$ distribution separately for the retained and AC-rejected stars. We find that $\sigma = 0.3430$~dex for the retained stars and 0.3576~dex for the AC-rejected stars, so the two samples have almost the same width for their $\DJtz$ distribution. The difference in mean values is more significant: the clipped mean is $-0.0066$~dex for the retained stars (indistinguishable from zero), but the AC-rejected stars have a clipped mean of $-0.065$~dex. The recentring therefore removes a relative displacement of $0.17\sigma$.

\begin{figure}
    \centering
    \includegraphics[width=\linewidth]{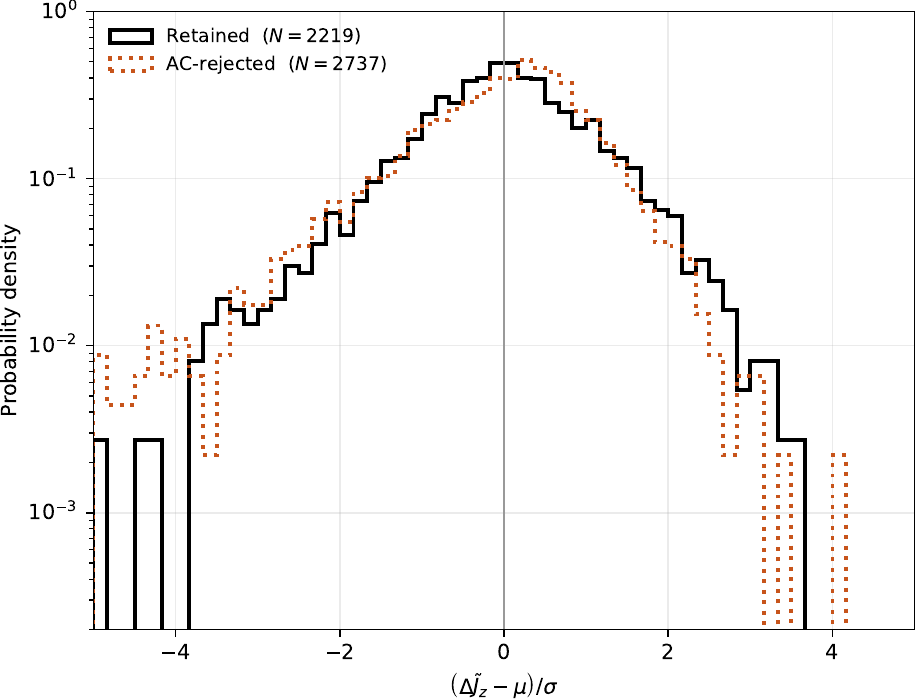} 
    \caption{The distribution of the residual $\DJtz$ relative to our model (Equation~\ref{Jz_terms}) for retained stars (solid black) and AC-rejected stars (dotted red), in both cases only for stars with $A = 13 - 14.5$~Gyr. This is similar to the top panel of Figure~\ref{jz_distributions}, but here the $\DJtz$ distribution is recentred to its mean value and rescaled to its standard deviation, both of which are found using iterative $3\sigma$ outlier rejection. Despite this, a clear difference is apparent at the left between the tails of the two distributions.}
    \label{jz_dimensionless_distributions}
\end{figure}

After the recentring and rescaling described above, the $\DJtz$ distributions of the retained and AC-rejected stars are shown in Figure~\ref{jz_dimensionless_distributions}, which focuses on stars with reported ages of $13-14.5$~Gyr. The dotted red AC-rejected distribution is still skewed towards negative residuals, suggesting that the cold-tail asymmetry remains despite subtracting the mean value of $\DJtz$ and dividing by its dispersion. To quantify this, we note that our previous threshold of 0.7~dex corresponds to $2.04\sigma$, so we evaluate our dimensionless version of $\Act$ using a threshold of $2\sigma$. We find that the retained stars have $\Act = 1.62 \pm 0.59\%$, but the AC-rejected stars have a much higher $\Act = 4.57 \pm 0.56\%$, which is $3.6\sigma$ higher. The difference is somewhat attenuated by the recentring and rescaling, which is to be expected because the impact of a contaminating population with $\DJtz \ll 0$ can be partly absorbed by shifting the mean value and inflating the dispersion. We find that this still leaves a significant difference in $\Act$, highlighting the robustness of this feature.

The asymmetry at a fixed number of dex becomes quite sensitive to small offsets in the mean or changes in the dispersion the further we go into the tail of the distribution. This is because the proportional impact on the number of stars in the tail is enhanced if we define the tail to be e.g. beyond 1~dex rather than 0.5~dex. But since changes in the mean and dispersion are now accounted for in our dimensionless cold-tail statistic, we can safely go further into the tail of the distribution. In what follows, we therefore also present results for a threshold of $2.5\sigma$, better isolating any feature at $\DJtz \ll 0$. We find that there are very few stars beyond $3\sigma$, so we omit these results.

We previously showed that $\Act$ sharply increases as a function of $\DFeH$, the metallicity excess over our nominal age--metallicity cutoff line (Equation~\ref{Age_metallicity_cutoff_line}), precisely upon crossing $\DFeH = 0$ (Figure~\ref{jz_cold_tail_vs_FeHoffset}). We now revisit this result using our dimensionless version of $\Act$. The results are shown in Figure~\ref{jz_dimensionless_cold_tail_vs_FeHoffset}, whose top panel uses a threshold of $2\sigma$. The bottom panel uses instead $2.5\sigma$, where we still expect reliable results for the reasons stated above. As done previously, the $\Jtz$ model used here is trained only on stars with $\DFeH < -0.2$, which we can think of as a control region where we expect the YY ages to be reliable. $\Act$ here is shown by the dashed black horizontal line, with the $1\sigma$ and $2\sigma$ uncertainties shown using shaded grey bands. Focusing on the $2.5\sigma$ threshold to better isolate the tail while retaining good statistics, we find that the control region has $\Act = 1.09 \pm 0.36\%$. This rises slightly to $1.56 \pm 0.53\%$ in the $\DFeH$ bin covering ($-0.2$, $-0.1$) and $1.88 \pm 0.62\%$ in the $\DFeH$ bin covering ($-0.1$, 0). In both cases, the results are not distinguishable from $\Act$ in the control region at $\DFeH < -0.2$, where the low $\Act$ may just reflect a mild intrinsic asymmetry of the $\DJtz$ distribution rather than a contaminating population at $\DJtz \ll 0$. The situation changes drastically upon considering the next equally wide $\DFeH$ bin covering (0, 0.1), where $\Act = 3.84 \pm 0.71\%$. This is higher than the control region by $3.5\sigma$ and the immediately preceding $\DFeH$ bin by $2.1\sigma$. The sharp jump in $\Act$ upon crossing the nominal age--metallicity cutoff line (Figure~\ref{jz_cold_tail_vs_FeHoffset}) persists when recentring and rescaling the $\DJtz$ distribution to its own mean and dispersion.

\begin{figure}
    \centering
    \includegraphics[width=\linewidth]{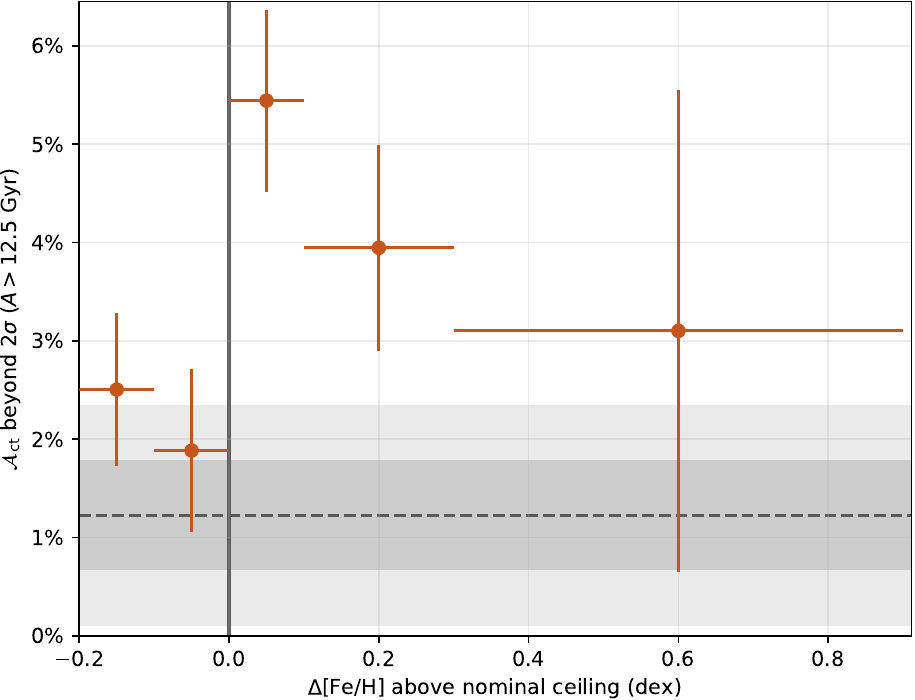} 
    \vspace{4pt}
    \includegraphics[width=\linewidth]{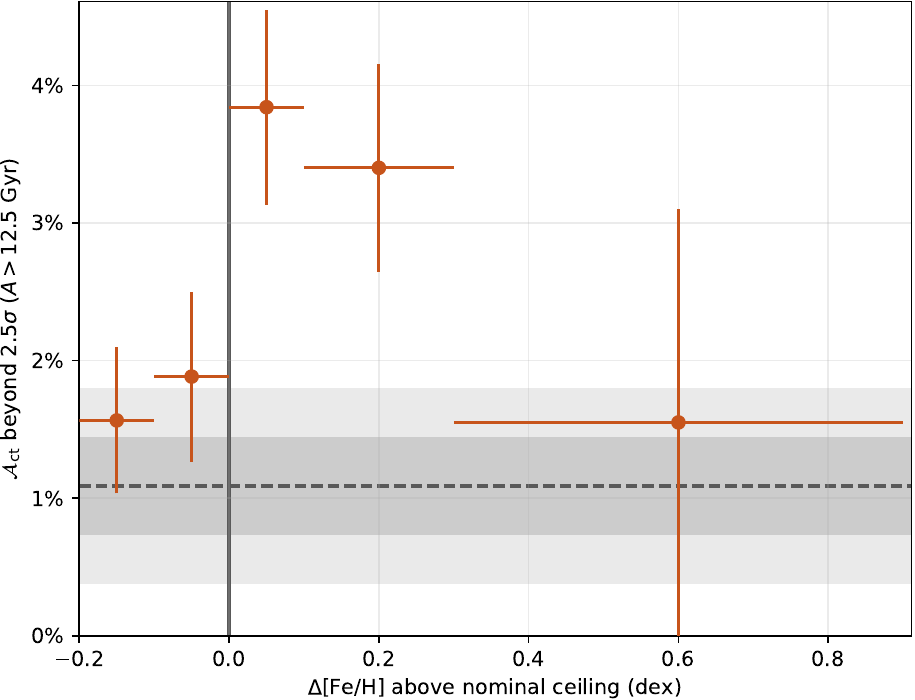}
    \caption{Similar to Figure~\ref{jz_cold_tail_vs_FeHoffset}, but here the $\DJtz$ distribution of stars older than 12.5~Gyr in each age--metallicity strip is recentred to the $3\sigma$-clipped mean of the distribution and rescaled to its standard deviation. The threshold used to calculate this dimensionless version of $\Act$ is $2\sigma$ (\emph{top}) or $2.5\sigma$ (\emph{bottom}). The $\Jtz$ model used here (Equation~\ref{Jtz_relation}) is fit to stars with $\DFeH < -0.2$. The stars in this training sample with $A > 12.5$~Gyr have a cold-tail asymmetry as indicated by the dotted black horizontal line, with grey bands showing the $1\sigma$ and $2\sigma$ Poisson uncertainties. Notice that $\Act$ still has a sharp increase upon crossing the nominal position of the cutoff line (vertical thick grey line), below which results are consistent with the training sample despite no overlap with it.}
    \label{jz_dimensionless_cold_tail_vs_FeHoffset}
\end{figure}

The only appreciable difference with our earlier result is that the final $\DFeH$ bin has $\Act = 1.55 \pm 1.55\%$, whereas our previous result using a 0.7~dex threshold had $\Act = 8.53 \pm 2.80\%$. Since there are only 129 stars in this strip of the age--metallicity diagram, the results are not very accurate. Moreover, the clipped standard deviation of $\DJtz$ for these stars is 0.380~dex against $0.335-0.363$~dex elsewhere, placing the $2.5\sigma$ threshold at $\left| \DJtz \right| = 0.95$~dex rather than the 0.84~dex of the strips with lower $\DFeH$. The higher threshold in dex inevitably reduces $\Act$. We also note that since we expect stars with $\DFeH > 0.3$ to be the most heavily contaminated by a population with overestimated ages, scaling the $\DJtz$ distribution to its own mean and standard deviation inevitably underestimates the extent of any feature at $\DJtz \ll 0$. In other words, the dimensionless version of our cold-tail statistic is conservative by construction. Even so, a sharp jump at $\DFeH = 0$ remains. This provides strong support for our nominal choice of the age--metallicity cutoff line, given also that Figure~\ref{age_feh_figure} provides little flexibility regarding its gradient.

\section{SNR distributions of the nominal and AC-rejected samples}
\label{snr_distributions_appendix}

In Section~\ref{Why_AC_cuts_help}, we argued that the AC-rejected stars typically have a lower SNR than the retained stars, increasing the likelihood of an inaccurate age estimate. Our main result used the ratio of SNR percentiles between the two samples (Equation~\ref{S_tilde_definition}). For completeness, we show the SNR distributions themselves in Figure~\ref{snr_histograms}, with the SNR shown on a logarithmic scale due to its large dynamic range. We use four panels to cover the same age range used in Figure~\ref{snr_ratio_vs_age}. Each panel in Figure~\ref{snr_histograms} shows the AC-rejected stars with a dotted red line, while the retained stars are shown with a solid black (grey) line if they are (are not) weighted to match the $m_\mathrm{G}$ distribution of the AC-rejected stars. Comparing the two solid lines therefore gives an idea of how important the magnitude-weighting is. We indicate the median along the bottom using an arrow in the same style and colour as the distribution.

\begin{figure*}
    \centering
    \includegraphics[width=0.49\linewidth]{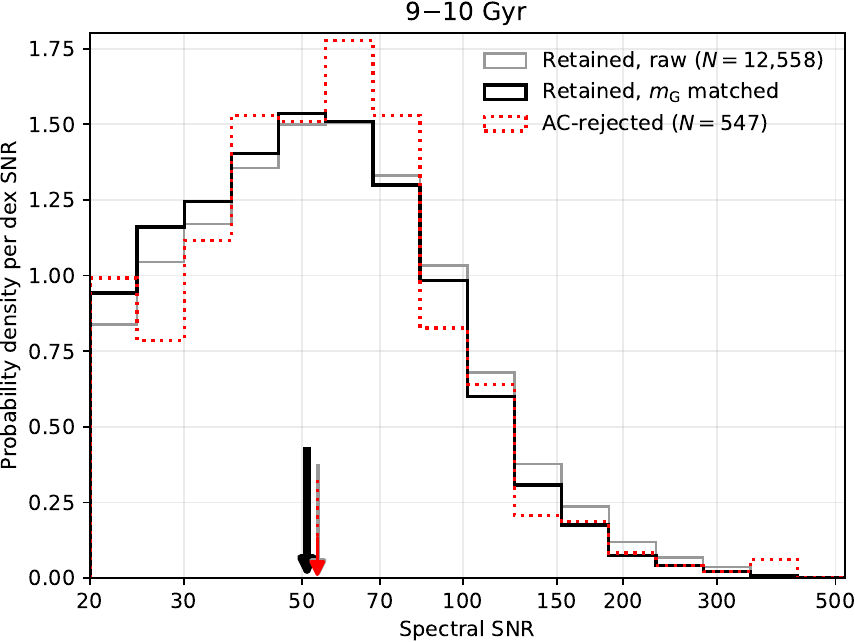} 
    \hfill
    \includegraphics[width=0.49\linewidth]{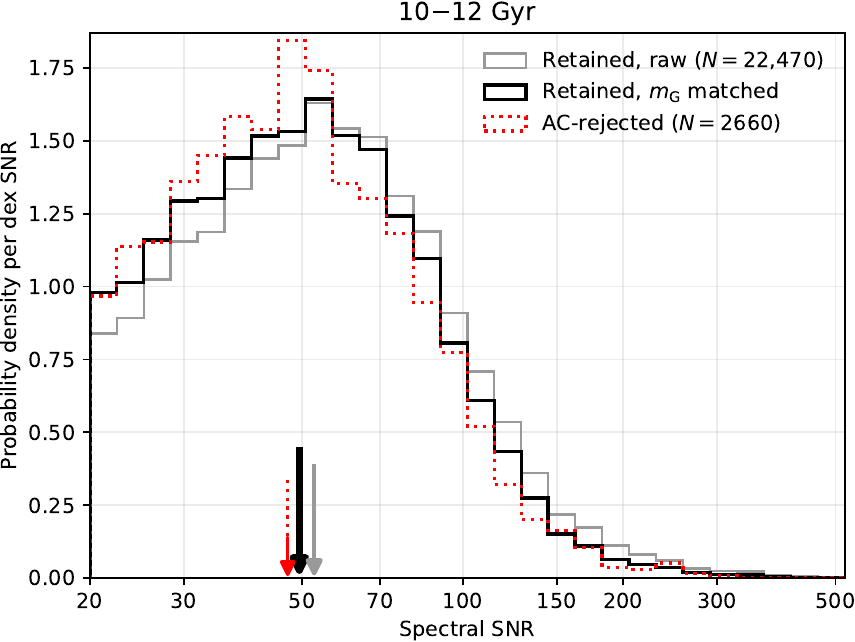}
    \vspace{4pt}
    \includegraphics[width=0.49\linewidth]{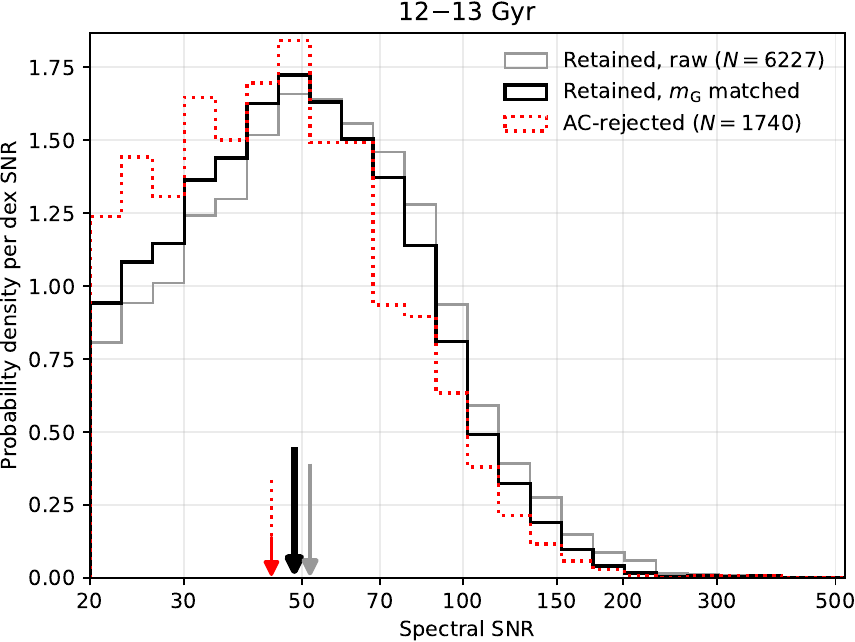}
    \hfill
    \includegraphics[width=0.49\linewidth]{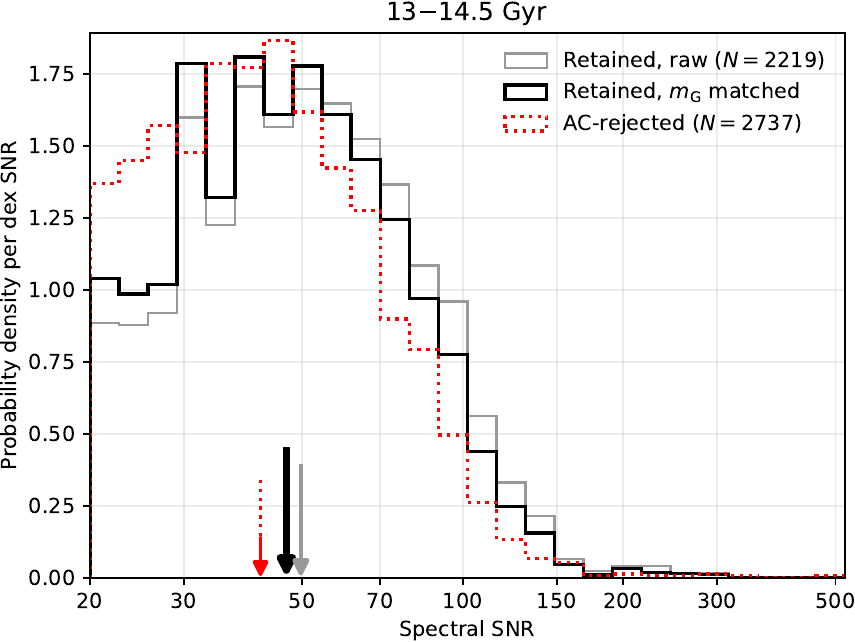}
    \caption{The SNR distribution of the AC-rejected sample (dotted red) and of the retained sample, either with no weighting (thin solid grey) or after weighting in $m_\mathrm{G}$ (solid black) to match the $m_\mathrm{G}$ distribution of the somewhat fainter AC-rejected sample. Results are shown in four age bins as indicated in the title, with a logarithmic scale used for the SNR. We mark the median of each distribution with a downward arrow in the same colour and a similar style. The SNR distributions shown here underlie the SNR ratios at fixed percentiles shown in Figure~\ref{snr_ratio_vs_age}.}
    \label{snr_histograms}
\end{figure*}

The top left panel of Figure~\ref{snr_histograms} shows stars with $A = 9-10$~Gyr, which we can think of as a control. In this age range, all three distributions roughly coincide, as indicated by their very similar medians. As we get to higher ages, the SNR distribution for the retained stars does not appreciably change and their median SNR remains very close to 50. However, the AC-rejected distribution progressively shifts towards lower SNR, causing the median to drop from 53 at $9-10$~Gyr down to 42 at $13-14.5$~Gyr. This leads to a growing gap with the retained distribution, whose median SNR after magnitude-weighting has a much more modest decrease from 51 at $9-10$~Gyr down to 47 at $13-14.5$~Gyr. The gap between the retained and AC-rejected SNR distributions is therefore widest in the bottom right panel of Figure~\ref{snr_histograms}, which covers the $13-14.5$~Gyr range that largely determines $A_\star$. We note that part of the decrease in median SNR towards higher ages for the magnitude-weighted retained sample is due to these stars being weighted onto the somewhat fainter magnitude distribution of the AC-rejected stars. Given that part of the reason for the AC-rejected stars having a higher incidence of overestimated ages seems to be their lower SNR, it may well be that part of the problem is their higher $m_\mathrm{G}$. We stress that even for the AC-rejected sample at the highest ages, the median SNR is well above the imposed floor of 20 and the SNR distribution is not that different to the retained sample, preventing the SNR being directly used as a quality cut.

The SNR's magnitude dependence is already accounted for in Figure~\ref{snr_histograms}, but we might expect that the metallicity is also relevant due to spectral lines affecting the flux ratio between the \emph{Gaia} and LAMOST filters. Since the AC-rejected stars by construction have a different metallicity than retained stars of the same age, we cannot control for metallicity in the AC-rejected sample as a whole. Controlling for metallicity is possible by focusing on those stars which pass the age--metallicity ceiling but are rejected by the age-dependent [$\alpha$/Fe] floor (c.f. Table~\ref{jz_cold_tail_table}). These stars have a median [Fe/H] of $-0.82$, which is practically identical to the $-0.79$ of the retained stars.

Using this technique to match in both $m_\mathrm{G}$ and [Fe/H] for stars with $A = 13-14.5$~Gyr, we find that $\widetilde{S}_{50} = 0.86 \pm 0.04$, which is $<1$ at $3.7\sigma$ confidence. If we expand the age range to $11-14.5$~Gyr, the sample size more than doubles and $\widetilde{S}_{50} = 0.84 \pm 0.01$, confirming that the AC-rejected stars indeed have a lower median SNR. The signal is slightly stronger than in the main article, where we control for age and magnitude but not metallicity (Figure~\ref{snr_ratio_vs_age}). We note that once we have controlled for $m_\mathrm{G}$, additionally controlling for metallicity only shifts $\widetilde{S}_{50}$ by 0.002, which is not distinguishable given the uncertainties.

These results confirm that the AC-rejected stars have a median SNR about 15\% below that of the retained stars for reasons unrelated to differences in $m_\mathrm{G}$ or metallicity. Since we compare the SNR in narrow age bins, differences in age are likewise not responsible. We argue instead that stars with lower SNR spectra are more likely to fail the AC cuts due to overestimated ages.

\end{appendix}

\bsp 
\label{lastpage}
\end{document}